\documentclass[epj,final,numbook]{svjour}

\usepackage{amsmath}
\usepackage{amssymb}
\usepackage{epsfig}
\usepackage{graphicx}
\usepackage{color}

\newcommand{\si}{\sigma}
\newcommand{\al}{\alpha}
\newcommand{\tet}{\theta}
\newcommand{\az}{\varphi}
\newcommand{\ro}{\rho}

\newcommand{\be}{\beta}
\newcommand{\ga}{\gamma}

\newcommand{\tro}{\tilde{\rho}}
\newcommand{\na}{\nabla}
\newcommand{\vsi}{\mbox{\boldmath{$\sigma$}}}
\newcommand{\vna}{\mbox{\boldmath{$\na$}}}
\newcommand{\Del}{\Delta}

\newcommand{\oeq}{\begin{equation}}
\newcommand{\ceq}{\end{equation}}
\newcommand{\oeqn}{\begin{eqnarray}}
\newcommand{\ceqn}{\end{eqnarray}}

\renewcommand{\>}{\rangle}
\newcommand{\<}{\langle}
\renewcommand{\(}{\left(}
\renewcommand{\)}{\right)}
\renewcommand{\[}{\left[}
\renewcommand{\]}{\right]}

\newcommand{\stf}{\,\,\,}
\newcommand{\sdf}{\,\,}
\newcommand{\stb}{\!\!\!}
\newcommand{\sdb}{\!\!}

\newcommand{\kfi}{|\phi \>}

\newcommand{\ksp}{|s'\>}
\newcommand{\kvac}{|-\>}

\newcommand{\bfi}{\<\phi |}

\newcommand{\bs}{\< s |}

\newcommand{\oQ}{\hat{Q}}

\newcommand{\oP}{\hat{P}}
\newcommand{\oM}{\hat{M}}
\newcommand{\oH}{\hat{H}}

\newcommand{\oV}{\hat{V}}

\newcommand{\oD}{\hat{D}}

\newcommand{\oro}{\hat{\rho}}
\newcommand{\oh}{\hat{h}}

\newcommand{\osi}{\hat{\sigma}}
\newcommand{\ovr}{\hat{\bf r}}
\newcommand{\ovp}{\hat{\bf p}}
\newcommand{\ovP}{\hat{\bf P}}
\newcommand{\ovR}{\hat{\bf R}}
\newcommand{\oT}{\hat{T}}

\newcommand{\op}{\hat{p}}

\newcommand{\of}{\hat{f}}

\newcommand{\oad}{\hat{a}^\dagger}
\newcommand{\oa}{\hat{a}}
\newcommand{\oA}{\hat{A}}
\newcommand{\oB}{\hat{B}}
\newcommand{\oF}{\hat{F}}
\newcommand{\oN}{\hat{N}}
\newcommand{\oX}{\hat{X}}
\newcommand{\oY}{\hat{Y}}
\newcommand{\oZ}{\hat{Z}}

\newcommand{\oL}{\hat{L}}
\newcommand{\ovsi}{\hat{\boldsymbol{\sigma}}}
\newcommand{\odel}{\hat{\delta}}
\newcommand{\ovk}{\hat{\bf k}}

\newcommand{\ov}{\hat{v}}

\newcommand{\del}{\delta\!}

\renewcommand{\d}{{\mbox d}}

\newcommand{\hb}{\hbar}

\newcommand{\vr}{{\bf r}}

\newcommand{\vj}{{\bf j}}
\newcommand{\vT}{{\bf T}}
\newcommand{\vJ}{{\bf J}}

\newcommand{\vR}{{\bf R}}

\newcommand{\vV}{{\bf V}}
\newcommand{\vW}{{\bf W}}
\newcommand{\vC}{{\bf C}}

\newcommand{\vS}{{\bf S}}

\newcommand{\ve}{{\bf e}}

\newcommand{\mH}{{\mathcal{H}}}

\newcommand{\mL}{{\mathcal{L}}}

\newcommand{\Tr}{\mathrm{Tr}}
\newcommand{\tr}{\mathrm{tr}}

\definecolor{gris}{gray}{0.3}

\begin{document}

\title{Nuclear Quantum Many-Body Dynamics}
\subtitle{From Collective Vibrations to Heavy-Ion Collisions}
\headnote{Review article}
\author{C\'edric Simenel}
\institute{CEA, Centre de Saclay, IRFU/Service de Physique Nucl\'eaire, F-91191 Gif-sur-Yvette, France, and\\  
 Department of Nuclear Physics, RSPE, Australian National 
University, Canberra, ACT 0200, Australia\\
\email{cedric.simenel@anu.edu.au}
}
\date{Received: date / Revised version: date}
\dedication{This review article is dedicated to the memory of Paul Bonche who pioneered the application of the time-dependent Hartree-Fock theory to nuclear systems.}
\abstract{\normalfont\small
A summary of recent researches on nuclear dynamics with realistic microscopic quantum approaches is presented. 
The Balian-V\'en\'eroni variational principle is used to derive the time-dependent Hartree-Fock (TDHF) equation describing the dynamics at the mean-field level, as well as an extension including small-amplitude quantum fluctuations which is equivalent to the time-dependent random-phase approximation (TDRPA). 
Such formalisms as well as their practical implementation in the nuclear physics framework with modern three-dimensional codes are discussed.
Recent applications to nuclear dynamics, from collective vibrations to heavy-ion collisions are presented. 
A particular attention is devoted to the interplay between collective motions and internal degrees of freedom.
For instance, the harmonic nature of collective vibrations is questioned.
Nuclei are also known to exhibit superfluidity due to pairing residual interaction. 
Extensions of the theoretical approach to study such pairing vibrations are now available. 
Large amplitude collective motions are investigated in the framework of heavy-ion collisions leading, for instance, to the formation of a compound system. 
How fusion is affected by the internal structure of the collision partners, such as their deformation, is discussed. 
Other mechanisms in competition with fusion, and responsible for the formation of  fragments which differ from the entrance channel (transfer reactions, deep-inelastic collisions, and quasi-fission)  are investigated. 
Finally, studies of actinide collisions forming, during very short times of few zeptoseconds, the heaviest nuclear systems available on Earth, are presented. 
}
\maketitle
\tableofcontents

\section{Introduction}

Nuclei are ideal to investigate fundamental  aspects of the quantum many-body problem. 
They exhibit collective motions built from coherent superpositions of the states of their constituents.
Examples range from collective vibrations to the formation of a compound system in collisions.
These features are common to other composite systems (atomic clusters, molecules, Bose-Einstein condensates...).
Their study in nuclear systems is obviously part of a wider physics field. 

An interesting feature of the dynamics of quantum many-body systems is the interplay between collective motion and single-particle degrees of freedom.
The latter is a source of complex and fascinating behaviours.
For instance, giant resonances are characterised by a collective vibration of many particles, but their  decay may occur  by the emission of a single nucleon.
Another example could be taken from the collision of composite systems where the transfer of few particles may have a strong impact on the final outcome of the reaction.

To describe these complex systems, one needs to solve the quantum many-body problem. 
The description of the dynamics of composite systems can be very challenging, especially when two such systems interact. 
An important goal of nuclear physics is then to find a unified way to describe the dynamics of nuclear systems.
Ultimately, the same theoretical model should be able to describe  vibration, rotation, fission, all the possible outcomes of heavy-ion collisions (elastic and inelastic scattering, particle transfer, fusion, and multifragmentation), and even the dynamics of neutron star crusts.

This desire for a global approach to nuclear dynamics has strongly influenced past research activities.
Beside the quest for a unified model of nuclear dynamics, possible applications of heavy-ion collisions to the formation of nuclear systems in extreme conditions (new exotic nuclei, large deformation and angular momentum, super-heavy elements...) are also strong motivations for the study of reaction mechanisms to optimise their production cross-sections. 

The Balian-V\'en\'eroni variational principle is a starting point leading to well-know formalisms, such as the time-dependent Hartree-Fock (TDHF) theory and the time-dependent random-phase approximation (TDRPA). 
Thanks to the recent increase of computational power, these approaches have been applied to investigate several aspects of the nuclear dynamics with modern numerical codes.
Recent applications to collective vibrations and heavy-ion collisions, as well as the underlying formalism and numerical details are discussed in this review article. 

The formalism and numerical developments are presented in section~\ref{chap:formalism}.
Studies of collective vibrations are then collected in section~\ref{chap:vib}, while heavy-ion collisions are investigated in section~\ref{chap:HIC}.

\section{The time-dependent Hartree-Fock theory and its extensions\label{chap:formalism}}

\subsection{Introduction}

The quantum many-body problem is common to many theoretical fields~\cite{neg98,bla86}.
It aims at describing the structure and dynamics of interacting particles. 
Electrons, atoms, molecules and nucleons are usual constituents of quantum many-body systems. 

In the non-relativistic regime, these systems obey the Schr\"odinger equation. 
\oeq
i \frac{d|\Psi(t)\>}{dt} = \oH |\Psi(t)\>,
\label{eq:schrod}
\ceq
where $|\Psi(t)\>$ describes the state of the many-body system and $\oH$ is the microscopic Hamiltonian. 
We use the notation $\hb\equiv1$.
This equation can be solved exactly for simple cases only. 
Generally, one has to rely on some approximations.

Variational principles offer an elegant starting point to build such approximations.
Indeed, they ensure an optimization of the equations of motion under the approximation that the variational space is limited to a sub-space of the full Hilbert (or Fock) space. 
Of course, without any restriction of the variational space, it is required that the chosen variational principle allows to recover the Schr\"odinger equation. 
However, their usefulness appears when restricting the variational space. 
Then, the validity of the approximation relies entirely on the choice of the variational space. 
In one hand, the latter has to be small enough so that the problem is numerically tractable.
In the other hand, the variational space should contain the relevant degrees of freedom to allow for a realistic description of the physical processes. 

Although the Schr\"odinger equation is unique,  different variational principles have been developed in the past. One of the mostly used action is
\oeq
S[t_0,t_1;\Psi(t)]=\int_{t_0}^{t_1}dt \stf \<\Psi(t)| \(i \frac{d}{dt}-\oH \)|\Psi(t)\>.
\label{eq:SDirac}
\ceq
The variational principle $\delta S=0$ is applied with the boundary conditions $\delta \Psi(t_0) = \delta \Psi(t_1) = 0$. 
If $\Psi$ is allowed to span the entire Hilbert space, one recovers the Schr\"odinger equation. 

In most practical applications, mean-field models are considered in a first approximation, 
and, eventually, serve as a basis for beyond-mean-field approaches~\cite{sim10a,lac04}. 
To construct such a mean-field theory from the above variational principle, one restricts the variational space by assuming that the $N$ particles (we consider fermions) are independent. 
In this case, they may be described by a Slater determinant
\oeq
\kfi=\prod_{i=1}^N\oad_i\kvac,
\ceq 
where $\oad_i$ creates a particle in the state $|\az_i\>$ when it acts on the particle vacuum~$\kvac$.
In such a state, all the information is contained in the one-body density-matrix $\ro$ 
associated to the single-particle operator 
\oeq
\oro=\sum_{i=1}^N|\az_i\>\<\az_i|.
\ceq
Solving the variational principle where the action defined in Eq.~(\ref{eq:SDirac}) is required to be stationary in the subspace of Slater determinants $\kfi$ with fixed boundary conditions at times $t_0$ and $t_1$ leads to (see appendix~\ref{annexe:standardTDHF})
\oeq 
i\frac{\partial \ro}{\partial t}=\[h[\ro],\ro\],
\label{eq:tdhf}
\ceq
where $h[\ro]$ is the Hartree-Fock (HF) single-particle Hamiltonian with matrix elements 
\oeq 
h_{\al\be}=\frac{\delta \bfi \oH \kfi}{\delta \ro_{\be\al}},
\label{eq:hHF}
\ceq
and 
\oeq
\ro_{\al\be}=\<\az_\al|\oro|\az_\be\>=\bfi \oad_\be\oa_\al \kfi.
\label{eq:roalbe}
\ceq
Eq.~(\ref{eq:tdhf}) is the  time-dependent Hartree-Fock (TDHF) equation. It was obtained by Dirac in 1930~\cite{dir30}.
It provides a self-consistent mean-field evolution where the interaction between the particles is replaced by a one-body mean-field potential generated by all the particles. 
It is, then, assumed that each particle evolves independently in this potential. 

We may question the validity of such a mean-field approximation in the nuclear physics context. 
This assumption could only give an approximation of the exact dynamics and we do expect, in 
general, that the system will deviate from the independent particle picture.
Indeed, the exact dynamics  is given by the time-dependent 
Schr\"odinger equation~[Eq.~(\ref{eq:schrod})], and, unless the Hamiltonian 
contains one-body operators only, the mean-field theory can only 
approximate the exact evolution of the system. 
The exact and mean-field Hamiltonians differ by the residual interaction.
The validity of the mean-field approximation depends, then, on the intensity of the residual interaction. 
The latter is a function of the state $\left| \Psi \right\rangle$ and, therefore, depends significantly  
on the physical situation. 

Starting from simple arguments \cite{lic76}, 
the time $\tau$ over which the Slater determinant picture breaks down could be expressed as:
\begin{eqnarray}
\tau &=& \frac{\hbar}{2}  \Big(\frac{1}{N}
\sum_{\bar \alpha \bar \beta \alpha \beta} 
|\left\langle \bar \alpha \bar \beta \left| \bar v \right| \alpha \beta \right\rangle|^2 \Big)^{-1/2},
\end{eqnarray} 
where $N$ is the number of particles, and $\alpha$ and $\beta$ denote hole states while $\bar{\alpha}$ and $\bar{\beta}$ denote particle states. 
In the nuclear physics context, typical values of the residual interaction lead to $\tau \simeq 100-200$~fm/$c$.
This gives an estimate of the time during which one can safely consider that the independent particle approximation is valid. 
For longer times, as encountered, e.g., in heavy-ion collisions, the mean-field approximation can only be verified by comparison with experiment and/or beyond mean-field calculations. 

One important aspect of the TDHF approach is the treatment of one-body dissipation mechanisms. 
In particular, one-body dissipation is crucial to properly describe low-energy heavy-ion collisions. 
Two kind of one-body dissipation mechanisms can easily be identified:
\begin{itemize}
\item {\it Coupling of collective motions with one-particle one-hole (1p1h) excitations.} In the case of giant resonances, this coupling leads to the so-called Landau damping (see Sec.~\ref{chap:vib}). In heavy-ion collisions, the field of the  collision partner generates a distortion of the single-particle wave-functions. This is particularly true when part of the wave functions are transferred from one fragment to another, leading to an excitation of the fragments, and, then, to a dissipation of the translational kinetic energy. Another example is the case of fusion reactions. Multiple reflexions of single-particle wave-functions on the wall of the mean-field dissipate original collective translational kinetic energy into single-particle excitations and collective vibrations of the compound system.
\item {\it Single-particle wave-function emission to the continuum.}  Emission of nucleons into the continuum is a natural cooling mechanism of excited nuclei. As an example, the direct decay by nucleon emission induces an escape width of giant resonances (see Sec.~\ref{sec:decay}).
\end{itemize}

Naturally, extensions of TDHF including pairing and in-medium two-body correlations should be considered, at least to determine the validity of the mean-field approximation in terms of energy and simulation time. 
Larger variational spaces could then be considered in order to incorporate some effects of the residual interaction which are missing at the TDHF level. 
This is the case, for instance, with the inclusion of pairing correlations by taking a variational space of quasi-particle vacua, leading to the time-dependent Hartree-Fock-Bogoliubov (TDHFB) formalism~\cite{bla81}. 
Another possible approach to include beyond mean-field correlations is to consider different levels of truncation of the Bogoliubov-Born-Green-Kirwood-Yvon (BBGKY) hierarchy~\cite{bog46,bor46,kir46}. 
The two first equations of this hierarchy read
\begin{equation}
\left\{ 
\begin{array}{cl}
i\hbar \frac{\partial }{\partial t}\rho _{1}
=&\left[ t_1,\rho _{1}\right] + \frac{1}{2}{\rm \Tr}_{2}\left[ \bar v_{12},\rho_{12}\right]  \\ 
&  \\ 
i \hbar \frac{\partial }{\partial t}\rho_{12} =& [t_1 + t_2 + \frac{1}{2} \bar v_{12}, \rho_{12}] 
+ \frac{1}{2} \Tr_3 \left[ \bar v_{13} + \bar v_{23} , \rho_{123} \right], \\ 
\end{array}
\right.  \label{eq:BBGKY}
\end{equation}
where $\rho_1$, $\rho_{12}$ and $\rho_{123}$ are the one-, two-, and three-body density matrices, respectively.
We see that $\rho_1$ and $\rho_{12}$ are coupled. 
In fact, each equation of the BBGKY hierarchy couples $\rho_{1\cdots k}$ to $\rho_{1\cdots k+1}$ and forms a closed system of coupled equations equivalent to the Schr\"odingier equation for a finite many-body system. 

The two-body  density matrix can be expressed as $\rho_{12}= \rho_1\rho_2 (1-P_{12})+C_{12}$ where $P_{12}$ is the permutation operator between particles 1 and 2, and $C_{12}$ is the two-body correlation matrix describing, e.g., pairing correlations and in-medium two-body collisions. 
Neglecting $C_{12}$ in the first equation of the BBGKY hierarchy leads to the TDHF equation [Eq.~(\ref{eq:tdhf})]. 
One could also include these two-body correlations by solving the coupled equations (\ref{eq:BBGKY})
and neglecting the three-body correlations.
The resulting set of coupled equations for the evolution of $\rho_1$ and $\rho_{12}$ is known as the time-dependent density-matrix (TDDM) formalism\footnote{Note that the TDDM formalism is not obtained by solving  a variational principle.}. 
It is obvious that solving the TDDM equations request much more computational efforts than standard TDHF calculations.
Only few applications have been made so far  \cite{cas90,bla92,luo99,toh01,toh02a,toh02b,ass09}. 

The difficulties of solving quantum many-body models have  sometimes been overcome by using their semi-classical limit.
The semi-classical limit of the TDHF approach is the Vlasov equation which
 is obtained by taking the Wigner transform of Eq.~(\ref{eq:tdhf}) and keeping only the first order in $\hbar$.
The in-medium two-body collision term can then be added, leading to the Landau-Vlasov equation which is the root of the Vlasov-Uehling-Uhlenbeck (VUU) \cite{kru85} and Boltzmann-Uehling-Uhlenbeck (BUU) \cite{aic85} transport theories. 
The latter approaches are common tools to interpret heavy-ion collision data at intermediate energy where the collision term is expected to play a significant role.  
However, at  energies closer to the barrier, the collision term is expected to be hindered by the Pauli blocking. 
In addition,  semi-classical approaches lead to a poor description of the ground state structure of the nuclei, such as their binding energies and deformations.
As one of the goal of the present review article is to study the interplay between nuclear structure and dynamics, it is a natural choice to focus on fully quantum approaches only and semi-classical models won't be further discussed here.

Several other variational principles have been proposed, depending on the physics one wants to address. 
For instance, Blaizot and Ripka introduced in 1981 a variational principle appropriate to the calculation of transition amplitudes~\cite{bla81}. 
Inspired by this work, Balian and V\'en\'eroni (BV) presented the same year their variational principle for the expectation value of an observable~\cite{bal81}. 
In the latter, both the state of the system and the observable of interest are allowed to vary in their own variational space. 

In particular, the BV variational principle allows a more transparent interpretation of the TDHF theory~\cite{bal81}: 
it is shown that the TDHF equation~(\ref{eq:tdhf}) is optimised for the {\it expectation value of one-body observables}. 
TDHF calculations are indeed successful to predict such quantities (several examples are presented in the following sections). 
It also explains why TDHF sometimes fail to reproduce other quantities such as fluctuations of one-body operators~\cite{koo77,dav78,das79}, which are outside the variational space used to derive the TDHF equation.
In fact, to predict such fluctuations, Balian and V\'en\'eroni proposed a prescription also based on their variational principle, but with a different variational space for the evolution of the observable~\cite{bal84,bal92}.

In the following part of this section, we first describe the Balian-V\'en\'eroni variational principle.
The latter is  used to derive the TDHF equation as well as fluctuations and correlations of one-body observables. 
Then, the Skyrme energy density functional describing the strong interaction between the nucleons is introduced. 
Numerical aspects are also discussed.
Finally, perspectives for beyond TDHF calculations are presented. 

\subsection{The Balian-V\'en\'eroni variational principle}

The BV variational principle has been applied to different problems 
in nuclear physics~\cite{tro85,mar85,bon85,zie88,bro08,bro09,sim11}, 
hot Fermi gas~\cite{mar91}, $\phi^4$-theory~\cite{mar95,mar99}, and Boson systems~\cite{ben99,bou10}.
The first realistic application of the BV prescription to fragment mass and charge distributions in heavy-ion collisions are reported in Ref.~\cite{sim11} and will be presented in section~\ref{sec:DIC}.
The importance of the BV variational principle for the interpretation of the TDHF theory, which will be discussed and applied in the following sections, as well as the derivation of the BV prescription for fluctuations and correlations of one-body observables, justify the more detailed discussion in this subsection. 

Let us first define two variational quantities: $\oD(t)$, describing the state of the system, and $\oA(t)$, describing the evolution of the observable in the Heisenberg picture. 
The application of the BV variational principle requires two boundary conditions:
\oeq
\oD(t_0)=\oD_0,
\label{eq:BCD}
\ceq
where the initial state of the system $\oD_0$ is assumed to be known, and
\oeq
\oA(t_1) = \oA_1,
\label{eq:BCA}
\ceq
where $\<\oA_1\>$ is the final expectation value we want to compute at $t_1>t_0$. 

The action-like quantity defined by Balian and V\'en\'eroni reads~\cite{bal81}
\oeqn
&J& = \Tr\[ \oA(t_1)\oD(t_1) \] \nonumber\\
&-& \int_{t_0}^{t_1} \stb dt \,\Tr\!\[ \oA(t) \(\frac{d\oD(t)}{dt} + i[\oH(t),\oD(t)]\) \]\sdb . 
\label{eq:JA}
\ceqn
 We see that, imposing $\delta_A J=0$, where $\delta_A$ induces small variations of $\oA(t)$, leads to the Liouville-Von Neumann equation
\oeq
i\frac{d\oD(t)}{dt}=\[\oH,\oD(t)\]
\label{eq:Liouville}
\ceq
which is fully equivalent to the Schr\"odinger equation. 

{
To get Eq.~(\ref{eq:Liouville}), we first note that, according to the boundary condition in Eq.~(\ref{eq:BCA}), $\oA(t_1)$ is fixed and we get
$$\delta_AJ= - \int_{t_0}^{t_1} dt\stf \Tr\[ \delta_A\oA(t)\, \(\frac{d\oD(t)}{dt} + i[\oH(t),\oD(t)]\) \].$$
To be equal to zero for any variation of $\oA$, the term inside the brackets must be zero, giving Eq.~(\ref{eq:Liouville}).
}

Variations of $\oD(t)$ should also be considered. 
It is easier to first rewrite Eq.~(\ref{eq:JA}) as
\oeqn
&J& = \Tr\[ \oA(t_0)\oD(t_0) \] \nonumber\\
&+& \int_{t_0}^{t_1} \stb dt\, \Tr\[ \oD(t) \(\frac{d\oA(t)}{dt} + i[\oH(t),\oA(t)]\) \]\sdb .
\label{eq:JD}
\ceqn
{
To get Eq.~(\ref{eq:JD}) we integrate by part the $\int dt \oA\partial_t\oD$ term in Eq.~(\ref{eq:JA}) and we use the relation $$\Tr(\oA[\oH,\oD])=-\Tr(\oD[\oH,\oA]).$$}

Equivalently to Eq.~(\ref{eq:Liouville}), with the boundary condition in Eq.~(\ref{eq:BCD}), requiring $\delta_DJ=0$ leads to 
\oeq
i\frac{d\oA(t)}{dt}=\[\oH,\oA(t)\],
\label{eq:Ehrenfest}
\ceq
which is also equivalent to the Schr\"odinger equation, and is expressed in the Heisenberg picture. 

\subsection{Derivation of the time-dependent Hartree-Fock equation}

The TDHF theory is obtained under the approximation that $\oA(t)$ is constrained to be a one-body operator  and that $\oD(t)$ is an independent particle state for all $t$. 
As a result, TDHF is optimised for the expectation value of one-body operators. 
On the contrary, other quantities, such as expectation values of two-body operators, 
 may be not well predicted because they are outside the  variational space of $\oA(t)$. 

To get the TDHF equation, we impose that the variation  $\delta_A$  leaves $\oA$ in the space of one-body operators. 
As we consider arbitrary variations, we can choose 
\oeq
\delta_A\oA(t)\equiv\oad_\al\oa_\be \mbox{ for } t_0\le t<t_1
\ceq
and $\delta\oA(t_1)=0$ due to the boundary contition in Eq.~(\ref{eq:BCA}).
Requiring $\delta_AJ=0$, we get from Eq.~(\ref{eq:JA}) 
\oeq
\Tr\[\oad_\al\oa_\be\(\frac{d\oD}{dt}+i[\oH,\oD]\)\]=0.
\ceq
In addition, the state of the system is constrained to be an independent particle state.
The variational space for $\oD(t)$ is then defined by $\oD(t)=|\phi(t)\>\<\phi(t)|$ where $|\phi(t)\>$ is a Slater determinant. 
Using the one-body density matrix defined in Eq.~(\ref{eq:roalbe}), we get
\oeq
i\frac{d\ro_{\be\al}(t)}{dt}=\<\phi(t)|\[\oad_\al\oa_\be,\oH\]|\phi(t)\>.
\label{eq:idrodt}
\ceq

Consider a Hamiltonian of the form
\oeq
\oH = \sum_{ij} \sdf t_{ij}\sdf \oad_i \oa_j + \frac{1}{4} \sum_{ijkl} \sdf  \bar{v}_{ijkl} \sdf \oad_i \oad_j \oa_l \oa_k 
\label{eq:oH2}
\ceq
where matrix elements associated to the kinetic energy and to the anti-symmetric two-body interaction are given, respectively, by 
\oeqn
t_{ij} &=& \frac{1}{2m} \sdf \<i | \op^2 | j \>\stf \mbox{ and }  \label{eq:tij}\\
\bar{v}_{ijkl} &=& v_{ijkl} - v_{ijlk}. \label{eq:vbar}
\ceqn

Reporting the Hamiltonian expression  
[Eq.~(\ref{eq:oH2})] in Eq.~(\ref{eq:idrodt}), we get
\oeqn
i\frac{d\ro_{\be\al}}{dt} &=& \sum_{kl} t_{kl} \< [\oad_\al\, \oa_\be , \oad_k \,\oa_l ]\,\>\nonumber\\
&+& \frac{1}{4}  \sum_{klmn}  \bar{v}_{klmn}  \< \, [  \oad_\al\, \oa_\be ,   \oad_k \,\oad_l \,\oa_n \,\oa_m ]\,\>,
\label{eq:erhenfest_detail}
\ceqn
where the time variable has been omitted and $\<\cdots\>$ denotes the expectation value on $|\phi(t)\>$ to simplify the notation.
Eq.~(\ref{eq:erhenfest_detail}) leads to the TDHF equation
 \oeq
i \frac{d{\ro}_{\be\al}}{dt} =\[ h[\ro], \ro \]_{\be\al} .
 \label{eq:TDHFbeal}
\ceq
The single-particle Hartree-Fock Hamiltonian reads
\oeq
h[\ro]=t+U[\ro]
\label{eq:h}
\ceq
with the self-consistent mean-field 
\oeq
U[\ro]_{ij} = \sum_{kl} \sdf\bar{v}_{ikjl} \sdf \ro_{{lk}}.
\label{eq:Uro}
\ceq

{
To show the equivalence between Eq.~(\ref{eq:erhenfest_detail}) and the TDHF equation, let us start with the  term associated to the kinetic energy:
\oeqn
\< \, [ \, \oad_i \, \oa_j \, , \, \oad_k \, \oa_l \, ] \, \>
&=& \del_{jk} \,  \<  \oad_i \, \oa_l \> -  \<  \oad_i \, \oad_k \, \oa_j  \,  \oa_l \>  \nonumber \\
&& - \del_{il} \,  \<  \oad_k \, \oa_j \>+   \<  \oad_k \, \oad_i \, \oa_l  \,  \oa_j \>  \nonumber \\
&=& \del_{jk} \,  {\ro}_{li} - \del_{il} \,  {\ro}_{jk} .\nonumber
\ceqn
The kinetic energy term reduces to 
$$
\sum_{kl} \sdf t_{kl} \< [\oad_i\, \oa_j\sdf ,\sdf \oad_k \,\oa_l ]\,\>
 = \sum_k \sdf \( t_{jk} \sdf {\ro}_{ki} - t_{ki} \sdf {\ro}_{jk} \)=\[t,\rho\]_{ji}.
\label{eq:terme_cinetique}
$$
For the two-body interaction, we need the expectation value of the commutator  

\oeqn
&&\stb\stb\stb\stb\stb\stb\stb\stb\<\, [ \, \oad_i \, \oa_j \, , \, \oad_k \, \oad_l  \, \oa_n \, \oa_m \, ] \, \>\nonumber\\
&=&\<  \oad_i \, \oad_l  \, \oa_n \, \oa_m  \> \sdf \del_{jk} -  \<  \oad_i \, \oad_k  \, \oa_n \, \oa_m  \> \sdf\del_{jl} 
\nonumber\\
&&+ \<  \oad_i \, \oad_k \, \oad_l  \, \oa_j \, \oa_n \, \oa_m  \>  
 - \<  \oad_k \, \oad_l  \, \oa_n \, \oa_j  \> \sdf \del_{mi} \nonumber\\
 && +  \<  \oad_k \, \oad_l  \, \oa_m \, \oa_j  \> \sdf \del_{ni} 
- \<  \oad_k \, \oad_l \, \oad_i  \, \oa_n \, \oa_m \, \oa_j  \>. \nonumber 
\label{eq:groscom}
\ceqn
The two terms with 6 annihilation/creation operators cancel out and we get
\oeqn
&&\stb\stb\stb\stb\stb\< \, [ \, \oad_i \, \oa_j \, , \, \oad_k \, \oad_l  \, \oa_n \, \oa_m \, ] \, \>\nonumber\\
&=& \( \ro_{{mi}}  \ro_{{nl}} -  \ro_{{ml}}  \ro_{{ni}} \) \del_{jk} + \( \ro_{{mk}}  \ro_{{ni}} -  \ro_{{mi}}  \ro_{{nk}}  \) \del_{jl} \nonumber \\
&&+ \( \ro_{{jl}}  \ro_{{nk}} -  \ro_{{jk}}  \ro_{{nl}} \) \del_{mi} + \( \ro_{{jk}}  \ro_{{ml}} -  \ro_{{jl}}  \ro_{{mk}}  \) \del_{ni} \nonumber 
\ceqn
Altogether, the two-body interaction contribution reduces to 
\oeqn
&&\stb\stb\stb\stb\stb\stb \stb \stb \frac{1}{4}\sum_{klmn}   \bar{v}_{klmn}  \< \, [  \oad_i\, \oa_j ,   \oad_k \,\oad_l \,\oa_n \,\oa_m ]\,\> \nonumber\\
&=& 
\frac{1}{2}  \sum_{klm}  \[ \bar{v}_{jklm} \( \ro_{{li}} \ro_{{mk}} - \ro_{{lk}}  \ro_{{mi}} \) \right.\nonumber \\
& &\left.\stf \stf \stf \stf \stf  +    \bar{v}_{klim} \( \ro_{{jl}} \ro_{{mk}} - \ro_{{jk}}  \ro_{{ml}}  \) \]\nonumber\\
&=& \sum_{klm} \[ \bar{v}_{jklm}  \(\ro_{{li}} \ro_{{mk}} \)- \bar{v}_{klim} \( \ro_{{jk}} \ro_{{ml}}\) \]\nonumber\\
&=&\sum_{k}  \( U[\ro]_{jk}  \ro_{{ki}} - U[\ro]_{ki}  \ro_{{jk}} \)=\[U[\rho],\rho\]_{ji}\nonumber
\ceqn
where we have used $\bar{v}_{klmn} = - \bar{v}_{klnm} = - \bar{v}_{lkmn}$.
Gathering the kinetic and interaction terms gives Eq.~(\ref{eq:TDHFbeal}).

We see that the TDHF equation is obtained by solving the BV variational principle with the variational spaces restricted to Slater determinants for the state of the system and to one-body operators for the observable. 
It is interesting to see that only the variation of $\oA$ is needed to get the TDHF equation. 
In addition, solving the TDHF equation allows to compute any one-body observable, and the equation does not depend on the final time $t_1$ entering the BV action. 
These properties are specific to the TDHF case. 
In a more general case, the resulting equations of motion are obtained from both the variation of $\oA$ and $\oD$, and the results are valid for only one observable $\oA_1$ and one final time $t_1$.

We recall that the TDHF equation is  optimised for the expectation value of one-body operators and its predictive power for other purposes may be questionable. 
In particular, two-body operators and fluctuations of one-body operators are outside the variational space of the observable used to derive the THDF equation.  
In the next section, we solve the BV variational principle in order to compute such fluctuations. (The case of more general two-body operators would be more complicated and the resulting equations of motions are not expected to be easily solvable numerically.) 

\subsection{Fluctuations and correlations of one-body observables \label{sec:fluccor}}

Let $\oX$ and $\oY$ be two observables. Their correlation in the state $|\Psi\>$ is defined as
\oeq
\sigma_{XY}=\sqrt{\frac{1}{2}\(\<\oX\oY\>+\<\oY\oX\>\)-\<\oX\>\<\oY\>}.
\ceq
The case $\oX=\oY$ defines the fluctuations of $\oX$:
\oeq
\sigma_{XX}\equiv\sigma_X=\sqrt{\<\oX^2\>-\<\oX\>^2}.
\ceq

If $\oX$ and $\oY$ are one-body operators, we see that $\sigma_{XY}^2$ includes the expectation value of the {\it square} of a one-body operator, which contains a two-body contribution.\\
{
Indeed,
\oeqn
\oX\oY&=&\sum_{\al\be}X_{\al\be}\oad_\al\oa_\be\sum_{\mu\nu}Y_{\mu\nu}\oad_\mu\oa_\nu\nonumber \\
&=&\sum_{\al\be\mu\nu} X_{\al\be} Y_{\mu\nu} (\delta_{\beta\mu}\oad_\al\oa_\nu-\oad_\al\oad_\mu\oa_\be\oa_\nu).\nonumber
\ceqn
The last term is clearly of a two-body nature. 
}
The TDHF theory is then not optimised for the prediction of correlations and fluctuations of one-body operators.
 
A possible improvement would be to solve the BV variational principle with a variational space for the observable which includes square of one-body operators~\cite{flo82}. This approach, however, leads to complicated equations of motion due to an intricate coupling between the evolution of the observable $\oA(t)$ and of the state $\oD(t)$. 

Fluctuations and correlations of one-body operators $\oQ_i=\sum_{\al\be}Q_{i_{\al\be}}\oad_{\al}\oa_\be$ can also be computed from the expectation value of 
\oeq
\oA_1\equiv e^{-\sum_i\varepsilon_i\oQ_i}
\ceq
in the limit $\varepsilon_i\rightarrow0$.
Indeed, 
\oeq
\ln\<\oA_1\> = -\sum_i\varepsilon_i\<\oQ_i\>+\frac{1}{2}\sum_{ij}\varepsilon_i\varepsilon_jC_{ij} +O(\varepsilon^3),\label{eq:lnA}
\ceq
where $C_{ij} =\sigma_{Q_iQ_j}^2$.
The linear and quadratic dependences in $\varepsilon$ of  $\ln\<\oA_1\>$ then lead to the expectation values and fluctuations/correlations of the one-body observables $\oQ_i$, respectively.

In Ref.~\cite{bal84}, Balian and V\'en\'eroni applied their variational principle with a variational space for the observable parametrised by
\oeq
\oA(t)=e^{-\oL(t)}=e^{-\sum_{\al\be}L_{\al\be}(t)\oad_\al\oa_\be}, 
\label{eq:defA}
\ceq
while the density matrix is constrained to be an independent particle 
state. 
For a pure state, the latter takes the form
\oeq
\oD(t) = |\phi(t)\>\<\phi(t)|,
\ceq
where $|\phi(t)\>$ is a Slater determinant. 
In fact, the original derivation of Balian and V\'en\'eroni~\cite{bal92} involves more general mean-field states of the form
\oeq
\oD(t) = e^{-m(t)-\oM(t)}=e^{-m(t)-\sum_{\al\be}M_{\al\be}(t)\oad_{\al}\oa_{\be}}.
\label{eq:defD}
\ceq
{
The particular case of a Slater determinant can be obtained with this parametrisation by letting the eigenvalues of the matrix $M(t)$ tend to $\pm\infty$, with the normalisation factor $m(t)$ also tending to $+\infty$ in such a way that the norm $z(t)=\Tr(\oD(t))$ is equal to~1~\cite{bal85}.
In this case, the Slater determinant $|\phi(t)\>$ is built from the eigenvectors of $M(t)$ associated with the eigenvalues~$-\infty$. }
Comparing Eqs.~(\ref{eq:defA}) and~(\ref{eq:defD}), we see that both variational spaces for the state and the observable are similar, being both composed of exponentials of one-body operators. 

The resolution of the BV variational principle with this choice of variational spaces and assuming the boundary conditions given in Eqs.~(\ref{eq:BCD}) and~(\ref{eq:BCA}) can be found in Refs.~\cite{bal92,bro09} 
and in Appendix~\ref{annexe:BV}.

Expanding the one-body density matrix in terms of $\varepsilon$,
\oeq
\rho(t)=\rho^{(0)}(t)+\rho^{(1)}(t)+O(\varepsilon^2),
\ceq
where $\rho^{(1)}$ is of the order $\varepsilon$,
the main results read:
\begin{itemize}
\item The expectation value of $\oQ_i$ is given by
\oeq 
\<\oQ_i\>(t_1)=\tr \(\ro^{(0)}(t_1)Q_i\)
\ceq
where $\rho^{(0)}$ is given by the TDHF equation~(\ref{eq:tdhf}) with the boundary condition
\oeq
\rho^{(0)}_{\al\be}(t_0)=\Tr\(\oD_0\oad_\be\oa_\al\)
\ceq
and $\oD_0$ is the initial density matrix. 
\item The fluctuations/correlations $\sigma_{Q_iQ_j}=\sqrt{C_{ij}}$ obey
\oeq
\sum_j \varepsilon_jC_{ij}(t_1)
=\sum_j \varepsilon_j C_{ij}^{TDHF}(t_1)-\tr\(\ro^{(1)}(t_1)Q_i\)
\label{eq:BVprescription1}
\ceq
where 
\oeqn
\stb \stb \stb C_{ij}^{TDHF}(t_1)&=&\frac{1}{2}\tr\(\ro^{(0)}(t_1)[Q_i,Q_j]\)\nonumber\\
&+&\tr\[Q_i\rho^{(0)}(t_1)Q_j(1-\rho^{(0)}(t_1))\]
\label{eq:CiiTDHF}
\ceqn
are the (square of the) fluctuations/correlations obtained from the standard TDHF approach. 
\end{itemize}
Eq.~(\ref{eq:BVprescription1}) gives fluctuations and correlations which differ from the standard TDHF result [Eq.~(\ref{eq:CiiTDHF})]. 
This is not surprising as only Eq.~(\ref{eq:BVprescription1}) is optimised for these fluctuations/correlations.
The result in Eq.~(\ref{eq:BVprescription1}) takes into account possible fluctuations around the TDHF mean-field evolution in the small amplitude limit, i.e., at the RPA level~\cite{bal84,bal92,ayi08}. 

The additional term in Eq.~(\ref{eq:BVprescription1}) involves $\rho^{(1)}$, i.e., the part of the one-body density matrix which is linear in~$\varepsilon$. However, the equation of motion for the latter is not trivial. Fortunately, for a Slater determinant, Eq.~(\ref{eq:BVprescription1}) can be re-written so that it is easier to implement. The final result reads (see  appendix~\ref{annexe:BV})
\oeqn
C_{ij}(t_1) = \lim_{\varepsilon_i,\varepsilon_j\rightarrow0} \frac{1}{2\varepsilon_i\varepsilon_j}
&\tr&\[\(\rho^{(0)}(t_0)-\eta_i(t_0,\varepsilon_i)\)\right.\nonumber\\
&&\stb \stb \stb\left.\(\rho^{(0)}(t_0)-\eta_j(t_0,\varepsilon_j)\)\],
\label{eq:CijBV}
\ceqn
where the single-particle matrices $\eta(t,\varepsilon)$  obey the TDHF equation~(\ref{eq:tdhf}) with a boundary condition defined at the final time $t_1$:
\oeq
\eta_j(t_1,\varepsilon_j)=e^{i\varepsilon_jQ_j}\rho^{(0)}(t_1)e^{-i\varepsilon_jQ_j}.
\label{eq:BCeta}
\ceq
The fluctuations $\sigma_{Q_i}=\sqrt{C_{ii}}$ are determined by taking $Q_i=Q_j$, leading to
\oeq
C_{ii}(t_1) = \lim_{\varepsilon_i\rightarrow0} \frac{1}{2\varepsilon_i^2}
\tr\[\(\rho^{(0)}(t_0)-\eta_i(t_0,\varepsilon_i)\)^2\],
\label{eq:CiiBV}
\ceq
with the boundary condition given in Eq.~(\ref{eq:BCeta}).

The fluctuations are generated by the boost in Eq.~(\ref{eq:BCeta}) and propagated in the backward Heisenberg picture from $t_1$ to $t_0$ according to the dual of the time-dependent RPA equation.
This is why $C_{ij}(t_1)$ is expressed as a function of density matrices at the initial time $t_0$. 
It is easy to show that, if the backward trajectories $\eta_i$ have the same mean-field as the forward evolution, then Eq.~(\ref{eq:CijBV}) leads to the TDHF expression in Eq.~(\ref{eq:CiiTDHF}).
If, however, deviations occur around the original mean-field, then additional terms appear and lead to an increase of $C_{ij}(t_1)$.

Eq.~(\ref{eq:BCeta}) imposes to solve the TDHF equation first for $\rho^{(0)}(t)$ forward in time, and then for $\eta(t)$ backward in time. 
Numerical applications solving Eq.~(\ref{eq:CiiBV}) with the boundary condition given in Eq.~(\ref{eq:BCeta}) are detailed in~\cite{tro85,mar85,bon85,bro08,bro09,sim11}.
In practice, several (typically $\sim5$ in~\cite{sim11}) backward TDHF trajectories with different but small values of $\varepsilon$ are performed to compute the limit in Eq.~(\ref{eq:CiiBV}). 
Fig.~\ref{fig:BVnum} gives a schematic illustration of the numerical technique.
In Ref.~\cite{sim11}, Eq.~(\ref{eq:CijBV}) is also solved. It is used to compute the correlations between proton and neutron numbers in fragments following deep-inelastic collisions in addition to their fluctuations  (see section~\ref{sec:DIC}). 

\begin{figure}
\begin{center}
\includegraphics[width=5cm]{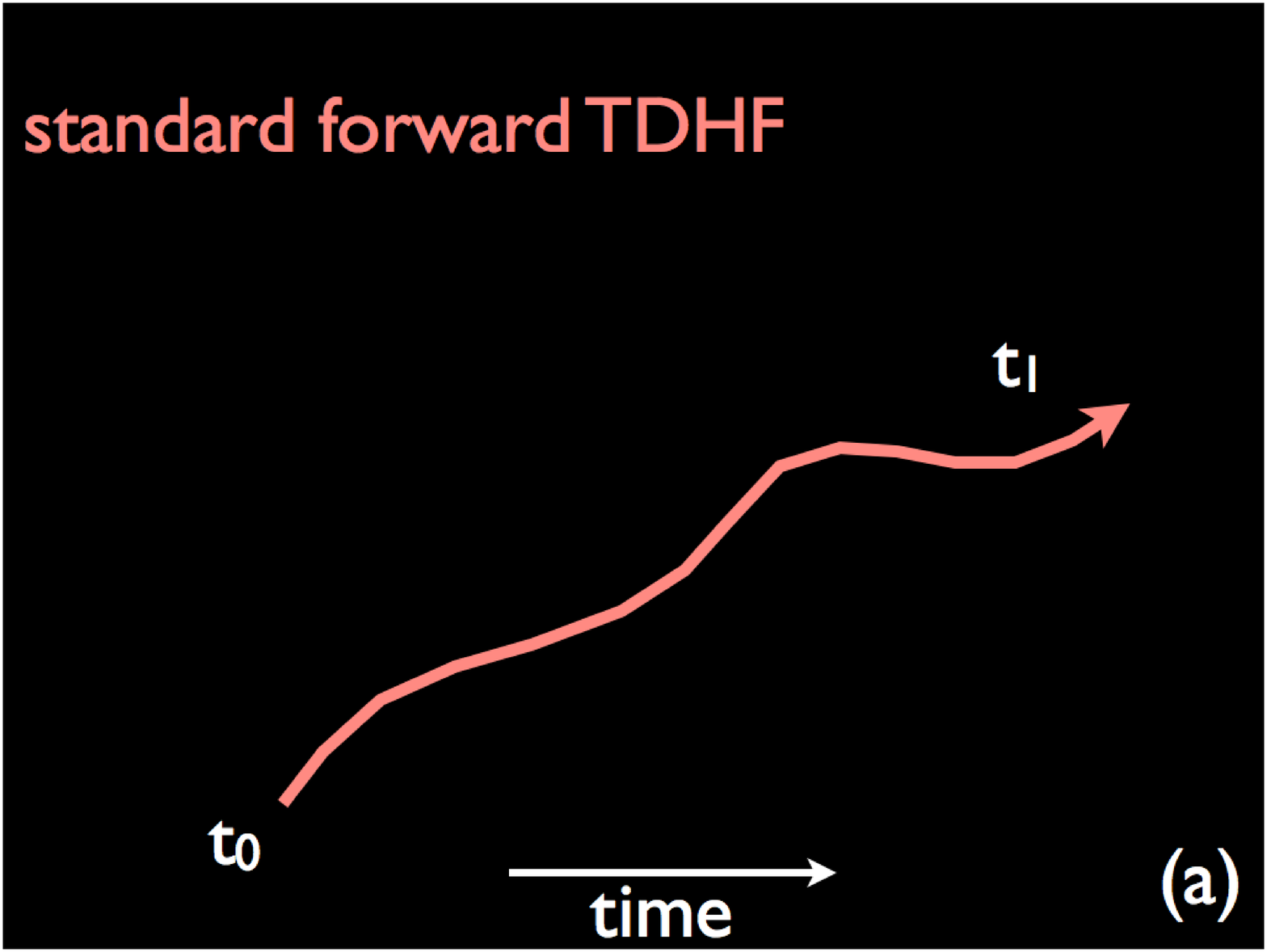}
\includegraphics[width=5cm]{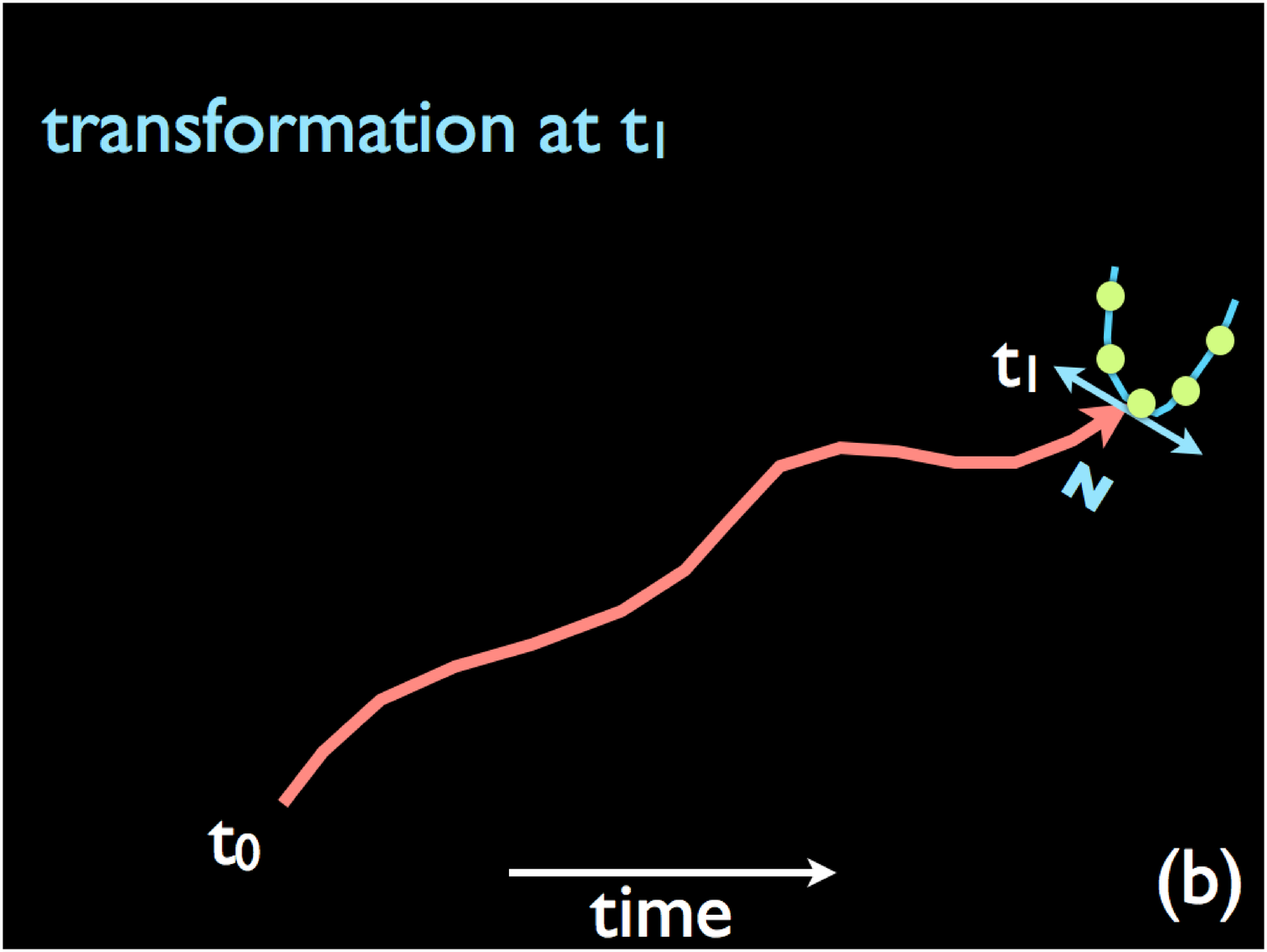}
\includegraphics[width=5cm]{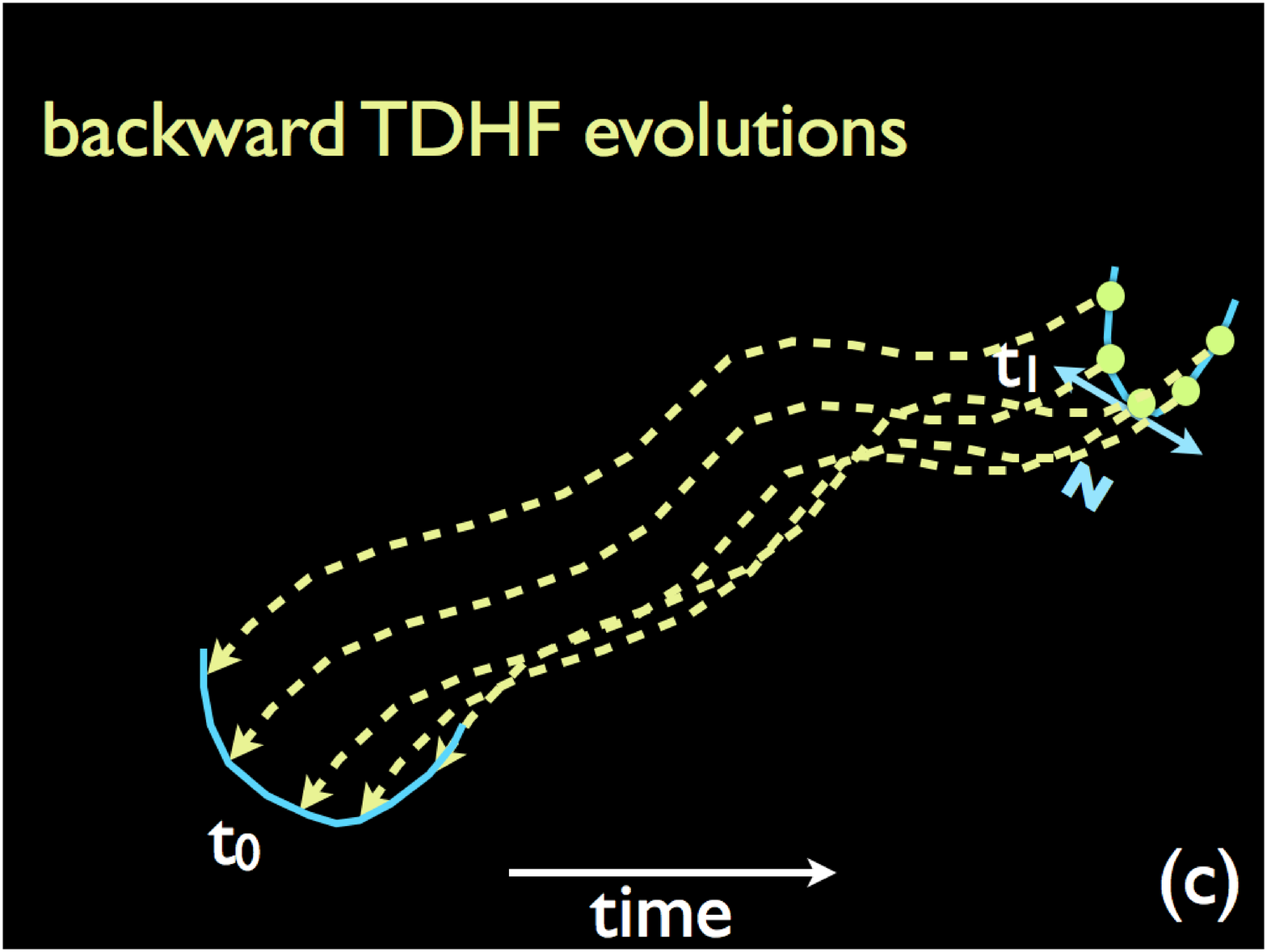}
\end{center}
\caption{Schematic illustration of the main numerical steps to compute the fluctuation of the one-body observable $\oN$ (e.g., neutron numbers in one fragment following a heavy-ion collision). (a) A standard TDHF evolution is performed from $t_0$ to $t_1$. (b) The transformation given in Eq.~(\ref{eq:BCeta}) is applied on the one-body density matrix at time $t_1$ with several (small) values of $\varepsilon$. (c) Backward TDHF evolutions from $t_1$ to $t_0$ are performed for each $\varepsilon$ and the resulting one-body density matrices at time $t_0$ are used to compute the (square of the) fluctuations from Eq.~(\ref{eq:CiiBV}).
\label{fig:BVnum}}
\end{figure}

It is important to note that, although Eqs.~(\ref{eq:CijBV}) and (\ref{eq:CiiBV}) are rather easy to solve numerically with a slightly modified TDHF code, they only provide the fluctuations and correlations for a specific choice of $\oQ_i$ and $\oQ_j$ at the time $t_1$. 
Another choice of operator(s) and/or final time implies to compute numerically another set of backward TDHF evolutions. 
This was not the case for the calculation of expectation values of one-body observables which only requires one forward TDHF evolution. 
Indeed, solving the TDHF equation is an initial value problem and the resulting one-body density matrix can be used to compute any one-body observable (i.e., the TDHF equation depends neither on $t_1$ nor on $\oQ_{i}$). 
As a result, the determination of fluctuations and correlations of one-body observables from the BV variational principle is much more computational time consuming than standard TDHF calculations of their expectation values. 

Finally, it is worth mentioning a recent alternative derivation of Eq.~(\ref{eq:CijBV}) considering small fluctuations in the initial state at time $t_0$ and propagated within the stochastic mean-field (SMF) approach~\cite{ayi08}.  
Numerical applications of the SMF method can be found in Refs.~\cite{ayi09,was09b,yil11}.

\subsection{Skyrme Energy-Density-Functional\label{sec:Skyrme}}

Early TDHF calculations~\cite{bon76,neg82} were based on simplified Skyrme effective interactions~\cite{sky56}.
Modern applications, however, use full Skyrme functionals including spin-orbit interactions to reach a realistic description of the strong interactions between the nucleons. 
Recent studies with the Gogny effective interaction~\cite{dec80} have also been reported~\cite{has12}.

Modern Skyrme energy density functionals (EDF) are usually derived from a Skyrme effective interaction of the form
\oeqn
\ov (1,2) &=& t_0 \sdf \( 1+x_0\, \oP_\si \) \sdf \odel \nonumber \\
&+& \frac{1}{2} \sdf t_1 \sdf \( 1+x_1\, \oP_\si \) 
\sdf \(\ovk'^2 \sdf  \odel + \odel \sdf \ovk^2 \) \nonumber \\
&+& t_2 \sdf \( 1+x_2\, \oP_\si \) 
\sdf \(\ovk' \cdot  \odel \sdf  \ovk \) \nonumber \\
 &+& \frac{1}{6} \sdf t_3 \sdf \( 1+x_3\, \oP_\si \) 
 \sdf \ro^\al\! (\ovR) \sdf \odel \nonumber \\
 &+& i\, W_0 \sdf \ovsi \cdot \(\ovk' \times \odel \, \ovk \) 
\label{eq:skyrme}
\ceqn
where $\odel = \del\(\ovr(1)-\ovr(2)\)$, $\ovk = \frac{1}{2i}\(\vna(1)-\vna(2)\)$ 
(relative momentum), $\ovk'$ complex conjugated of $\ovk$ acting on the left, and  $\ovR = \(\ovr(1)+\ovr(2)\)/2$.
The operators $\ovsi = \ovsi(1)+ \ovsi(2)$, with
$\ovsi(i) = \osi_x\!(i) \, \ve_x+ \osi_y\!(i) \, \ve_y + \osi_z\!(i) \, \ve_z$, 
are expressed in terms of the Pauli matrices $\osi_{x/y/z}(i)$ 
acting on the spin of the particle $i$.
 $\oP_\si = \[1+  \ovsi(1) \cdot \ovsi(2) \]/2$ corresponds to the exchange 
of the spins.  The particle density in~$\vr$
is noted $\ro(\vr) \equiv \sum_{sq} \ro(\vr s q ,\vr s q )$ where $\rho$ is the one-body density matrix, $s$ the spin and $q$ the isospin.
The "$t_1$" and "$t_2$" terms are non-local in space and simulate the short range part of the interaction.
Finally the last term accounts for the spin-orbit interaction.

The very interesting aspect of this interaction is its zero range nature, which greatly 
simplifies the mean-field expression in coordinate space (see below).
Parameters ($t_{0-3}$, $x_{0-3}$, $W_0$ and $\al$) are generally adjusted to 
reproduce nuclear properties like saturation and incompressibility of nuclear matter 
and selected properties of finite nuclei (see for instance \cite{cha98}).

In the following of this section, when there is no ambiguity, the spin and isospin indices, as well as the indices $\al$ denoting single-particle states  are omitted in the notation for clarity.
We also  assume that $\ro$ is diagonal in isospin. 
Let us define the usual densities entering the Skyrme EDF:
\oeqn
\ro(\vr) &=& \sum_{\al s}\sdf  \az_{\al }^*(\vr s)\,  \az_{\al }(\vr s) \equiv  \sum_{\al s}\sdf  \az^*\,  \az   \\
\tau(\vr) &=&   \sum_{\al s}\sdf  |\vna \az|^2  \\
\vj(\vr) &=& \frac{1}{2\,i} \sum_{\al s} \az^* \sdf \vna \sdf \az \stf +c.c. \\
\vna . \vJ(\vr) &=& -i\sum_{\al ss'}   \vna\az^*(\vr s) \times \vna \az (\vr s')\cdot \bs \vsi \ksp \\
\vS(\vr) &=& \sum_{\al s} \sdf  \az^*(\vr s) \sdf \az (\vr s')\sdf \bs \vsi \ksp ,
\ceqn
where $c.c.$ means "complex conjugated", are the local, kinetic, current, (gradient of) spin-orbit, and spin densities, respectively. 
The $\vj$ and $\vS$ densities are time-odd and vanish in time-reversal invariant systems. 
They are, however, important in time-dependent calculations to ensure Galilean invariance~\cite{eng75}. 

The total energy of an interacting system can be written as an integral of a local energy density
\oeq
E=\int d\vr \mH(\vr).
\ceq
Within the framework of the Skyrme EDF, we have~\cite{bon87}
\oeqn
\mH(\vr) &=& \frac{\hb^2}{2m}\tau + B_1 \rho^2 + B_2 \sum_q\rho_q^2 \nonumber \\
&&+B_3(\rho \tau -\vj^2)+B_4\sum_q(\rho_q \tau_q -\vj_q^2)\nonumber \\
&&+B_5\rho\Delta\rho+B_6\sum_q\rho_q\Delta\rho_q+B_7\rho^{2+\alpha}+B_8\rho^\al \sum_q\rho_q^2\nonumber\\
&&+B_9(\rho\vna\!\cdot\!\vJ + \vj\!\cdot\! \vna\!\times \!\vS+\sum_q\rho_q\vna\!\cdot\!\vJ_q+\vj_q\!\cdot\!\vna\!\times\!\vS_q)\nonumber\\
&&+B_{10} \vS^2+B_{11}\sum_q\vS_q^2 +B_{12}\rho^\al \vS^2 +B_{13} \rho^\al \sum_q\vS_q^2,\nonumber \\
&&\label{eq:Hr}
\ceqn
where the coefficients $B_i$ are the usual Skyrme functional coefficients:
\oeqn
B_1&=&\frac{t_0}{2}\(1+\frac{x_0}{2}\)\nonumber\\
B_2&=&-\frac{t_0}{2}\(x_0+\frac{1}{2}\)\nonumber\\
B_3&=&\frac{1}{4}\[t_1\(1+\frac{x_1}{2}\)+t_2\(1+\frac{x_2}{2}\)\]\nonumber\\
B_4&=&-\frac{1}{4}\[t_1\(x_1+\frac{1}{2}\)-t_2\(x_2+\frac{1}{2}\)\]\nonumber\\
B_5&=&-\frac{1}{16}\[3t_1\(1+\frac{x_1}{2}\)-t_2\(1+\frac{x_2}{2}\)\]\nonumber\\
B_6&=&\frac{1}{16}\[3t_1\(x_1+\frac{1}{2}\)+t_2\(x_2+\frac{1}{2}\)\]\nonumber\\
B_7&=&\frac{t_3}{12}\(1+\frac{x_3}{2}\)\nonumber\\
B_8&=&-\frac{t_3}{12}\(x_3+\frac{1}{2}\)\nonumber\\
B_9&=&-\frac{1}{2}W_0\nonumber\\
B_{10}&=&\frac{t_0x_0}{4}\nonumber\\
B_{11}&=&-\frac{t_0}{4}\nonumber\\
B_{12}&=&\frac{t_3x_3}{24}\nonumber\\
B_{13}&=&-\frac{t_3}{24}.\nonumber\\
\ceqn 

In fact, the complete Skyrme functional is more general and contains other terms of the form $\vS\cdot\Delta\vS$ and with other densities, i.e., the spin-current pseudo-tensor $\stackrel{\leftrightarrow}{J}$ and the spin kinetic energy density $\vT$~\cite{eng75,uma06a}. All or some of these additional terms are sometimes included in TDHF calculations~\cite{uma06a,mar06,loe12}. In Eq.~(\ref{eq:Hr}), only the anti-symmetric part of $\stackrel{\leftrightarrow}{J}$, which is the spin-orbit density $\vJ$, is  included. The spin-orbit energy is indeed expected to be more important (by about one order of magnitude) than the other spin-gradient terms~\cite{cha98}.

The Skyrme-HF mean-field is derived from Eq.~(\ref{eq:hHF}) by replacing the expectation value of the Hamiltonian on the Slater determinant by the expression of the Skyrme EDF. 
The action of this field on single-particle wave functions is then given by \cite{bon87}
\oeqn
&&\(h[\ro]  \az_\al \)(\vr, s ) = \nonumber\\
&&\sum_{s'} \!\[\! \(\!-\vna \frac{\hb^2}{2m^*_{q_\al}\!(\vr)} \vna \!+\! U_{q_\al}\!(\vr)\!  +\! i\vC_{q_\al}\!(\vr)\! \cdot\! \vna\! \)\!\delta_{ss'}\right. \nonumber \\
&&\left. + \vV_{q_\al}\!(\vr)\cdot  \bs \vsi \ksp 
 +i\vW_{q_\al}\!(\vr) \cdot \( \bs \vsi \ksp \times \vna \)\frac{}{} \] \az_\al\! (\vr,s').
 \nonumber \\
\label{eq:HFskyrme}
\ceqn
The derivatives act on each term sitting on their right, including the wave function.
The fields (functions of $\vr$) read
\oeqn
\frac{\hb^2}{2\,m_q^*} &=& \frac{\hb^2}{2\,m} + B_3 \,\ro + B_4\, \ro_q \\
U_q &=& 2B_1\ro+2B_2\ro_q+B_3(\tau+i\vna\cdot\vj)+B_4(\tau_q+i\vna\cdot\vj_q)\nonumber\\
&&+2B_5\Delta\ro+2B_6\Delta\ro_q+(2+\al)B_7\ro^{1+\al}\nonumber\\
&&+B_8[\al\ro^{\al-1}\sum_q\ro_q^2+2\ro^\al\ro_q]+B_9(\vna\cdot \vJ +\vna \cdot \vJ_q)\nonumber \\
&&+\al\ro^{\al-1}(B_{12}\vS^2+B_{13}\sum_q\vS_q^2)\\
\vV_q&=&B_9\vna\times (\vj+\vj_q)+2B_{10}\vS+2B_{11} \vS_q\nonumber\\
&&+2\rho^\al(B_{12}\vS+B_{13}\vS_q)\\
\vW_q &=& -B_9 \, \vna \, \(\ro+ \ro_q \)\label{eq:W}\\
\vC_q &=& 2 \, B_3 \, \vj + 2 \, B_4 \, \vj_q - B_9 \,  \vna \times \(\vS + \vS_q \) ,\label{eq:C} 
\ceqn
where the derivatives act on the first term sitting on their right only. 

In addition to this mean-field potential, protons are also affected by the Coulomb interaction.
The direct part of the Coulomb energy reads
\oeq
E_c^{dir}=\frac{ e^2}{2} \int d^3r \int d^3r' \frac{\rho_p(\vr)\rho_p(\vr')}{|\vr-\vr'|}.
\ceq
The latter is usually computed by solving first the Poisson equation to get the Coulomb potential $V_c(\vr)$, and, then, by evaluating the integral $\frac{1}{2}\int d^3r \rho_pV_c$.
The exchange part of the Coulomb energy is usually determined within the Slater approximation as
\oeq
E_c^{ex}=\frac{-3e^2}{4}\(\frac{3}{\pi}\)^{\frac{1}{3}} \int d^3r \rho_p(\vr)^\frac{4}{3}.
\ceq
As a result, the contribution of the Coulomb interaction to the proton mean-field reads 
\oeq
U_c=V_c-e^2\(\frac{3\rho_p}{\pi}\)^{\frac{1}{3}}.
\ceq

\subsection{Numerical solution of the TDHF equation}


As we saw in section \ref{sec:fluccor},  the calculations of both expectation values and fluctuations/ correlations of one-body observables imply to determine the time evolution of the one-body density matrix with the TDHF equation~(\ref{eq:tdhf}). 
Few numerical codes solving the TDHF equation in three dimensions with a full Skyrme EDF including spin-orbit terms \cite{sky56,cha98} are now available~\cite{kim97,mar05,nak05,uma05,seb09}. 
They were  used in many studies of giant resonance properties~\cite{sim03,cho04,ste04,nak05,uma05,rei07,bro08,sim09,ste11,sca12}, neutron star crust properties~\cite{seb09,seb11} 
and heavy-ion collisions. Amongst the latter, fusion and nucleus-nucleus potentials~\cite{kim97,sim01,sim04,mar06,uma06a,uma06b,uma06c,uma06d,guo07,sim07,uma07,sim08,uma08b,was08,uma09a,ayi09,uma09b,uma09c,was09a,uma10b,obe10,loe11,iwa11,leb12,uma12a,obe12,kes12,uma12b}, isospin equilibration \cite{sim01,sim07,uma07,iwa09,iwa10a,iwa10b,iwa12,obe12,sim11,sim12}, Coulomb excitation \cite{sim04,uma06c,uma06d,uma07}, rotational properties \cite{guo08}, quasi-elastic transfer \cite{sim08,uma08a,was09b,sim10b,eve11,yil11,sim12}, breakup \cite{ass09}, deep-inelastic collisions \cite{sim11}, as well as the dynamics of $\alpha$-clusters \cite{uma10a} and actinide collisions \cite{gol09,ked10} have been investigated.

\subsubsection{Numerical method}

Most of these applications were performed on a three dimensional cartesian grid using a time iterative method. 
We give here a summary of the main steps which are usually followed to treat the collision of two nuclei:
\begin{enumerate}
\item Static Hartree-Fock (HF) calculations are performed to determine the initial condition where the nuclei are usually assumed to be in their HF ground state. 
\item The nuclei are placed in a larger box, avoiding any overlap of the HF solutions. This latter condition allows to construct a single Slater determinant from the two initial HF states.
\item A Galilean boost\footnote{In case of a single nucleus, for instance to study its response to a specific excitation, the Galilean boost is replaced by the appropriate velocity boost generating the excitation (examples are given in section~\ref{chap:vib}). Alternatively, one can start with a constrained Hartree-Fock (CHF) solution obtained with an external constraint in the HF calculation. The response to the excitation is then studied by relaxing the constraint in the TDHF calculation~\cite{blo79}.}
 is applied at the initial time assuming that the nuclei followed a Rutherford trajectory prior to this time.
\item The TDHF equation is solved iteratively in time and expectation values of one-body observables are eventually computed at each time step to get their time evolution.
\end{enumerate}
Of course, variations of the main numerical steps described above are possible. 
For instance, one can question the validity of the assumption that the nuclei are in their HF ground state at initial time. In particular, heavy nuclei generate strong Coulomb fields which may induce long range excitations of the collision partners~\cite{sim04}. 


\subsubsection{Center of mass corrections in heavy-ion collisions}

HF calculations are usually performed with center of mass corrections to improve the description of the nucleus in its intrinsic frame. This is done by removing spurious center of mass motion, i.e., by replacing the total kinetic energy $\oT$ by 
\oeqn
\oT-\frac{\ovP^2}{2Am}&=& \oT-\frac{\(\sum_{i=1}^{A}\ovp_i\)^2}{2Am}\nonumber\\
&=&\oT-\frac{1}{2Am} \[\sum_i\ovp_i^2+\sum_{i\ne j}\ovp_i\ovp_j\].
\ceqn
This correction contains a one-body and a two-body contributions. 
Usually, the two-body term is neglected and only the one-body part of the correction is included in standard HF calculations~\cite{cha98}. 

The initial condition of a TDHF calculation uses static Hartree-Fock (HF) or constrained Hartree-Fock (CHF) solutions.
However, center of mass corrections are difficult to incorporate in TDHF calculations of heavy-ion collisions~\cite{uma09c}. 
They are usually neglected to allow a consistent treatment of colliding partners. 
Indeed, these corrections are explicitly dependent on the number of nucleons $A_i$ of the collision partner $i$ and would induce a different treatment of the single-particle wave-functions depending on which nucleus they come from. 

To treat structure and dynamics on the same footing, one can also neglect the center-of mass corrections in the initial HF calculations~\cite{kim97}. However, one should then use an EDF which has been fitted {\it without} these corrections.
This is the case of the SLy4$d$ parametrisation~\cite{kim97} of the Skyrme EDF which is widely used in this review article.

\subsubsection{Numerical approximations and algorithm}

Most of the numerical approximations and techniques used in modern TDHF codes are based on similar algorithms, but may contain differences such as in the calculation of spatial derivatives.  For instance, the latter are computed with finite-difference formulae in the \textsc{tdhf3d} code~\cite{kim97}, while spline and fast-Fourier transform (FFT) techniques are used in Refs.~\cite{uma05} and \cite{mar05}, respectively.  
Typical regular mesh spacing with $\Delta x\simeq0.6$~fm \cite{sim09}, 0.8~fm \cite{kim97} and 1.0~fm \cite{uma05,mar05} are used, although adaptive grids have been also considered~\cite{nak05}.

The \textsc{tdhf3d} code contains all the time-odd and even terms of standard Skyrme EDF (see Sec.~\ref{sec:Skyrme}).
The inclusion of time-odd terms is indeed crucial for a proper description of translational motion and to avoid spurious excitations~\cite{mar06}.
As time-reversal symmetry is not assumed, the code contains no degeneracy of single-particle wave-functions. 
This means that up to $\sim500$ (for, e.g., two actinides) wave-functions are evolved in time. 

Due to the self-consistency of the  mean-field, the TDHF equation needs to be solved iteratively in time 
with a small time step increment $\Delta t$ with typical values ranging from $\sim5\times10^{-25}$~s \cite{nak05,sim09} to $\sim1.5\times10^{-25}$~s \cite{kim97,uma05}.
Over small time intervals $\[t,t+\Del t\]$, the HF Hamiltonian is assumed to be constant.
However, to conserve  energy, the numerical algorithm should be symmetric with respect to time-reversal operation.
This implies to consider the Hamiltonian  at time $t+\frac{\Del t}{2}$ for the evolution of single-particle wave-functions from $t$ to $t+\Del t$~\cite{bon76}
\oeq
|\nu(t+\Del t)\> \approx e^{-i \frac{\Del t}{\hb} \oh\(t+\frac{\Del t}{2}\)} \sdf |\nu(t)\>.
\ceq
A schematic illustration of the real time propagation could be written as: 
\oeq
\begin{array}{ccc}
\{ |\nu_1^{(n)}\> \cdots |\nu_N^{(n)} \>\} & \Rightarrow &\ro^{(n)}\\
\Uparrow && \Downarrow\\
|\nu_i^{(n+1)}\> =  e^{-i\frac{\Delta t}{\hb}  \oh^{(n+\frac{1}{2})}}  |\nu_i^{(n)}\>
 && \oh^{(n)}\equiv\oh[\ro^{(n)}] \\
\Uparrow && \Downarrow\\
 \oh^{\(n+\frac{1}{2}\)}\equiv\oh\[\ro^{\(n+\frac{1}{2}\)}\] &&
   |\tilde{\nu}_i^{(n+1)}\> = e^{-i\frac{\Delta t}{\hb} \oh^{(n)}}  |\nu_i^{(n)}\> \\ 
\Uparrow & &   \Downarrow \\
\ro^{\(n+\frac{1}{2}\)}= \frac{\ro^{(n)} + \tro^{(n+1)}}{2} & \Leftarrow & \tro^{(n+1)} \\
\end{array} 
\label{eq:algo}
\ceq
where $|\nu^{(n)}\>$ corresponds to an approximation of $|\nu(t_n=n\Del t)\>$.
In this algorithm, starting from the density at time $t$, a first estimate of the density at time $t+\Del t$, denoted by  $\tro^{(n+1)}$ is obtained.
The Hamiltonian used in the propagator is computed using the average density obtained from  $\ro^{(n)}$ and $\tro^{(n+1)}$. 
Then, the new density at time $t+\Del t$ is obtained using this Hamiltonian. 
An approximate form of the exponential function is generally used which in some cases, breaks the unitarity, and orthonormalisation of the single particle states must be controlled.

\subsection{Perspectives for beyond TDHF calculations\label{sec:persp}}

\begin{figure}
\includegraphics[width=8.8cm]{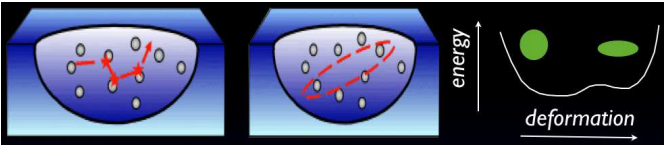}
\caption{(courtesy of D. Lacroix). Illustration of the different kinds of correlations. From left to right: in medium particle-particle collisions, pairing correlations, and large amplitude dynamical correlations.
\label{fig:correlations}}
\end{figure}

The TDHF theory is a microscopic quantum mean-field approach of independent particles. 
Extensions of the TDHF theory include correlations which are not present at the TDHF level. 
These correlations are typically of three kinds illustrated in Fig.~\ref{fig:correlations}:
\begin{itemize}
\item {\it In medium particle-particle collisions.} In a first approximation, 
this collision term can be neglected thanks to the Pauli blocking. However, in violent collisions, where the Pauli principle is less efficient to block collisions between nucleons of the two colliding fragments, this term is expected to affect the dynamics.  It is also responsible for the thermalisation of the compound nucleus and for the spreading width of giant resonances. 
\item {\it Pairing correlations.} They generate a superfluid phase in nuclei. They have a strong effect on the structure of mid-shell nuclei (e.g., odd-even mass staggering) and in transfer reactions where they favour the transfer of paired nucleons. 
\item {\it Large amplitude dynamical correlations.} Unlike a single Slater determinant which is usually localised in a potential energy surface (PES), the correlated state may be described by a configuration mixing of localised states across the entire PES. A typical example is the zero point motion along a collective coordinate. These correlations allow also for the state to be in a classically forbidden region of the PES, and, then, are necessary to treat quantum tunnelling of the many-body wave function, e.g. in sub-barrier fusion.
\end{itemize}

The Balian-V\'en\'eroni prescription discussed before is an example where fluctuations and correlations of specific observables, not included at the TDHF level, are described dynamically. 
In the classification used above, the included correlations belong to the class of dynamical correlations, although, in this case, they are of small amplitude nature. 
In fact, the fluctuations obtained in Eq~(\ref{eq:CiiBV}) are those included in the time-dependent random phase approximation (TDRPA) which is obtained assuming small fluctuations of the density matrix around the average evolution~\cite{bal92,ayi08}. 

A possible extension of TDHF  including dynamical correlations of large amplitude involves the path integral technique with the stationary phase approximation (SPA) \cite{neg82}.
The SPA assumes that the path integral is dominated by the classical action (for a single particle), or, equivalently, by the TDHF action for a many-fermion system. 
The method provides an elegant way to include fluctuations around the mean-field trajectory. 
Unlike the BV prescription, however, these fluctuations are not limited to the small amplitude limit.  
This approach leads to a self-consistent eigenvalue problem in four space-time dimensions. 
It is in fact similar to a HF eigenvalue problem with time as an additional dimension. 
In case of a vibrational motion, the problem involves the (difficult) task to find periodic solutions of the TDHF equations. 
Another important possible application of this path integral approach is to treat quantum tunnelling of the many-body wave function through a barrier. 
In this case,  quantum many-particle closed trajectories in imaginary time need to be found.
Up to now, realistic applications in imaginary time have faced the difficulties brought by the limitations of computational power, and only simple cases, such as the spontaneous fission of the $^{8}$Be in two~$\alpha$ have been studied~\cite{neg82}.
However, the recent increase of computational power, and the strong interests in understanding deep-sub-barrier fusion~\cite{das07}, should lead to a revival of these techniques.

An alternative way to include large amplitude dynamical correlations is to derive collective Hamiltonians, such as the Bohr Hamiltonian, from microscopic calculations using the generator-coordinate method (GCM)~\cite{hil53,gri57} and its time-dependent extension (TDGCM)~\cite{rei83}. Recent applications of the GCM to low-lying collective excitations can be found in Refs.~\cite{ben08,del10,rod10}. The TDGCM has also been applied with the Gaussian overlap approximation (GOA) to investigate the fission process \cite{gou05}.

Other examples of beyond-TDHF approaches include the time-dependent Hartree-Fock-Bogoliubov (TDHFB) theory~\cite{bla86,ave08,eba10,ste11}, the time-dependent density-matrix (TDDM) theory \cite{cas90,bla92,luo99,toh01,toh02a,toh02b,ass09}, the extended TDHF \cite{won78,won79,dan84,bot90,ayi80,lac99} and stochastic TDHF theories \cite{rei92,ayi01,lac01,lac04}, and stochastic mean-field based on functional integrals \cite{car01,jui02}. 
Table~\ref{tab:sum_micro} gives a summary of these approaches. 
Numerical applications of the   TDHF, BV, and TDHFB approaches are presented in this review.

\begin{table*}[th]
    \begin{center}
{ 
\begin{tabular}{|l|l|l|l|} \hline
{\it Name} & {\it Approximation} & {\it Variational}  & {\it Associated}  \\
&&{\it space}&{\it observables}\\
\hline
TDHF & mean-field (m.-f.) & indep. part. & one-body \\ 
&&&\\
\hline
BV prescription & m.-f. + small ampl.  & indep. part & one-body \\ 
&fluctuations&&+fluctuations\\
\hline
Path integrals & m.-f. + fluctuations  & indep. part. & one-body \\ 
&&&+fluctuations\\
\hline
(TD)GCM & m.-f. + fluctuations & correlated & one-body \\ 
&&states&+fluctuations\\
\hline
TDHF-Bogoliubov & m.-f. + pairing  & indep. & generalised  \\
&&quasipart.  &one-body\\
\hline
Extended-TDHF & m.-f. + collision  & correlated  & one-body  \\ 
&(dissipation)&states&\\
\hline
Stochastic-TDHF & m.-f. + collision  & correlated  & one-body   \\ 
&(dissipation+fluctuations)&states&\\
\hline
Time-dependent & m.-f. + two-body & correlated  & one- and \\ 
density matrix & correlations&states&two-body\\
\hline
Stochastic m.-f. & Exact & correlated  & all   \\ 
(Functional integrals) &(within statistical errors)&states&\\
\hline
\end{tabular}
}
\end{center}
\caption{The TDHF approach and several possible extensions.}
\label{tab:sum_micro}
\end{table*}

\section{Collective vibrations \label{chap:vib}}

\subsection{Introduction\label{sec:intro_vib}}

A particular interest in strongly interacting systems is their ability to present disorder or chaos, 
and, in the same excitation energy range, well-organised motion.
Atomic nuclei are known to show both behaviors~\cite{boh75}.
In particular, they exhibit a large variety of  vibrations,
from low-lying collective modes to giant resonances (GR) with excitation energy usually above the particle emission threshold~\cite{har01}.

Baldwin and Klaiber observed the isovector giant dipole resonance (GDR) 
in photofission of uranium nuclei~\cite{bal47}, 
interpreted as a vibration of neutrons against protons~\cite{gol48}. 
Other kinds of GR have been discovered, such as the isoscalar giant quadrupole resonance (GQR) 
associated with an oscillation of the shape between a prolate and an oblate deformation~\cite{fuk72},
and the isoscalar giant monopole resonance (GMR) corresponding to a breathing mode~\cite{mar76,har77,you77}. 

GR are usually associated with the first phonon of a small-amplitude harmonic motion.
In the harmonic picture, it corresponds to a coherent sum of one-particle one-hole ($1p1h$) states~\cite{boh75}.
GR are usually unbound and may decay by particle ($p$, $n$, $d$, $\alpha$, $\gamma$...) emission, leading to an escape width in the GR spectra. 
The escape width is then due to a coupling of the correlated $1p1h$ states to the continuum. 
Other contributions to the width of GR are the Landau damping and the spreading width. 
Landau damping occurs due to a one-body coupling to non-coherent $1p1h$ states~\cite{pin66}. 
The spreading width is due to the residual interaction coupling $1p1h$ states to $2p2h$ states.
The $2p2h$ states can also couple to $3p3h$ states and more complex $npnh$ configurations until an equilibrated system is reached. 
These contributions to the GR width are illustrated in Fig.~\ref{fig:damping}.

\begin{figure}
\includegraphics[width=8.8cm]{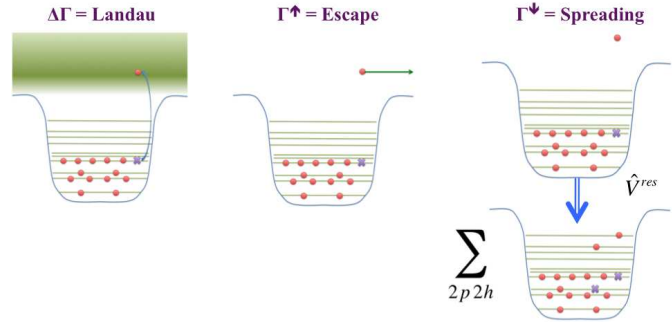}
\caption{Illustration of the three contributions to the width of the GR (see text). (left) Landau damping due to coupling to incoherent $1p1h$ states. (middle) Escape width due to direct decay. 
(right) Spreading width due to coupling to $2p2h$ states.
\label{fig:damping}}
\end{figure}

The proof of the vibrational nature of GR came with the observation 
of their two- and three-phonon states~\cite{cho95,aum98,sca04}.
Multiphonon studies also provided a good test to the harmonic picture. 
In particular, anharmonicity was found in an abnormally large excitation probability of these states~\cite{cho95}.
The coupling between different phonon states is predicted to be an important source of this anharmonicity~\cite{vol95,bor97,sim03,fal03,cho04,sim09,lan06}. 

Coherent motion of fermions such as collective vibrations in nuclei
can be modeled by time-dependent mean-field approaches like the 
TDHF theory. 
In fact, in its linearised version, TDHF is equivalent to the
Random Phase Approximation (RPA)~\cite{rin80} which is the basic tool to understand the
collective vibrations in terms of independent phonons.
In particular, time evolution of one-body (collective) observables, which can be estimated using a TDHF code, contains relevant informations on the vibration properties, such as their energy spectra.

Direct decay contributing to the escape width can be studied within the continuum-(Q)RPA model\footnote{The quasiparticle-RPA (QRPA) is an extension of the RPA including pairing correlations.}~\cite{kre74,liu76,kam98,mat01,hag01,kha02}. 
GR direct decay can also be investigated within the TDHF framework~\cite{cho87,pac88,ave12,sca12}.
Indeed, the TDHF  theory is able to treat such process as it allows for the evaporation of unbound components of the single particle wave-functions.
Due to its one-body nature, Landau damping is also included in RPA and TDHF.
However, they do not contain the residual interaction responsible for the spreading width.

An interesting feature of TDHF applications is that they are not limited to small amplitude vibrations, unlike RPA, allowing for investigations of non linear effects in collective motions. 
In particular, couplings between collective modes, which is a possible source of the anharmonicity discussed above, has been investigated~\cite{sim03,cho04,sim09}. 

In principle, TDHF and RPA codes can be used to study the vibrational spectra of any nucleus. 
However, most of the applications have focused on doubly magic nuclei. 
The main reason is that these nuclei are well described at the HF level, as there is usually no pairing correlation in their ground-state. 
Mid-shell nuclei, however, are better described with the HF+BCS or Hartree-Fock-Bogoliubov (HFB) approach. 
Recently,  TDHF codes have been extended to study the role of pairing correlations on collective vibrations at the BCS level~\cite{eba10,sca12} and within the TDHFB theory. 
Indeed, TDHFB calculations with full Skyrme EDF are now possible in spherical symmetry~\cite{ave08}, and more recently in three dimensions~\cite{ste11}.

In this section, we first briefly introduce the linear response theory. 
For an illustrative purpose, applications of the latter to both low-lying vibrational states and GR  are given within the TDHF framework. 
These standard applications to the linear response theory are followed by an investigation of the direct decay by nucleon emission and its link to the GR microscopic structure.
Then, non-linearities in collective vibrations are studied to investigate a possible source of anharmonicity in GR multiphonon spectra. 
In the last application, we come back to the linear response theory and study pairing vibrations with a TDHFB code.
Finally we conclude this section and present some perspectives to the study of collective vibrations.

\subsection{Linear response theory}

The linear response theory has been widely used with TDHF
to study collective vibrations in nuclei~\cite{blo79,str79,uma86,cho87,pac88,chi96,sim03,nak05,uma05,mar05,alm05,rei07,ste07}.\\

In this theory, one computes the time evolution of an observable $Q(t)$
after an excitation induced by a small boost on the ground state $|\Psi_0\>$,
\oeq
|\Psi(0)\> = e^{-i\epsilon \oQ /\hbar} |\Psi_0\>=|\Psi_0\>-\frac{i\epsilon}{\hb}\sum_\nu q_\nu|\Psi_\nu\>+O(\epsilon^2),
\ceq
where  $q_\nu=  \<\Psi_\nu| \oQ |\Psi_0\>$ is the transition amplitude between 
the ground state and the eigenstate $|\Psi_\nu\>$ of the Hamiltonian with eigenenergy $E_\nu$.
The time-evolution of the state reads
\oeqn
|\Psi(t)\> &=& e^{-i\oH t/\hb}|\Psi(0)\> \nonumber \\
&=&e^{-iE_0t/\hb} \( |\Psi_0\> - \frac{i\epsilon}{\hb}\sum_\nu q_\nu e^{-i\omega_\nu t} |\Psi_\nu\>  \) + O(\epsilon^2),\nonumber \\
&&
\ceqn
with $\hb\omega_\nu = E_\nu-E_0$.

The response $Q(t)=\<\Psi(t)|\oQ|\Psi(t)\>-\<\Psi_0|\oQ|\Psi_0\>$ to this excitation can be written
\oeq
Q(t)=-\frac{2\epsilon}{\hb}\sum_\nu |q_\nu|^2\sin \omega t+O(\epsilon^2).
\label{eq:linresp}
\ceq
The latter can be decomposed
into various frequencies $\omega$, giving the strength function
\begin{eqnarray}
R_{Q}(\omega) &=&\lim_{\epsilon\rightarrow 0} \frac{-\hbar }{\pi \epsilon}\,
\int_{0}^{\infty}  dt\, {Q}(t) \, \sin (\omega t). \label{eq:strengthlin} \\
 &=& \sum_\nu \, |q_\nu |^2 
 \delta (\omega - \omega_\nu). \label{eq:strengthfinal}
\end{eqnarray}

We see in Eqs.~(\ref{eq:strengthlin}) and (\ref{eq:strengthfinal})  that the strength function is obtained for small $\epsilon$.
In practice, it is sufficient to check that the amplitude $Q_{max}$ of $Q(t)$ evolves linearly with $\epsilon$ [see Eq.~(\ref{eq:linresp})]. 
An example of evolution of $Q_{max}$ as a function of $\epsilon$ is shown in Fig.~\ref{fig:quadratic2}-a in the case of a dipole response of the $^{132}$Sn nucleus. 

It is interesting to note that the first phonon energy is obtained with an amplitude of the oscillation which is much smaller than the one associated to the first phonon. 
Indeed, in a coherent picture, and for a single mode with transition amplitude $q$, the number of excited phonons reads~\cite{sim03}
$$n=\(\frac{2Q_{max}}{q}\)^2.$$
In particular, the amplitude associated to the first phonon, $Q_{max}^{1ph}=q/2$ may be well beyond the linear regime. 

\begin{figure}
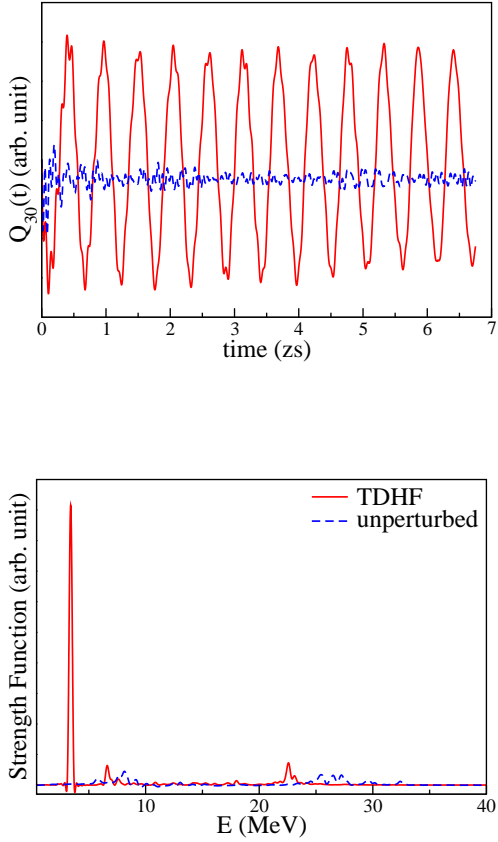

\begin{center}
\includegraphics[width=6.5cm]{Pb3-.eps}\\
\vspace{1.5cm}
\includegraphics[width=6.5cm]{TFPb3-.eps}
\caption{(top) Time evolution of the octupole moment in $^{208}$Pb after an octupole boost obtained with the \textsc{tdhf3d} code for both TDHF (solid line) and unperturbed (dashed line) responses (see text). (bottom) Associated strength function.
\label{fig:PbQ3}}
\end{center}
\end{figure}

Fig.~\ref{fig:PbQ3} gives an example for the octupole modes in $^{208}$Pb obtained with the \textsc{tdhf3d} code~\cite{kim97} (see also Ref.~\cite{nak05} for a study of this mode in $^{16}$O).
An octupole boost is applied at initial time on the HF ground state of $^{208}$Pb. 
The latter has spin-parity $J^\pi=0^+$ and the boost induces a transition with $\Delta L=3$, exciting vibrational states with $J^\pi=3^-$. 
We can see on the upper panel of Fig.~\ref{fig:PbQ3} the oscillation of the octupole moment induced by the boost (solid red line). 
The associated strength function, shown in the lower panel, exhibits a strong peak at an energy of $\sim3.4$~MeV. 
It corresponds to the main oscillation seen in the time evolution of $Q_{30}(t)$. 
Note that this state is clearly bound, as can be seen from the undamped nature of the oscillation. 
This peak is associated to the low-lying $3^-$ state in $^{208}$Pb. 
The energy of this state is overestimated with the SLy4 parametrisation of the Skyrme EDF, as the experimental value gives 2.6~MeV.
However, its collective nature is unambiguous. 
This can be seen from a comparison with the unperturbed response of the same boost.
The latter is obtained by freezing the mean-field in its initial HF value, i.e., neglecting the self-consistency of the mean-field in the dynamics. 
This procedure removes the residual interaction which is responsible for the collectivity of vibrations in TDHF (and RPA). 
We see that this peak disappears in the unperturbed spectrum, proving its collective nature. 

\begin{figure}
\begin{center}
\includegraphics[width=7cm]{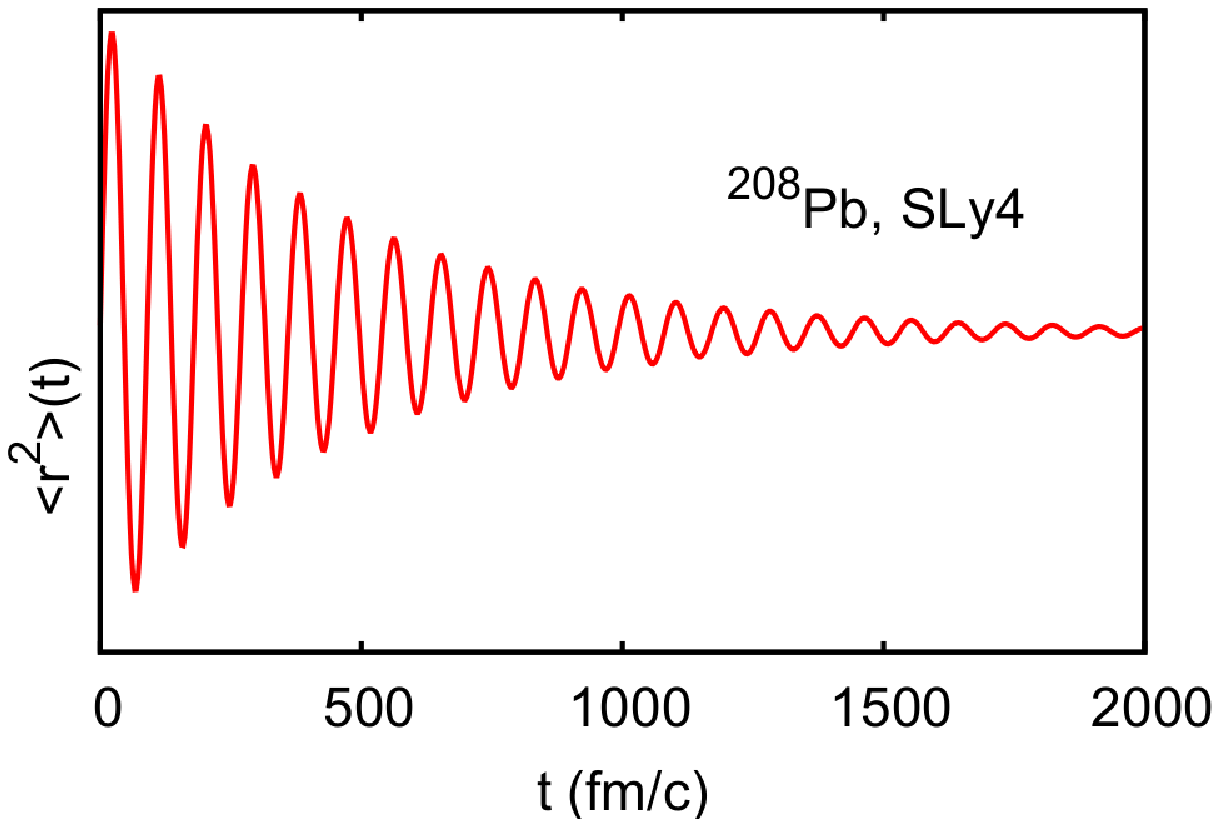}
\includegraphics[width=6cm]{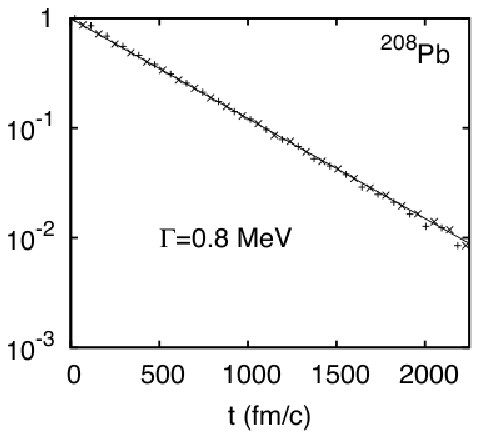}
\includegraphics[width=7cm]{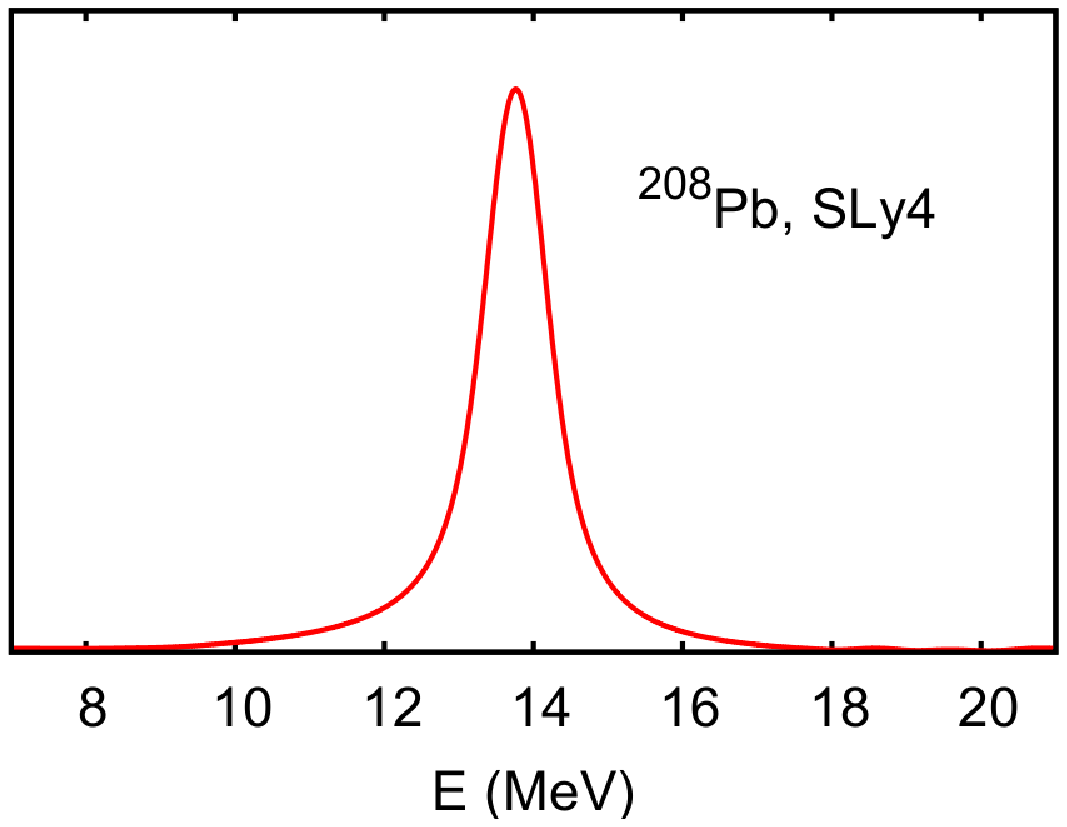}
\caption{(top) Time evolution of the mean square radius in $^{208}$Pb after a monopole boost obtained with the \textsc{tdhfbrad} code (without pairing). (middle) Time evolution of the oscillation amplitude (log scale). 
(bottom) Associated strength function. From Ref.~\cite{ave09}.
\label{fig:PbQ0}}
\end{center}
\end{figure}

Another example  is shown in Fig.~\ref{fig:PbQ0}.
Here, a monopole boost is applied on the $^{208}$Pb ground-state.
The evolution has been obtained with a TDHF code\footnote{This is the \textsc{tdhfbrad} code, developed by B. Avez, which solves the TDHFB equation. However, the pairing interaction is set to zero in the calculation shown in Fig.~\ref{fig:PbQ0}.} in spherical symmetry (allowing for large boxes)~\cite{ave09}. 
In this case, almost all the strength (if not all) goes into the GMR which is unbound.
Such a resonance can then decay by particle emission.
In particular, the direct decay induces an escape width.
The latter is included in the TDHF framework thanks to evaporation of unbound components of the single particle wave functions~\cite{cho87,pac88,ave12}.
Note that the emission of nucleons in the continuum is quadratic with the boost intensity $\epsilon$. 
As a result, it does not affect the evolution of $Q(t)$ in the linear regime. 
However, special care should be taken with possible spurious numerical effects due to the reflection of the nucleons on the box edges~\cite{rei07}.

The decay induces an exponential decrease of the oscillation amplitude (see middle panel of Fig.~\ref{fig:PbQ0}). 
The corresponding escape width is $\Gamma^\uparrow=0.8$~MeV.
This escape width is responsible for the main part of the width of the peak in the lower panel of Fig.~\ref{fig:PbQ0}.
Indeed, the total width of the peak is $\Gamma^{tot.}\simeq1.1$~MeV.
The difference between the TDHF predictions of $\Gamma^{tot.}$ and $\Gamma^\uparrow$ could be attributed to Landau damping. 

Experimentally, the energy of the GMR in $^{208}$Pb is~\cite{you04}
$$E_{GMR}^{exp.}\equiv \frac{m_1}{m_0}= 14.0\pm0.2 \mbox{ MeV,}$$
where the energy weighted moment $m_k$ is defined as~\cite{rin80}
\oeq
m_k = \sum_\nu \omega_\nu^k |q_\nu|^2.
\ceq
This value is in excellent agreement with the TDHF result determined from the $m_1/m_0$ ratio: $E_{GMR}^{TDHF}\simeq13.9$~MeV~\cite{ave09}.
However, the total width obtained from TDHF, $\Gamma^{tot.}_{TDHF}\simeq1.1$~MeV,  underestimates the experimental value $\Gamma^{tot.}_{exp.}=2.9\pm0.2$~MeV \cite{you04}. 
This is due to the fact that TDHF does not take into account the spreading width (see Fig.~\ref{fig:damping} and the discussion in the introduction of this section).  
To go beyond and describe the spreading width, one would need to include the residual interaction.
Calculations at the semi-classical level~\cite{sur89} and with the TDDM approach~\cite{luo99,bla92} have indeed shown an increase of the damping thanks to the introduction of the collision term.

TDHF calculations of giant-resonances are not limited to the study of monopole modes with spherical symmetry. Indeed, 3-dimensional TDHF codes have also been used to study other modes, such as the GDR and the GQR~\cite{sim03,nak05,mar05,uma05,sim09}.
An example of such calculations is shown in Fig.~\ref{fig:U-GDR}, where the TDHF dipole strength function is plotted in $^{238}$U for different excitation modes~\cite{mar05} and compared with experimental data from Ref.~\cite{die88}. 
The $^{238}$U being prolately deformed, the dipole response is investigated along and perpendicular to the deformation axis, exhibiting the well known splitting of the GDR peak, i.e., a lower (higher) energy along (perpendicular to) the deformation axis~\cite{har01}. 
The qualitative agreement between the theoretical prediction of the peak positions and experimental data is good. However,  part of the TDHF width is due to the finite time window used in the Fourier transform. It is then difficult to draw any conclusion on the width of the GDR.
\begin{figure}
\begin{center}
\includegraphics[width=6cm]{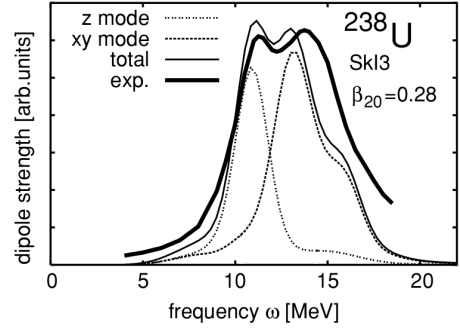}
\caption{Dipole strength in the prolately deformed $^{238}$U nucleus with the SkI3 parametrisation of the Skyrme functional. The  strength is shown along the symmetry axis ($z$ mode) and perpendicular to the symmetry axis ($xy$ mode). The total strength is shown with the thin-solid line and compared with experimental data from Ref.~\cite{die88} (thick-solid line). Adapted from Ref.~\cite{mar05}.
\label{fig:U-GDR}}
\end{center}
\end{figure}

\subsection{Direct decay of giant resonances\label{sec:decay}}

Giant resonances usually lie above the proton and neutron emission thresholds.
As mentioned above, their direct decay induces an escape width contributing to the total width of the GR. 
Such escape widths can be studied with the TDHF formalism. 
Indeed, TDHF codes have been used in the past to investigate the direct decay of GMR~\cite{cho87,pac88}.
In addition to its contribution to the escape width, the GR direct decay is particularly interesting as it brings informations on the microscopic structure of the GR~\cite{har01}. 
This is the main purpose of the present section.

\begin{figure}
\includegraphics[width=8.8cm]{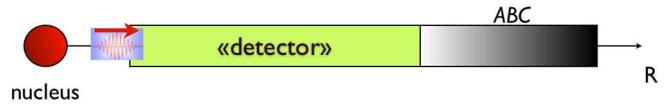}
\caption{Schematic description of the spatial repartition of numerical elements to compute spectra of emitted nucleons. The excited nucleus is in the center of a (spherical) box. Unbound parts of single-particle wave-functions are emitted in the continuum. The ''detector''  shows the region of space where the energy of the emitted wave functions is computed. Absorbing boundary conditions (ABC) are used to absorb particles leaving the detector and to avoid spurious reflection on the box boundary.
\label{fig:detector}}
\end{figure}

TDHF calculations of GR direct decay are performed with large spatial grids to construct spectra of emitted nucleons with a  good precision. 
It is then easier to use spherical TDHF codes, although the applications are limited to monopole vibrations only.  
Let us introduce a numerical ''detector'' corresponding to the region of space where the energy of the emitted nucleons is computed from Fourier transform of their spatial wave-functions. 
This detector should be away from the center of the box to avoid any nuclear interaction of the emitted nucleons with the nucleus. 
Fig.~\ref{fig:detector} shows a schematic representation of this numerical setup. 
Absorbing boundary conditions (ABC) with an imaginary potential may be used to avoid any spurious interaction with particles reflected on the box boundary~\cite{nak05,rei06}.

Fig.~\ref{fig:spectra_ivGMR} shows an example of a calculation of emitted nucleon spectra for the isovector GMR in $^{40}$Ca.
The calculations have been performed with the same spherical code and simplified Skyrme EDF as in Ref.~\cite{sim05}.
The upper panel of Fig.~\ref{fig:spectra_ivGMR} shows the proton spectra at different times. 
The first protons to reach the ''detector'' are obviously  the fastest, i.e., with the highest kinetic energy. 
They also leave quickly the detector while slower protons reach it. 
Put together, these proton spectra form an envelope\footnote{This envelope is  defined by the maxima of the spectra obtained at different times.}, shown in black thick solid line in the lower panel of Fig.~\ref{fig:spectra_ivGMR}.
The neutron spectrum is also shown (dashed line). 
The fact that neutrons are more bound than protons in $^{40}$Ca, together with the absence of Coulomb barrier for neutrons, explain that the neutron spectrum shows a more important contribution at low energy than the proton one.

\begin{figure}
\begin{center}
\includegraphics[width=7.5cm]{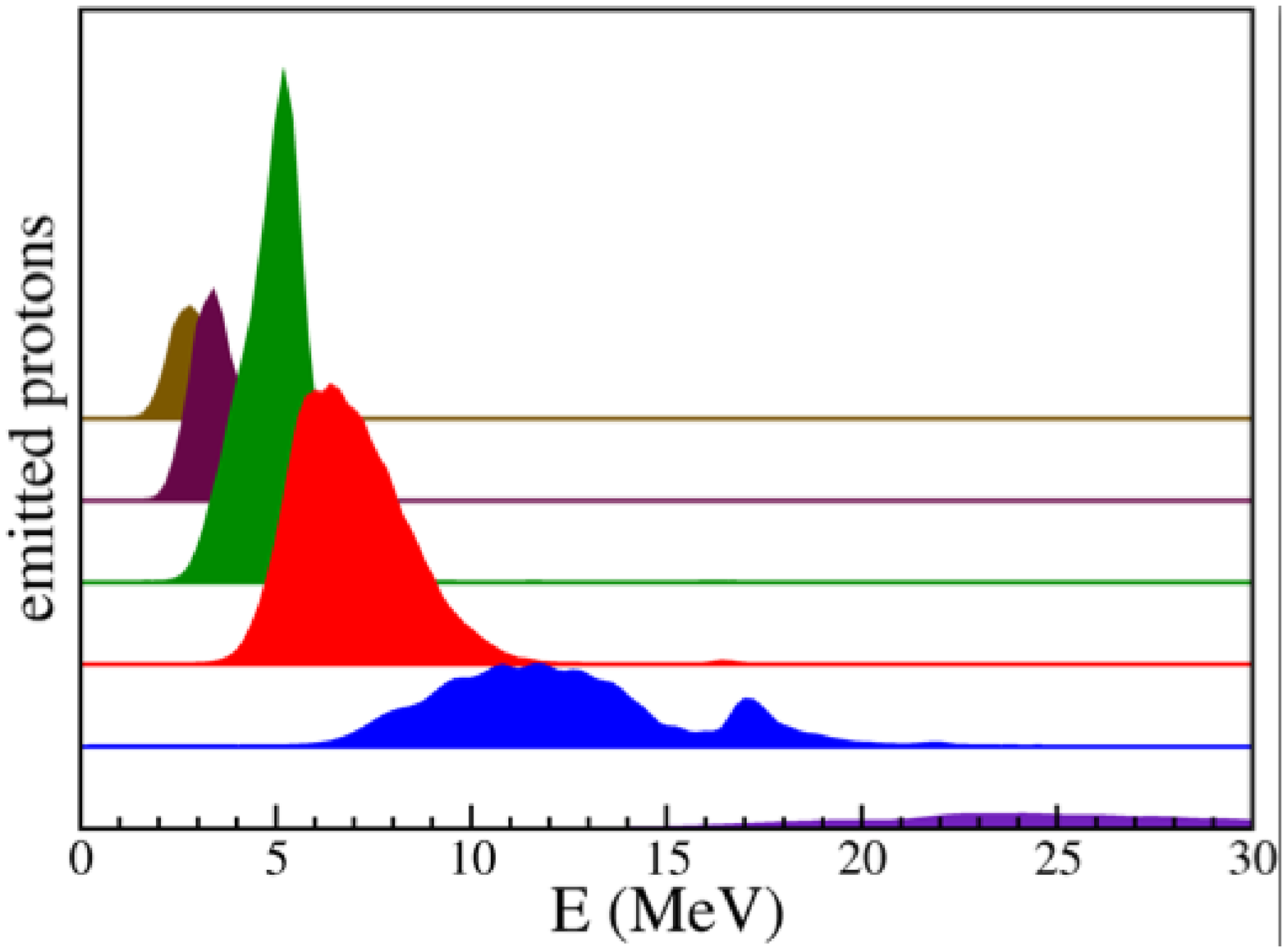}
\includegraphics[width=7.5cm]{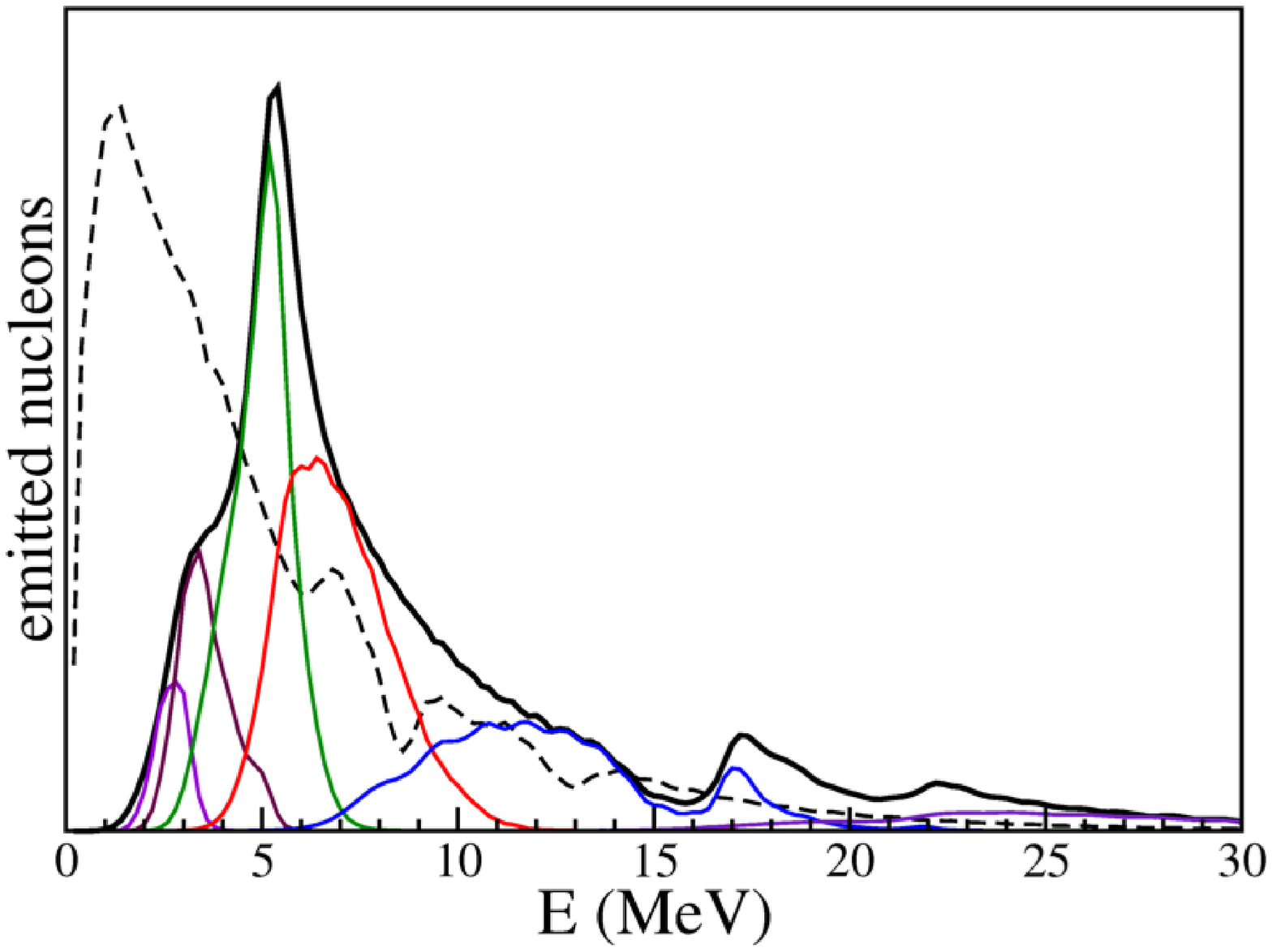}
\caption{Spectra of protons in the ''detector'' (see text and Fig.~\ref{fig:detector}) at different times following an isovector monopole boost in $^{40}$Ca. (top) Each spectrum is shifted vertically for clarity (time increases from bottom to top). The time delay between two consecutive spectra is $\Delta T=5$~zs. (bottom) The proton spectra at different times form an envelope (thick solid black line). The similar envelope obtained for neutron is also shown with a dashed line. 
\label{fig:spectra_ivGMR}}
\end{center}
\end{figure}

We can see in Fig.~\ref{fig:spectra_ivGMR} that both proton and neutron spectra exhibit some structures which cannot be explained by a simple hydrodynamical model.
Instead, one should seek for an explanation in terms of the microscopic structure of the GR. 
This motivated a more detailed investigation with a realistic TDHF spherical code and a full Skyrme EDF~\cite{ave12}.
A brief summary of the results for the GMR in $^{16}$O is presented here (see Ref.~\cite{ave12} for more details and for more results on, e.g., tin isotopes).
In this study, the \textsc{tdhfbrad} code~\cite{ave08} is used with the SLy4 parametrisation~\cite{cha98} of the Skyrme EDF without pairing. 
Fig.~\ref{fig:16Oa} shows the time evolution (top) of the monopole moment after an isoscalar monopole boost, and the associated spectrum (bottom) obtained within the linear response theory. 
The GMR spectrum exhibits structures which are associated to different single-particle orbitals. 
For instance, the high energy shoulder around 31~MeV is due to $s_{1/2}$ particle-hole excitations\footnote{The monopole excitations is associated to a $\Delta L=0\hb$ angular momentum transfer so that particle and hole have the same quantum numbers at the time of the excitation.}.

\begin{figure}
\begin{center}
\includegraphics[width=8.5cm]{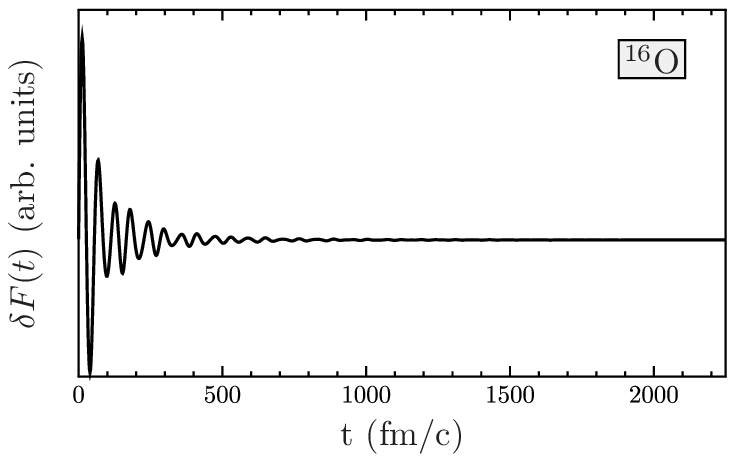}
\includegraphics[width=8.5cm]{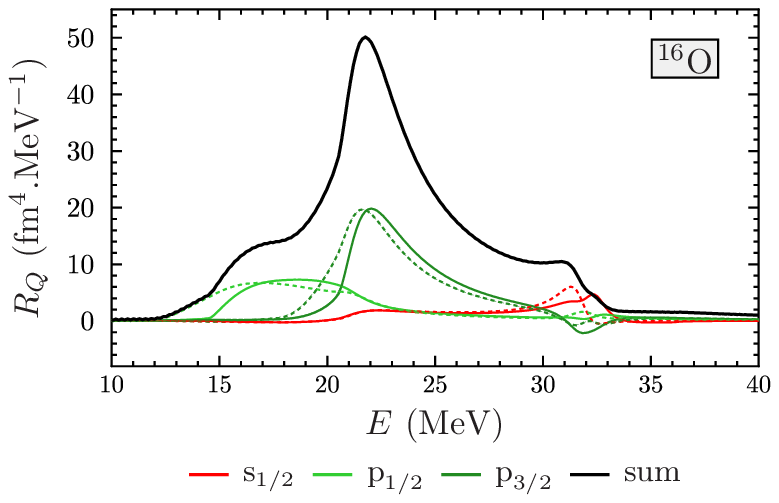}
\caption{(top) Time evolution of the monopole moment in $^{16}$O after a monopole boost obtained with the \textsc{tdhfbrad} code~\cite{ave08}. (bottom) Associated strength function and its decomposition onto single-particle quantum numbers $l$ and $j$ (spectroscopic notation).
Solid (dashed) lines show neutron (proton) contributions. 
\label{fig:16Oa}}
\end{center}
\end{figure}

The spectra of emitted protons and neutrons are shown in the upper panel of Fig.~\ref{fig:16Ob}.
The latter depend strongly on the associated single-particle quantum numbers. 
In particular, no $s_{1/2}$ nucleons are emitted. 
This is due to the fact that the  $1s_{1/2}$ hole state is deeply bound (-32.4~MeV for  protons and -36.2~MeV for neutrons according to the HF initial configuration~\cite{ave12}).
In fact, the high energy shoulder of the GMR spectrum (see Fig.~\ref{fig:16Oa}-bottom) does not have enough energy to bring the initial $1s_{1/2}$ particle into the continuum.

\begin{figure}
\begin{center}
\includegraphics[width=8.5cm]{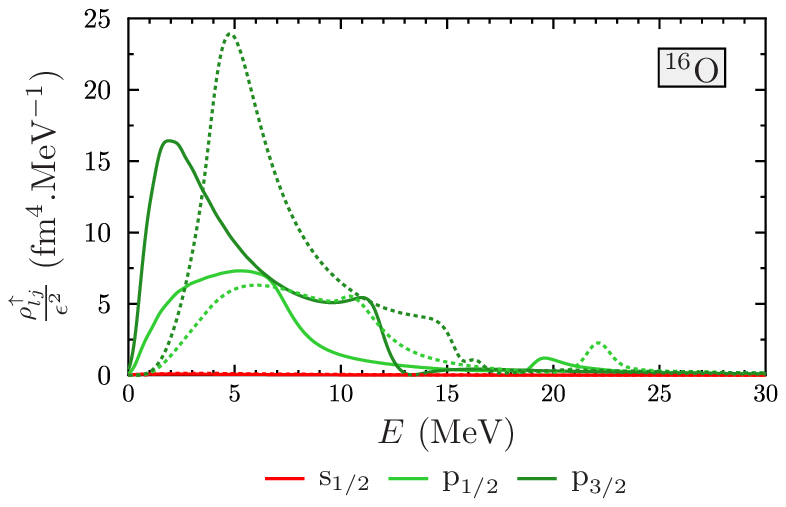}
\includegraphics[width=8.5cm]{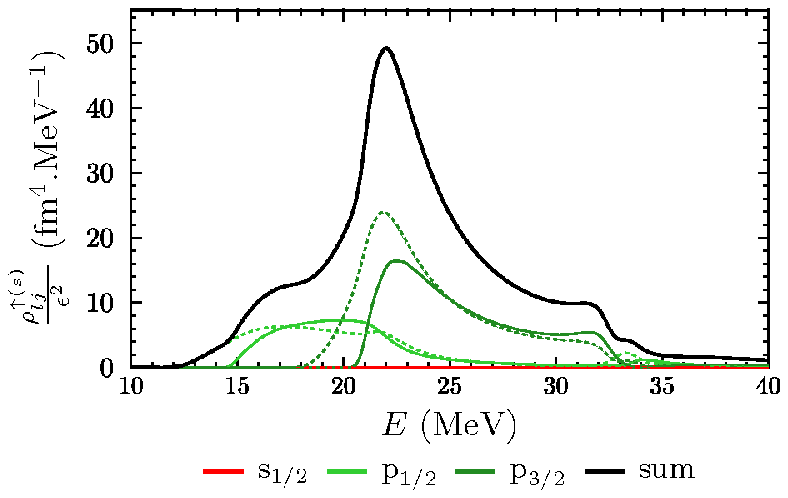}
\caption{(top) Neutron (solid lines) and proton (dashed lines) direct-decay spectra. (bottom) 
Same spectra ''shifted'' by the energy of 
the initially occupied single-particle state. Their 
sum is shown in black solid line. 
\label{fig:16Ob}}
\end{center}
\end{figure}

The lower panel of Fig.~\ref{fig:16Ob} shows the same quantity as the upper panel, with a shift in energy (different for each $l_j$ contribution) corresponding to the binding energy of the hole state. 
The sum of each shifted $l_j$ contribution gives a spectrum which is very close to the GMR spectrum (compare with Fig.~\ref{fig:16Oa}-bottom). 
The agreement is excellent for both the shape and the magnitude of the spectra. 
The origin of the structures in the direct  emission spectra is entirely due to the shell structure of the nucleus. 

It is also interesting to note that, although the high energy shoulder in the $^{16}$O GMR spectrum is due to the excitation of a {\it bound} $1p1h$ state, it appears in the ''shifted'' spectrum (Fig.~\ref{fig:16Ob}-bottom) due to the {\it emission} of particles in $p$-states.
In fact, the TDHF (or RPA) residual interaction is responsible for the coupling  between the bound  $1p1h$ $s_{1/2}$-state and unbound $1p1h$ $p$-states. 
Note that similar couplings have been obtained in tin isotopes~\cite{ave12}.
For instance, the GMR in $^{100}$Sn decays by protons only, while it is associated to a collective oscillation of both protons and neutrons. 

Coincidence experiments between particles emitted in the GR decay and the ejectile resulting from the GR excitation process have been performed in the past to investigate GR properties~\cite{har01}. 
The present theoretical analysis of GR direct decay allows, in principle, a direct comparison between theoretical and experimental spectra. 
However, for quantitative comparisons, one should use a more elaborated approach than the TDHF theory.
Indeed, the fact that TDHF does not include $2p2h$ residual interaction is a strong limitation, as the latter has been shown to be crucial to reproduce the width and the fragmentation of GR spectra~\cite{lac04}.
The present analysis of GR decay should then be repeated with, e.g., the extended-TDHF (ETDHF) or the time-dependent density-matrix (TDDM) approaches (see table~\ref{tab:sum_micro} and Ref.~\cite{sim10a}). 
Note also that calculations should be performed with 2D or 3D codes in order to study the decay of GR with higher multipolarity than the GMR. 

\subsection{Anharmonicity of giant resonances\label{sec:anharm}}

In the harmonic picture, a GR is the first phonon eigenstate of an harmonic oscillator describing the collective motion, and corresponds to a coherent sum of  $1p1h$ states \cite{boh75}.
Experimental observations of 2 and 3-phonon states proves the vibrational nature of GR.
However, they also show limitations of the harmonic picture~\cite{cho95,aum98,sca04}.
In particular, the excitation probability of multi-phonon states is  larger than predicted by the harmonic picture.
This indicates that different phonon states may be coupled by the residual interaction~\cite{vol95,bor97,sim03,fal03,cho04,lan06,sim09}. 

The TDHF approach takes into account some effects of the residual interaction if the considered phenomenon can be observed in the time evolution of a one-body observable. 
In particular, the nonlinear response in TDHF contains the couplings between one- and two-phonon states coming from the $3p1h$ and $1p3h$ residual interaction~\cite{sim03}.
In that sense, it goes beyond the RPA, which is a harmonic picture and contains only $1p1h$ residual interaction. 

The couplings leading to the excitation of a GMR or a GQR (resp. a GMR) on top of a GDR (resp. a GQR) were investigated in Ref.~\cite{sim03} using the nonlinear response to an external field in the TDHF theory.  
As a continuation to this work, different techniques to compute the matrix elements of the residual interaction responsible for these couplings were introduced in Ref.~\cite{sim09}.

These matrix elements can be written $v_{\mu}=\<\nu|\oV|\nu\mu\>$ where the residual interaction $\oV=\oH-\oH_0$ is the difference between the full Hamiltonian $\oH$ and the HF+RPA Hamiltonian $\oH_0$. 
$|\nu\>$ and $|\nu\mu\>$ are 1 and 2-phonon eigenstates of $\oH_0$ with eigenenergies $E_\nu=E_0+\hb\omega_\nu$ and $E_{\nu\mu}=E_0+\hb\omega_\nu+\hb\omega_\mu$, where $\omega_{\nu,\mu}$ denote the collective frequencies and $E_0$ is the ground state energy. 
The state $|\nu\mu\>$ can be seen as one phonon of the GR $\mu$ (e.g., a GQR) excited on top of one phonon of the GR $\nu$ (e.g., a GDR).

In addition to the original technique based on the non-linearities of the time-dependent response~\cite{sim03},  two other methods were introduced in Ref.~\cite{sim09}.
A brief summary of these three methods is given below (see Ref.~\cite{sim09} for more details):
\begin{itemize}
\item {\it method 1:} A boost $e^{-ik_\nu\oQ_\nu}$ applied on the ground state induces, at lowest order in $k_\nu$, an oscillation of $Q_\nu(t)=\<\oQ_\nu\>(t)$  linear in $k_\nu$, and an oscillation of $Q_\mu(t)$ quadratic in $k_\nu$ and proportional to $v_\mu$. Computing the response $Q_\mu(t)$ to such a boost with TDHF gives then access to~$v_\mu$.
\item {\it method 2:} The same boost can be applied on a HF state obtained with a small constraint $\lambda\oQ_\mu$. The linear response $Q_\nu(t)$ oscillates then with a frequency $\omega_\nu(\lambda)$. The variation $\frac{\partial\omega_\nu}{\partial\lambda}$ is proportional to $v_\mu$. Computing  $\frac{\partial\omega_\nu}{\partial\lambda}(\lambda=0)$ with a TDHF or a deformed RPA code\footnote{This technique involves linear response only. The matrix element $v_\mu$ can then be computed with a RPA code  allowing initial deformations (generated by the constraint $\lambda\oQ$) of the vacuum.} gives also access to $v_\mu$.
\item {\it method 3:} A ''double'' boost  $e^{-ik_\nu\oQ_\nu}e^{-ik_\mu\oQ_\mu}$ is applied on the ground state $|0\>$. Define the ''coupling response function'' as\footnote{This function differs from the standard response function [see Eq.~(\ref{eq:strengthlin})] essentially by the cosine function instead of a sine function.} 
\oeq
R_\nu^c(\omega) = \frac{-1}{\pi k_\nu k_\mu}\int_0^\infty dt\, \cos(\omega t) Q_\nu(t).
\label{eq:R_nu}
\ceq
At lowest order in $k_{\mu,\nu}$, we can show that $R_\nu^c(\omega)$ is proportional to $v_\mu$ and  exhibits peaks at $\omega_\nu$ and $|\omega_\nu\pm\omega_\mu|$ with opposite signs.
The amplitude of these peaks provides  a third way to extract $v_\mu$.
\end{itemize}

\begin{figure}
\begin{center}
\includegraphics[width=7.5cm]{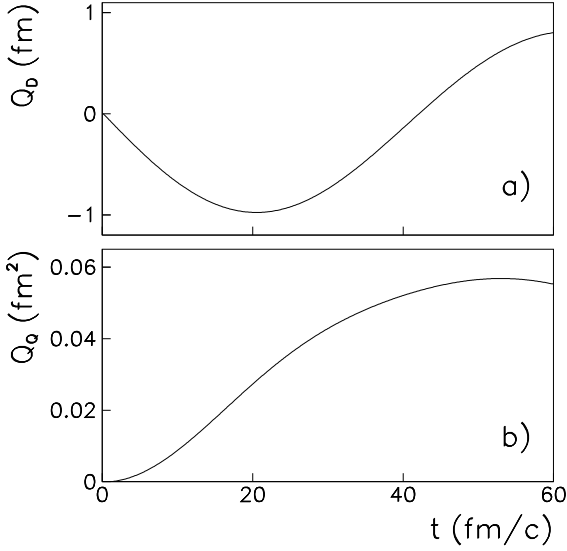}
\caption{Time evolution of the dipole (a) and quadrupole 
(b) moments in $^{132}$Sn after a dipole boost.
\label{fig:quadratic1}}
\end{center}
\end{figure}

\begin{figure}
\begin{center}
\includegraphics[width=7.5cm]{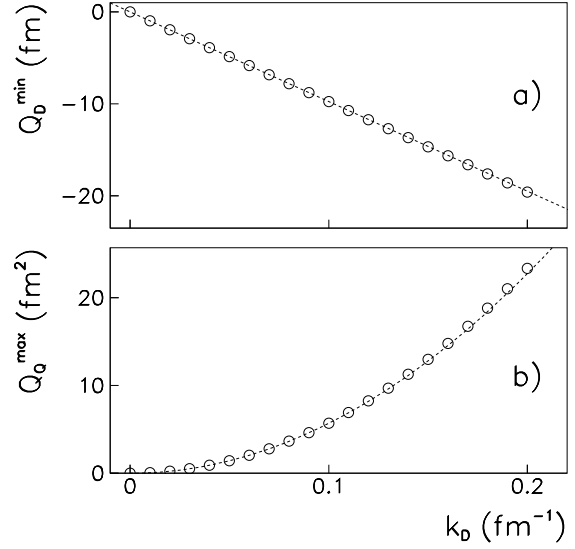}
\caption{Circles indicate the first minimum and maximum of the dipole (a) and quadrupole (b) moment evolution, respectively, following a dipole boost in $^{132}$Sn with a boost velocity~$k_D$. 
Dashed lines show linear and quadratic extrapolations at $k_D\rightarrow0$ of the dipole (a) and quadrupole (b) amplitudes, respectively.
\label{fig:quadratic2}}
\end{center}
\end{figure}

An illustration of the first method applied to the $^{132}$Sn nucleus with $\nu\equiv$GDR and $\mu\equiv$GQR is given in Figs.~\ref{fig:quadratic1} and~\ref{fig:quadratic2}.
On Fig.~\ref{fig:quadratic1}, we observe an oscillation of both the dipole and quadrupole moments, although the boost contains only the dipole moment. 
The oscillation of the quadrupole moment is, in fact, induced by the residual interaction. 
The right panel shows that, as expected, the amplitude of the dipole (resp. quadrupole) oscillation is linear (quadratic) in the boost velocity~$k_D$. 
Numerical application gives a matrix element of the residual interaction $v_Q^{(1)}\simeq-0.61$~MeV. 
The two other methods give $v_Q^{(2)}\simeq-0.56$ and $v_Q^{(3)}\simeq-0.68$~MeV, respectively, showing a relatively good agreement between the three methods~\cite{sim09}.

Couplings have been computed in other tin isotopes~\cite{sim09} and in other nuclei ($^{40}$Ca, $^{90}$Zr, and $^{208}$Pb) with TDHF~\cite{sim03} and with a boson mapping method~\cite{fal03}.
Refs.~\cite{sim03,fal03} also discuss couplings involving the GMR built on top of the GQR or the GDR. 
The TDHF results provide a confirmation to the amplitude of the couplings computed with the  boson mapping method (see discussion in Ref.~\cite{sim03}).

Another conclusion of Ref.~\cite{sim09} is that there is no (or little) dependence of the coupling between dipole and quadrupole motion with isospin.
However,  an overall decrease of the coupling is obtained with increasing mass, indicating that the couplings are mediated by the surface~\cite{sim03,fal03,sim09}. 

Overall, the couplings are small but significant compared to the GR energies (e.g., $v/\omega\sim5\%$ for $^{132}$Sn). 
Their effect on the first phonon is negligible, but becomes sensible on the second and third phonon, with a typical shift in $\hb\omega$ of the order of $\sim0.5$~MeV as compared to the harmonic picture~\cite{fal03}.
How the anharmonicities induced by these couplings affect the excitation probability to the multiphonon states have been investigated by Lanza {\it et al.} within a semiclassical coupled-channels formalism~\cite{lan06}.
This model, based on the boson mapping method for the multi-phonon properties, allows for calculations of inelastic cross sections for the multiple excitation of giant resonances induced by heavy-ion probes.
Their calculations show that these anharmonicities induce an increase of the inelastic cross-section (as compared to the harmonic model) in the multi-phonon region, in good agreement with experimental data.

The role of pairing correlations, neglected so far, should also be considered. 
For instance, fully self-consistent quasi-particle-RPA (QRPA) codes allowing for static deformations (see, e.g., the code developped by S. P\'eru~\cite{per08}) could be used to obtain the couplings between the GQR and the GMR built on top of it. 
Couplings with exotic modes such as the pygmy dipole resonance~\cite{lan09} should also be investigated with the present methods.

\subsection{Pairing dynamics\label{sec:pairing}}

The TDHF calculations presented in the previous sections were applied to ''normal'' vibrations, i.e., vibrations of the one-body density (also called {\it normal} density) $\rho(t)$ with matrix elements 
\oeq
\rho_{\al\be}(t)=\<\Psi(t)|\oad_\be\oa_\al|\Psi(t)\>.
\ceq
These vibrations do not probe directly the pairing correlations between nucleons. 
Inclusion of pairing is possible in the small amplitude limit with the QRPA based on HFB vacua. 
The HFB +QRPA has been widely used in nuclear structure studies~\cite{eng99,kha02,fra05,per08}.

Similarly to the fact that the TDHF approach is an extension to the HF+RPA, a natural extension to the HFB+QRPA is the time-dependent Hartree-Fock-Bogoliubov (TDHFB) theory~\cite{bla86}.
In particular, the TDHFB theory provides a fully self-consistent\footnote{Here, the self-consistency refers to the fact that the HFB vacuum and the residual interaction inducing the collective dynamics are derived from the same EDF.} response to an external excitation including pairing dynamics and non-linearities. 

The development of a realistic TDHFB spherical code with a full Skyrme EDF and a density-dependent pairing effective interaction was  initiated in Ref.~\cite{ave08} and applied to the study of pairing vibrations. 
Recently, 3-dimensional codes have been developed to study the effect of the pairing interaction on ''normal'' vibrations at the BCS level~\cite{eba10,sca12} and solving the  TDHFB equation with a Skyrme functional~\cite{ste11} and with the Gogny effective interaction~\cite{has12}.
The present section introduces briefly the TDHFB formalism and discusses the application to pairing vibrations presented in Ref.~\cite{ave08}.

\subsubsection{The TDHFB theory}

Pairing correlations are essentially due to an attractive short range contribution of the residual interaction in the $^1S_0$ channel\footnote{This notation means that the two nucleons are coupled to produce a total isospin 1, a total orbital angular momentum $L=0$ ($S-$wave), and a total spin 0.}~\cite{may50,dea03}.
Pairing correlations affect then mostly (but not only) time-reversed states. 
The pairing residual interaction induces a scattering of a pair of nucleons across the Fermi surface.
As a result, the ground-state with pairing correlations is a sum of $2p2h$ states where the $2p$ ($2h$) are essentially time-reversed states. 
Such a state is represented schematically in the upper part of Fig.~\ref{fig:pairing}.

\begin{figure}
\includegraphics[width=8.8cm]{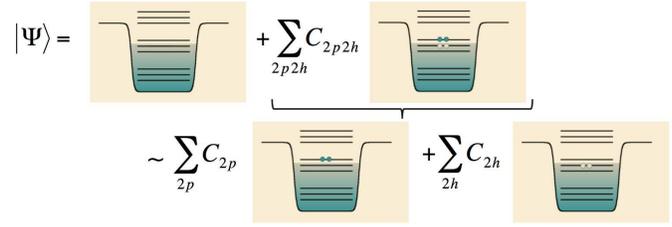}
\caption{Illustration of the configurations used to generate a ground state with pairing correlations. (top) Exact case. (bottom) BCS approximation. 
\label{fig:pairing}}
\end{figure}

The treatment of pairing correlations in finite nuclei is simplified with the Bardeen-Cooper-Schrieffer (BCS) approximation initially developed to interpret supraconductivity in metals~\cite{bar57}. 
In this approximation, the $2p2h$ states are replaced by a sum of $2p$ and $2h$ states (see bottom part of Fig.~\ref{fig:pairing}).
The resulting approximation of the ground-state can then be written as a vacuum of quasiparticles, allowing for the application of the Wick theorem~\cite{rin80}. 
As a result, the BCS approximation leads to a {\it generalised} mean-field theory. 

As we can see in Fig.~\ref{fig:pairing}, the ''price to pay'' is that the BCS ground-state is not an eigenstate of the particle number operator anymore. 
Pairing correlations are  included thanks to a gauge symmetry breaking. 
The Hartree-Fock-Bogoliubov (HFB) theory is more general but shares the same features with the BCS approach. 
In fact, it is an extension to the BCS theory where pairs are not limited to time-reversed states. 

In the (TD)HFB theory, all the information on the state of the system is contained in the {\it generalised} density matrix ${\cal R}$ defined as 
\begin{eqnarray}
{\cal R}  = \left( 
\begin{array} {cc}
\left( \langle \oad_j \oa_i \rangle \right) & \left(\langle \oa_j^{ } \oa_i \rangle \right)\\
&\\
\left(\langle \oad_j \oad_i \rangle \right) &  \left(\langle \oa_j \oad_i \rangle   \right)
\end{array} 
\right)
 = \left( 
\begin{array} {cc}
\rho & \kappa \\
- \kappa^* & 1-\rho^*  
\end{array} 
\right),
\label{eq:Rmatrix}
\end{eqnarray}
where $\kappa$ is the so-called {\it pairing tensor}.
$\kappa$ and $\kappa^*$ contain the pairing correlations (at the HFB level). 

The time evolution of the generalised density matrix is given by the TDHFB equation 
\begin{eqnarray}
i \hbar \frac{d{\cal R}}{dt} = \left[{\cal H} , {\cal R} \right] 
\label{eq:tdhfbR}
\end{eqnarray}
which has the same form than the TDHF equation~(\ref{eq:tdhf}). 
The generalised HFB Hamiltonian reads
\begin{eqnarray}
{\cal H}  \equiv \left( 
\begin{array} {cc}
h & \Delta \\
- \Delta^* & - h^*
\end{array} 
\right),
\end{eqnarray}
where 
\begin{eqnarray}
h_{\mu\nu}=\frac{\delta \mathcal{E}[\rho,\kappa,\kappa^*]}{\delta \rho_{\nu\mu}} \mbox{~~and~~} 
\Delta_{\mu\nu}=\frac{\delta \mathcal{E}[\rho,\kappa,\kappa^*]}{\delta
 \kappa^*_{\mu\nu}} 
 \label{eq:field}
\end{eqnarray}
are the HF Hamiltonian and the pairing field, respectively, and $ \mathcal{E}[\rho,\kappa,\kappa^*]$ is the EDF including pairing. 

\subsubsection{Application to pairing vibrations}

Pairing vibrations are a particular manifestation of the dynamics of pairing correlations~\cite{boh75,rin80,bes66}.
They are probed in two-nucleon transfer reactions~\cite{rip69,oer01,kha04,pll11,shi11}. 
Pairing correlations are then expected to induce a collectivity which manifests itself as an increase of transition amplitude toward states associated to pairing vibrations. 
Starting with an even-even nucleus ground-state with $A$ nucleons and spin-parity $0^+$, and assuming a $\Delta L=0$ direct pair transfer reaction, pair vibration states with $J^\pi=0^+$ are populated in the $A+2$ (pair addition) and/or $A-2$ (pair removal) nuclei. 

Such a process can be simulated within the TDHFB formalism using an initial boost with a Hermitean pair-transfer operator~\cite{bes66}
\begin{eqnarray}
\oF = 
\int d\mathbf{r} \, f(r) \left( \oad_{\mathbf{r},\downarrow} 
\oad_{\mathbf{r},\uparrow} 
+ \oa_{\mathbf{r},\uparrow} \oa_{\mathbf{r},\downarrow} \right),
\label{eq:pairtrans}
\end{eqnarray}
where the arrows label the spin of the single-particles (we omit the isospin to simplify the notation). 
In the present application, $f(r)$ is a Fermi-Dirac spatial distribution containing the nucleus and cutting at 4~fm outside the nucleus. 
Its role is to remove unphysical high energy modes associated to pair creation outside of the nucleus.

In this approach, it is assumed that the spectroscopy of the $A-2$, $A$ and $A+2$ nuclei can be obtained from the same quasiparticle vacuum (the $A$ ground-state). 
Note that recent improvements have been proposed by Grasso~{\it et al.} where this limitation is overcome for ground-state to ground-state transitions by using different vacua for the parent and daughter nuclei~\cite{gra12}.

\begin{figure}
\begin{center}
\includegraphics[width=4.5cm,angle=-90]{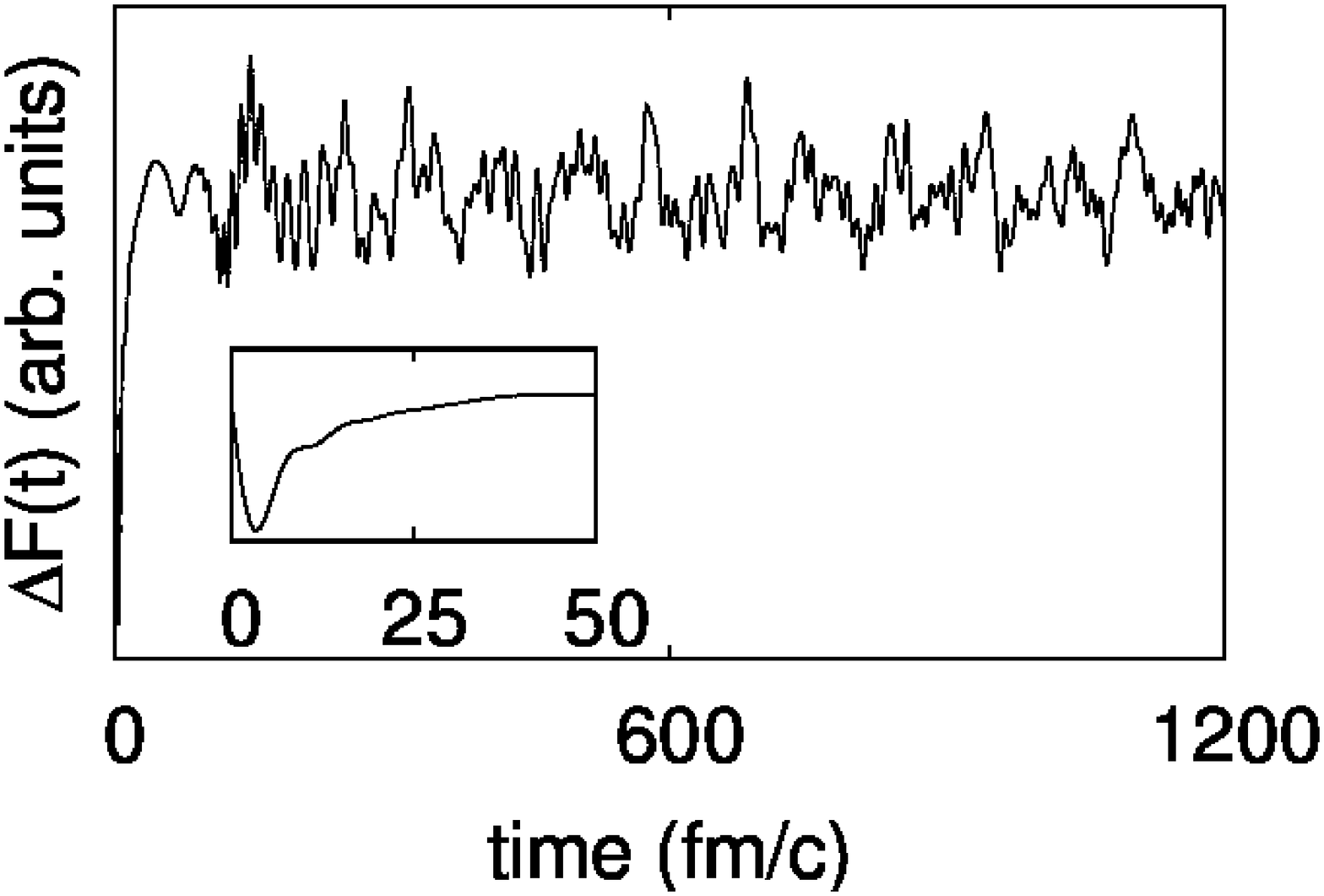}
\includegraphics[width=8cm,angle=-90]{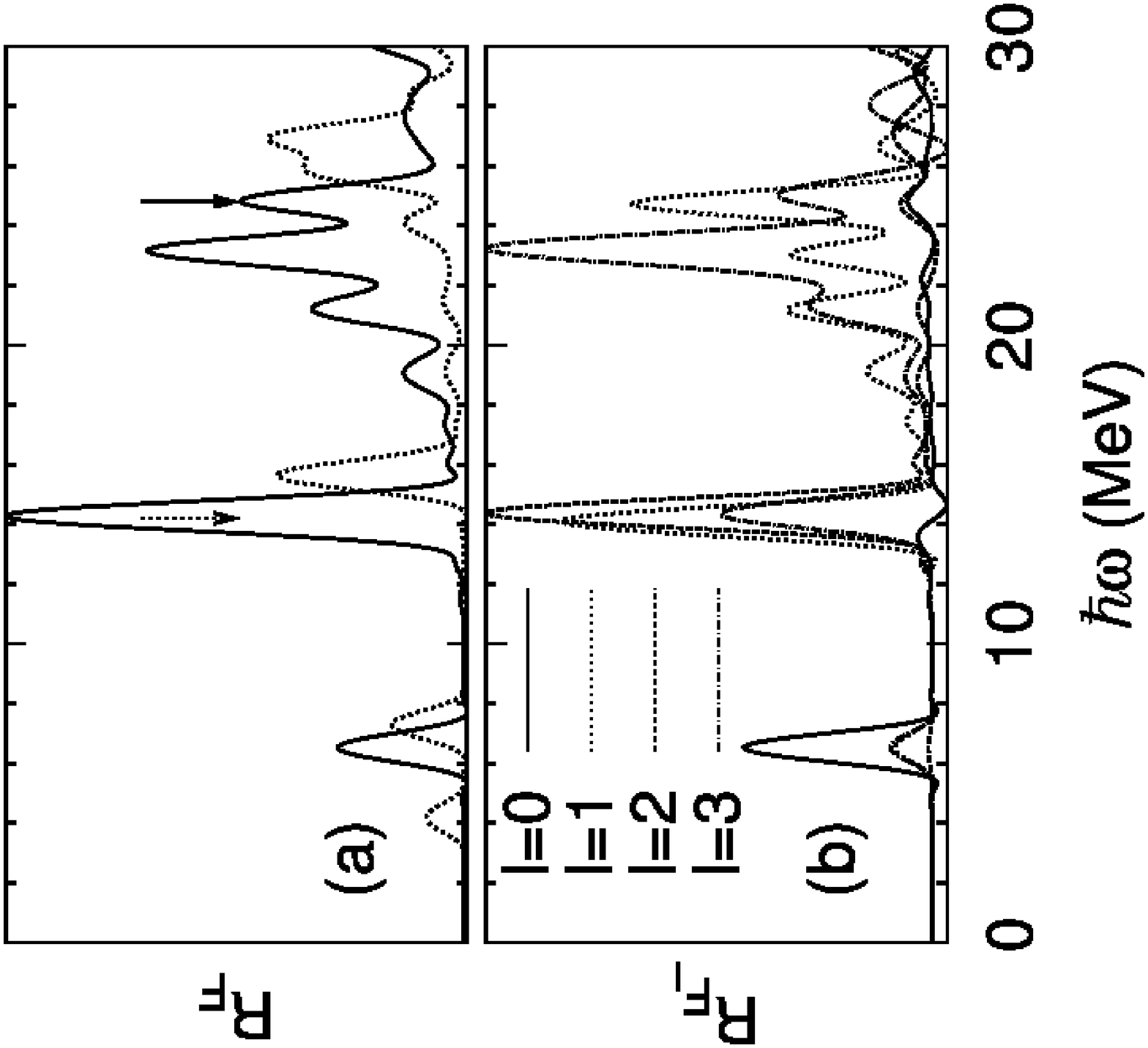}
\caption{(top) Evolution of $\<F\>(t)$ after a pair transfer type excitation on $^{18}$O. The inset shows the same quantity at early times. (middle) Associated TDHFB strength function (solid) compared with the unperturbed spectrum (dashed). 
The arrows indicate pair removal transitions from the $1p_{3/2}$ (solid) and $1p_{1/2}$ (dotted) deep hole states. 
(bottom) TDHFB strength function decomposed into single-particle orbital angular momentum $l$-components. 
\label{fig:pairvib}}
\end{center}
\end{figure}

The \textsc{tdhfbrad} code has been developed to solve the TDHFB equation in spherical symmetry with a full Skyrme EDF and density-dependent pairing effective interaction. 
As a first application, the linear response of $\<\oF\>(t)$ has been computed in several oxygen and calcium isotopes~\cite{ave08}.
The time-evolution of $\<\oF\>(t)$ is shown in the upper panel of Fig.~\ref{fig:pairvib} for a $^{18}$O vacuum.
The apparent chaotic behaviour of $\<\oF\>(t)$ is due to the simultaneous excitation of several pair vibrations, as we can see from the strength function (solid line) in the middle panel. 
Both pair additional and pair removal (indicated by the arrows) modes are present. 
A comparison with the unperturbed strength function (dashed line) obtained by removing the self-consistency of the generalised mean-field shows two features in the TDHFB spectrum:
\begin{itemize}
\item an increase of the strength,
\item and a lowering of the transition energies. 
\end{itemize}
Both are compatible with the attractive nature of the dynamical pairing residual interaction. 
In particular, the increase of the strength is a clear signature for collective effects. 
Note that similar conclusions were drawn from  continuum QRPA calculations by Khan {\it et al.}~\cite{kha04}.

The bottom panel of Fig.~\ref{fig:pairvib} shows a decomposition of the response in terms of single-particle orbital angular momentum $l$. 
Together with the structure of the initial HFB vacuum, this decomposition allows for an understanding of each peak in terms of their main particle and hole contributions. 
See Ref.~\cite{ave08} for a detailed microscopic analysis. 
See also Refs.~\cite{ave08,ave10} for an analysis of other nuclei (oxygen, calcium and tin isotopes).

This first realistic application of the TDHFB theory to nuclear systems has confirmed previous QRPA calculations of pairing vibrations~\cite{kha04}. 
Future applications of the \textsc{tdhfbrad} code to study the decay of GMR, and to investigate non-linear effects in pairing dynamics are envisaged. 
The recent development of 3-dimensional codes~\cite{ste11,eba10,has12} opens also new perspectives for the study of $L\ne0$ vibrations. 

\subsection{Conclusions and perspectives}

Real time mean-field calculations have been performed to investigate collective vibration properties. 
The response to an external excitation has been obtained with 3D and spherical TDHF codes and associated strength functions have been computed within the linear response theory. 

Direct decay of GR have been analysed from energy spectra of emitted nucleons. 
Within the TDHF approach, the latter contains enough information to reconstruct the strength function if the hole structure of the nucleus is known. 
A comparison between the microscopic decompositions of the strengths obtained from the time response of the excitation operator and from the emitted nucleon spectra shows that the residual interaction couples bound $1p1h$ states with unbound ones, allowing for particle emission in the continuum. 

Non linear vibrations were also studied within the TDHF framework. 
They are used to quantify the coupling between one GR phonon and two (different) GR phonon states.
The large values of the couplings which have been obtained in different nuclei confirm that these kind of couplings is a probable source of the anharmonicities observed in GR multiphonon experiments. 

Recents works were devoted to  the inclusion of pairing correlations in the mean-field dynamics. 
To that purpose, a realistic spherical TDHFB code has been built and applied to the study of pairing vibrations excited in pair-transfer reactions.
Comparisons with unperturbed calculations show that the dynamical pairing residual interaction included in TDHFB is attractive and induces some collectivity to the pairing vibrations. 

Several possible extensions to these works have already been discussed, such as the study of non-linear vibrations and particle decay with pairing, and the study of $L\ne0$ vibrations with 3-dimensional TDHFB codes.

Other perspectives  could be considered.

In the present studies, we focused on collective motion at zero temperature, in particular vibrations built on top of the ground state. 
The role of finite temperature on collective motion (e.g., the so-called ''hot GR'') has been widely discussed in the nuclear physics community~\cite{bor98}.
Questions such as the effect of temperature on pairing dynamics and on the couplings between GR multiphonon states could be addressed with an extension of the present calculations to finite temperature systems. 
TDHF studies of giant resonances at finite temperature are indeed possible~\cite{vau87} starting from an initial hot HF solution~\cite{bon84}.
In section \ref{sec:charge-eq}, we also discuss the particular case of GDR excited in the fusion of two nuclei. 

Recently, it has been shown with a molecular dynamics approach that GR in light nuclei ($^{12}$C, $^{16}$O, $^{24}$Mg) were also affected by $\alpha$-clustering~\cite{fur10}. 
In particular, these $\alpha$-clusters induce new vibrational modes which could couple to ''standard'' GR and produce a new source of anharmonicity. The coupling between these new modes and standard vibrations could be investigated in the future with the present techniques.

The calculations presented in this section and the possible perspectives discussed above are based on a mean-field approach. 
For a more realistic comparison to experimental data, extensions to theories going beyond the one-body limit are mandatory, in particular to reproduce the GR fragmentations and widths~\cite{lac04}.
Possible approaches include extended-TDHF~\cite{lac98}, second RPA~\cite{dro90,lac00}, time-dependent density matrix~\cite{toh01,toh02a,toh02b,wan85},
or stochastic one-body transport~\cite{rei92,lac01,jui02} theories.
Although they all face their own technical difficulties, these approaches could benefit from  the recent increase of computational power. 

\section{Heavy-ion collisions \label{chap:HIC}}

The previous section dealt with small amplitude vibrations. 
We now investigate the collision of two atomic nuclei which is by nature a large amplitude collective motion.
We first start in section~\ref{sec:motivations} by describing some motivations to study heavy-ion collisions. 

The outcome of a heavy-ion collision depends essentially on few properties of the entrance channel: energy, masses, angular momentum, deformation and orientation, $N/Z$ asymmetry, and internal structure (e.g., magicity). 
Thus, we discuss different reaction mechanisms and their dependence on entrance channel properties.
In sections~\ref{sec:fusion}, \ref{sec:transfer}, \ref{sec:DIC}, and \ref{sec:QF}, we present studies on fusion, transfer, deep-inelastic, and quasi-fission reactions, respectively. 
In section~\ref{sec:actinides} we present a theoretical study of actinide collisions. 
Note that only a brief summary of each study is given, in particular when more details can be found in the associated publications. 
Finally, we conclude and give some perspectives.

\subsection{Motivations\label{sec:motivations}}

There are several motivations to achieve a good understanding of heavy-ion reaction mechanisms.
Two of them are particularly relevant to the work presented in this section:
\begin{itemize}
\item To investigate quantum phenomena with complex systems,
\item To optimize the production of specific nuclei.
\end{itemize}
Some examples are also discussed below. 

\subsubsection{Unique perspectives brought by the nuclear case to study quantum phenomena}

A huge difference between the quantum treatment of a macromolecule, such as a fullerene, and an atomic nucleus is their interaction with their environments. 
Indeed, large molecules interact strongly with the surrounding gas and photon bath, while a system of colliding nuclei is free of such interactions. 
This is because $(i)$ nuclei do not interact with most photons in nature due to their small size (few fm), and $(ii)$ nuclear excited states have much longer lifetimes ($\sim10^{-18}$~s) than typical reaction times in heavy-ion collisions\footnote{This point is crucial. Indeed, if an excited state decays by $\gamma$-emission during the collision, then the state of the environment changes from the vacuum $0\gamma$-state to $1\gamma$-state. If such an emission occur, the system of nucleons would then be strongly coupled to its environment.} ($<10^{-20}$~s). 
Being entirely isolated during the reaction, heavy-ion collisions could then be considered as 
ideal to investigate the role of the composite nature of nuclei on quantum processes such as tunnelling\footnote{If the system was not isolated from its environment, then decoherence would occur and the system would have a classical behaviour~\cite{joo03}.}.  

As an example, one of the most striking signatures of quantum phenomena in nuclear collisions is the coherent quantum superposition of reaction channels leading to fusion barrier distributions~\cite{das98} which are interpreted within the coupled-channel framework~\cite{hag99}. 
Fusion via tunnelling is effectively a powerful microscope, magnifying quantum effects in nuclear collisions through the exponential dependence of tunnelling probability on the potential, which is modified by quantum coherent effects. 

\subsubsection{Production of specific nuclear systems: case of the heaviest nuclei\label{sec:introSHE}}

Stellar nucleosynthesis has produced a great variety of nuclei thanks to various reaction mechanisms. 
Of course all that remains on Earth are the stable and very long-lived by-products of these reactions, and in order to perform experimental nuclear studies, these must serve as our starting points. 
The internal structure of nuclei is of considerable interest in their own right and we can use the nuclei left by nature to create others, like new super-heavy elements (SHE) or isotopes which may or not be produced at the far end of the astrophysical r-process~\cite{lan11}. 

Transfermium nuclei are of great interests for our understanding of the quantum nature of the nuclear many-body problem as most of them are stable against fission thanks to quantum shell effects only. 
In particular, an island of stability is predicted in the super-heavy region with $Z=114-126$ and $N=172-184$ (see, e.g., Ref.~\cite{bender99} for mean-field calculations).

The discovery of SHE is important for various fundamental research fields: 
\begin{itemize}
\item {\it Astrophysics:} Does the r-process reach the super-heavy island of stability, or does it stop with spontaneous fissile lighter nuclei (see, e.g.,~\cite{lan11})? 
\item {\it Atomic physics and chemistry:} Relativistic effects on the electrons play a role on the atomic structure of SHE. As a result, SHE chemical properties may deviate from the standard classification suggested by the Mendeleiev table (see, e.g.,~\cite{eic07,gag11}). 
\item {\it Nuclear physics:} The precise location of the island of stability at the top of the nuclear chart would constrain the theoretical models (see, e.g.,~\cite{bender99}). 
\end{itemize}

The most important breakthrough in the quest for SHE came from heavy-ion fusion-evaporation reactions~\cite{hof00}.
Elements up to Z=118 have been produced with this technique~\cite{oga06}. 
However, the cross-sections for the production of SHE in their ground state are extremely small. 
The smallest cross-section measured is 30 femtobarns (fb) with the production of two $Z=113$ nuclei, after more than half-year irradiation of $^{209}$Bi targets with $^{70}$Zn beam~\cite{mor07}. 

Two main factors are  hindering SHE formation by fusion: 
\begin{itemize}
\item The quasi-fission mechanism strongly dominates~\cite{boc82,tok85,she87}. 
It leads to a fast re-separation of the projectile and target-like fragments after an exchange of nucleons from the heavy fragment to the light one. 
\item In case of fusion and formation of a compound nucleus at finite energy, the survival probability against fission is very small due to the small fission-barriers of SHE. 
\end{itemize}

\begin{figure*}
\includegraphics[width=18cm]{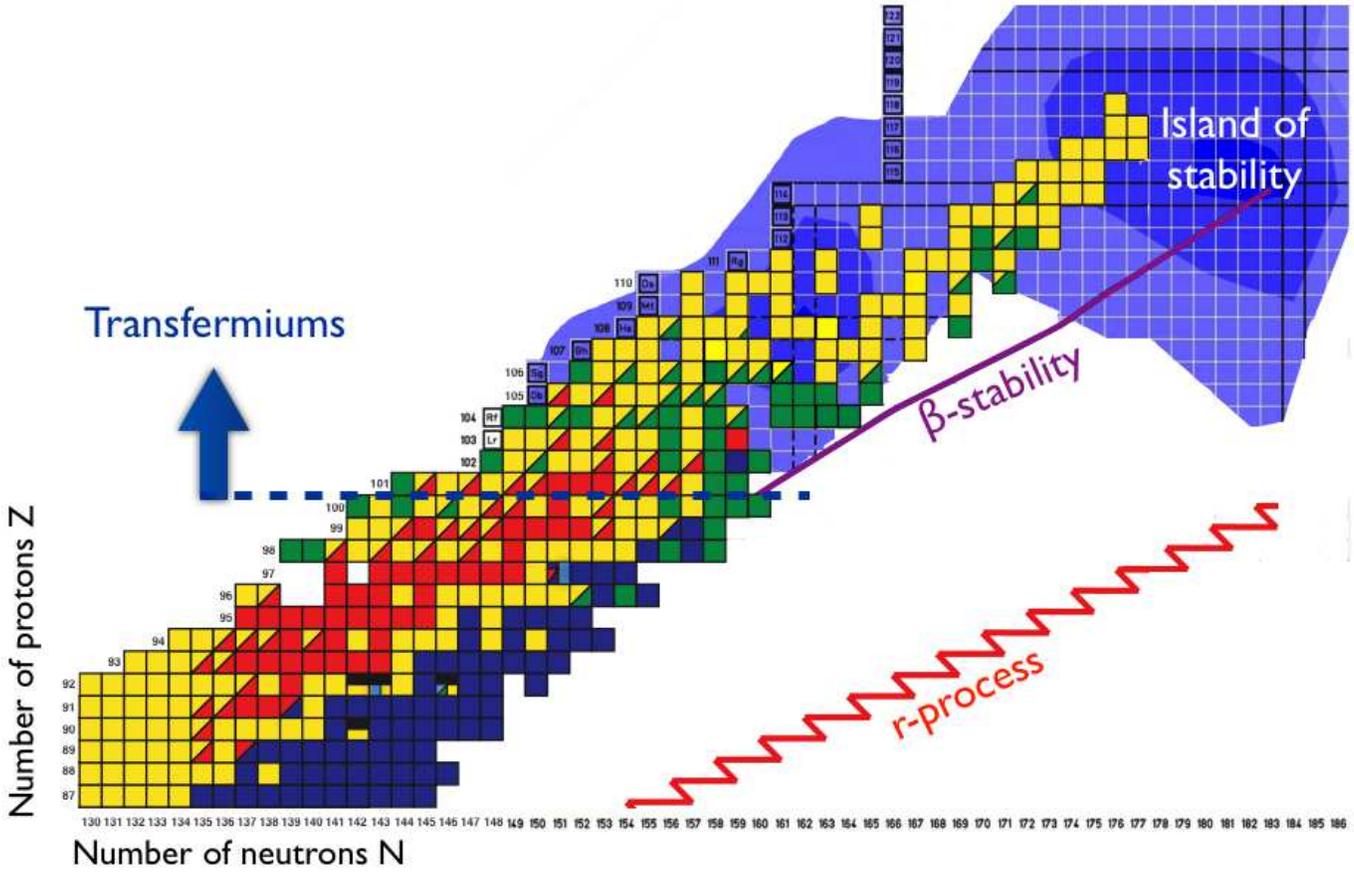}
\caption{Upper part of the nuclear chart. The dark blue area indicates quantum shell correction energy leading to more bound, and then more stable nuclei, according to a microscopic-macroscopic approach~\cite{smo97}. The purple line shows the expected $\beta$-stability line. The expected path followed by the r-process is shown with a red line.
\label{fig:SHE}}
\end{figure*}

A summary of the heaviest nuclei, most of them produced artificially on Earth, is given in Fig.~\ref{fig:SHE}. 
All transfermiums formed by fusion of stable nuclei are neutron-deficient. 
As a consequence, nothing is known about $\beta$-stable or neutron-rich transfermiums. 

There are three possible mechanisms which could produce neutron-rich transfermium nuclei and, then, get closer to the island of stability: $(i)$ a rapid neutron capture process, $(ii)$ fusion with neutron-rich nuclei, and $(iii)$ multi-nucleon transfer in actinide collisions. 
Neutron fluxes in nuclear reactors are not sufficient to use $(i)$. 
A critical limitation of the upcoming exotic beam facilities is the beam intensities, without which the feasibility of $(ii)$ is questionable. 
However, the process $(iii)$, which is discussed in Sec.~\ref{sec:actinides}, seems promising. 
The main advantages of using actinides are: 
\begin{itemize}
\item They exist in nature (e.g., $^{238}$U) or can be produced as radioactive targets (e.g., $^{248}$Cm). 
\item They have $\sim50\%$ more neutrons than protons, which is of great help for the production of new isotopes along the $\beta$-stability line. 
\end{itemize}

\subsection{Fusion with medium mass systems\label{sec:fusion}}

Dynamical mean-field theories like TDHF are well suited to the study of low-energy reaction  mechanisms, such as fusion, at energies around the barrier.
Indeed, at low energies, the Pauli principle blocks collisions between nucleons, increasing their mean-free path to the order of the size of the nuclear system. 
In addition, fusion occurs by transferring relative motion into internal excitation via one-body mechanisms well treated by the TDHF approach.

Early TDHF codes have been successfully applied to describe above-barrier fusion reactions in light systems~\cite{bon78}.
However, these calculations also predicted a lower limit to the angular momentum for fusion. 
For smaller angular momenta, a ''transparency'' was observed in the calculations (two fragments are emitted along the collision axis). 
This prediction was never confirmed experimentally.
In fact, it was shown by Umar and collaborators that this so-called ''fusion-window'' problem was solved with the inclusion of the spin-orbit interaction~\cite{uma86}.
Indeed, the latter was shown to be an important source of dissipation in heavy-ion collisions. 
Modern TDHF calculations are now performed with a full Skyrme EDF including spin-orbit terms~\cite{kim97,mar05,nak05,uma05}. 

The first observable we propose to study is the capture threshold between two nuclei.
We consider here systems with $Z_1Z_2<1600$ to avoid the well known fusion hindrance observed in very heavy systems\footnote{The case of heavier systems exhibiting fusion hindrance is discussed in section~\ref{sec:QF}.}. 
We also discuss above barrier fusion cross-sections, and investigate the path to fusion from the charge equilibration process. 

\subsubsection{Fusion of spherical nuclei: the example of the $^{16}$O+$^{208}$Pb system\label{sec:spher}}

A reference nucleus-nucleus potential could be obtained in the frozen approximation with HF (or HFB) densities~\cite{den02}, where the collision partners are assumed to keep their ground-state density during the approach. 
The frozen potential can be computed with the same Skyrme EDF as in the TDHF calculations by translating the nuclei in their HF state~\cite{sim09b}. 
Comparisons between TDHF and frozen fusion barriers allow to identify the role of dynamical effects, which are included in TDHF but absent from the frozen approach. 

Writing the HF energy $E[\rho]$ as an integral of an energy density $\mH[\rho(\vr)]$, i.e., 
\oeq
E[\rho]=\int \d\vr \sdf \mH[\rho(\vr)],
\ceq
we get the expression for the frozen potential 
\oeq
V(\vR)=\int \d\vr \sdf \mH[\rho_1(\vr)+\rho_2(\vr-\vR)] - E[\rho_1] -E[\rho_2],
\label{eq:frozen}
\ceq
where $\vR$ is the distance between the centers of mass of the nuclei, and $\rho_{1,2}$ are the densities of their 
HF ground-state. 
In order to take into account the Pauli principle between the nucleons of one nucleus and the nucleons of the other one in Eq.~(\ref{eq:frozen}), a proper treatment of the kinetic energy has to be considered, for instance using the Thomas-Fermi approach~\cite{den10a,den10b}.
However, for light and medium heavy systems, the barrier radius is large enough to neglect the Pauli principle between the two reactants at the barrier.
We then neglect the latter in  the determination of the HF-frozen barrier height. 

An example is shown in Fig.~\ref{fig:frozen} for the $^{16}$O+$^{208}$Pb system which could be considered as a benchmark in low energy reaction studies (see, e.g., Refs.~\cite{vid77,tho89,mor99,das07,eve11,sim10b}).
A comparison with the Wong formula~\cite{won73} is given. 
We see that, although the potential heights agree, differences appear at short distances. 
In fact, inside the barrier, the parametrisations of nucleus-nucleus potentials are less constrained by experimental data. 
In addition, as mentioned before, the frozen approximation neglects the Pauli principle for the nucleons of different collision partners, which may affect the inner barrier region. 

The height of the barrier obtained with the frozen approximation\footnote{The same value has been obtained, independently, by Washiyama {\it et al.}~\cite{was08} and Guo {\it et al.}~\cite{guo12}.} is $V_B^{frozen}\simeq76.0$~MeV. 
This value is close to the barrier obtained with the Wong formula~\cite{won73}, $V_B^{Wong}\simeq75.9$~MeV, while it is 1 MeV smaller than the Bass barrier~\cite{bas77}, $V_B^{Bass}\simeq77.0$~MeV. 
All these barriers overestimate the experimental value obtained from the centroid of the barrier distribution (see Fig.~\ref{fig:dist_barr}) $V_B^{exp.}\sim74.5$~MeV.

\begin{figure}
\begin{center}
\includegraphics[width=7.5cm]{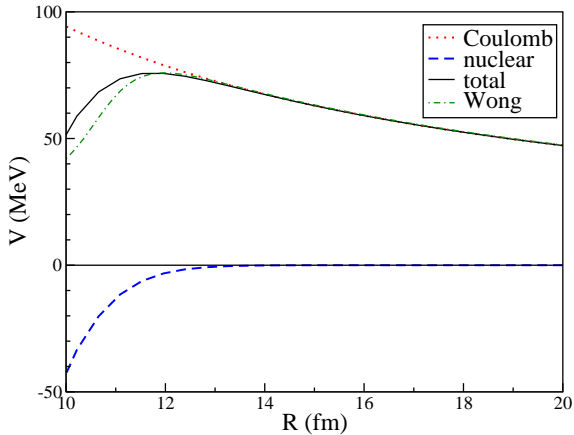}
\caption{Nucleus-nucleus potentials obtained with the frozen approximation in the $^{16}$O+$^{208}$Pb system. Nuclear (dashed blue line) and Coulomb (dotted red line) contributions, and their sum (solid line) have been obtained with the frozen approximation. 
For comparison, the Wong potential~\cite{won73} is shown in green dot-dashed line. The latter is obtained with a potential depth  $V_0=70$~MeV, a potential diffuseness $a=0.48$~fm, and nuclear radii $R_i=1.25A_i^{1/3}$~fm. 
\label{fig:frozen}}
\end{center}
\end{figure}

\begin{figure}
\begin{center}
\includegraphics[width=7.5cm]{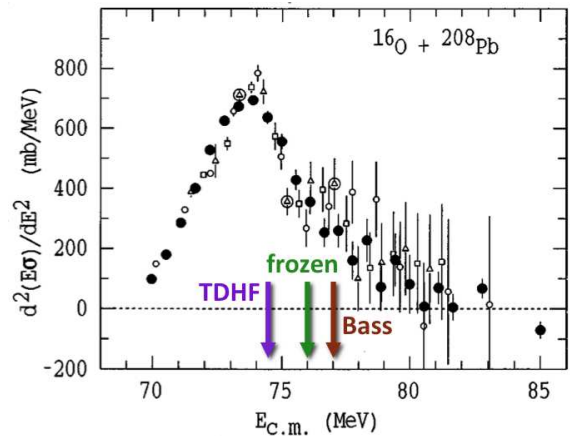}
\caption{Experimental fusion barrier distribution of the $^{16}$O+$^{208}$Pb system from Ref.~\cite{mor99}. The frozen, Bass and TDHF barriers (see text) are shown with arrows. 
\label{fig:dist_barr}}
\end{center}
\end{figure}

To investigate the possible role of dynamical effects on the fusion barrier, the latter has been computed with the \textsc{tdhf3d} code~\cite{kim97}. 
Because the TDHF theory does not allow for quantum tunnelling of the many-body wave function, the TDHF fusion barrier can be identified as the capture threshold for central collisions, above which a compound system is formed and below which the two fragments re-separate. 
Due to the finite time of the TDHF evolutions, one has to define a maximum computational time\footnote{This time may depend on the system. For medium mass systems such as $^{16}$O+$^{208}$Pb, a typical time of $10^3$~fm/$c$ (1~zs=300~fm/$c$) is used.} above which the final configuration (i.e., one compound system or two fragments) is assumed to be reached. 

\begin{figure}
\includegraphics[width=8.8cm]{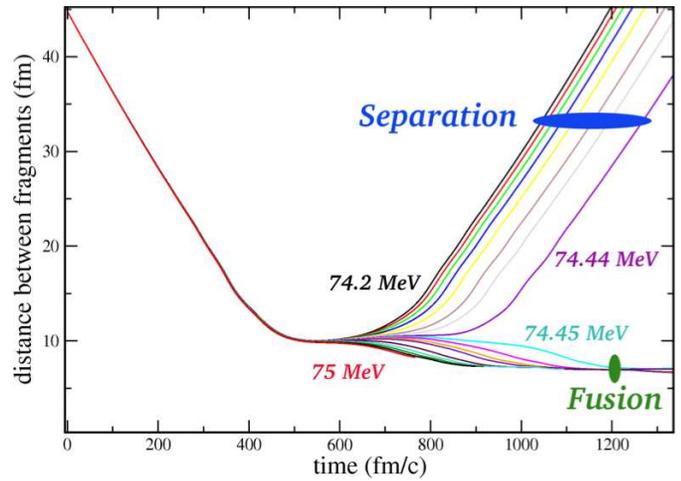}
\caption{Relative distance between the fragments as a function of time for head-on 
$^{16}$O+$^{208}$Pb reactions computed with the \textsc{tdhf3d} code~\cite{kim97}.\label{fig:distances}}
\end{figure}

Figure~\ref{fig:distances} shows the evolution of the relative distance between fragment centers of mass in central $^{16}$O+$^{208}$Pb collisions at different energies around the capture threshold. 
We clearly identify two sets of trajectories associated to capture (fusion) and to re-separation of the fragments. 
These calculations predict a fusion threshold of $V_B^{TDHF}=74.445\pm0.005$~MeV. 
As a result, the dynamical effects included in TDHF calculations lower the barrier by $\sim1.5$~MeV for this system as compared to the HF-frozen calculation. 
We observe in Fig.~\ref{fig:dist_barr} a good agreement between the TDHF prediction and the centroid of the experimental barrier distribution. 
Other methods based on a macroscopic reduction of the mean-field dynamics, namely the dissipative-dynamics TDHF \cite{was08} and the density-constrained TDHF \cite{uma09b},
also find similar results with an energy dependence to the
barrier heights ranging from 74.5 at low energies to 76 MeV
at higher energies where the frozen approach is expected to be more reliable. 

\begin{figure}
\includegraphics[width=8.8cm]{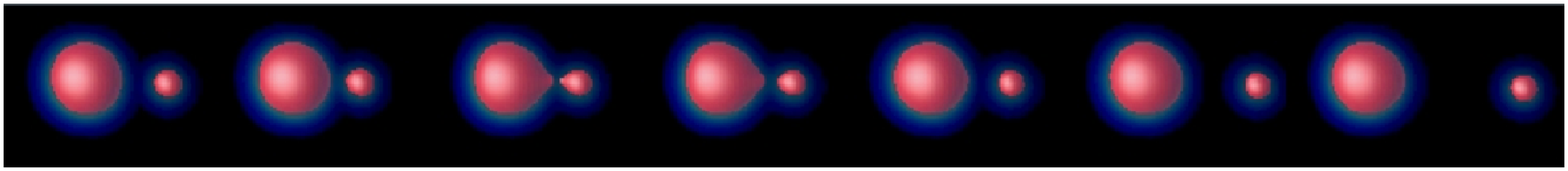}
\includegraphics[width=8.8cm]{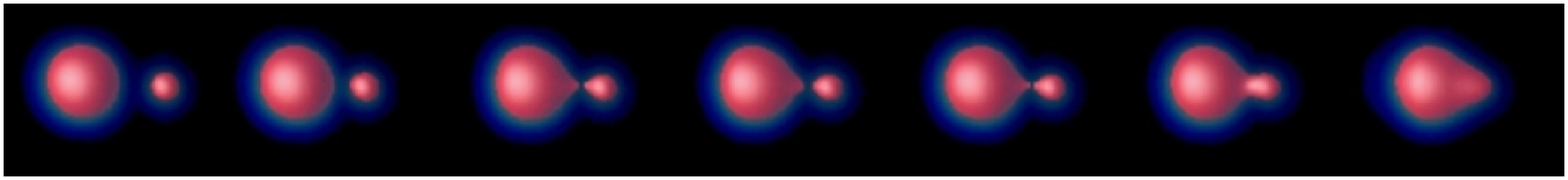}
\caption{(top) Density evolution for the reaction $^{16}$O+$^{208}$Pb corresponding to a head-on collision at a center of mass energy $E_{c.m.}=74.44$~MeV (just below the fusion barrier). The red surfaces correspond to an iso-density at half the saturation density ($\rho_0/2=0.08$~fm$^{-3}$). Each figure is separated by a time step of 135~fm/c. Time runs from left to right. (bottom) Same at $E_{c.m.}=74.45$~MeV, i.e., just above the fusion threshold.\label{fig:dens}}
\end{figure}

To get a deeper insight into these dynamical effects,  the density evolutions at $E_{c.m.}=74.44$ and 74.45~MeV are plotted in Fig.~\ref{fig:dens}.
In the first case, a ''di-nuclear'' system is formed during a relatively long time ($\sim500$~fm/c) before re-separation. 
In the second case, the system overcomes the fusion barrier. 
More generally, the two figures illustrate the richness of physical phenomena contained in TDHF calculations: surface diffuseness, neck formation, quadrupole/octupole shapes of the compound system... 

The observed lowering of the fusion barrier due to dynamical effects could be partly explained by a coupling of the relative motion to a transfer mechanism~\cite{sim08}.
In fact, the outgoing channel of $^{16}$O+$^{208}$Pb at $E_{c.m.}=74.44$~MeV (see top of Fig.~\ref{fig:dens}) is, in average, $^{14}$C+$^{210}$Po. 
This two-proton transfer channel effectively lowers the barrier by decreasing $Z_1Z_2$ and, then, the Coulomb repulsion. 
Transfer reactions in the $^{16}$O+$^{208}$Pb system are discussed in more details in section~\ref{sec:transfer}. 
Note  that low-lying collective vibrations, such as the first $3^-$ state in $^{208}$Pb (see Fig.~\ref{fig:PbQ3}) also affect the fusion barrier distribution~\cite{mor99}.

\begin{figure}
\includegraphics[width=8.8cm]{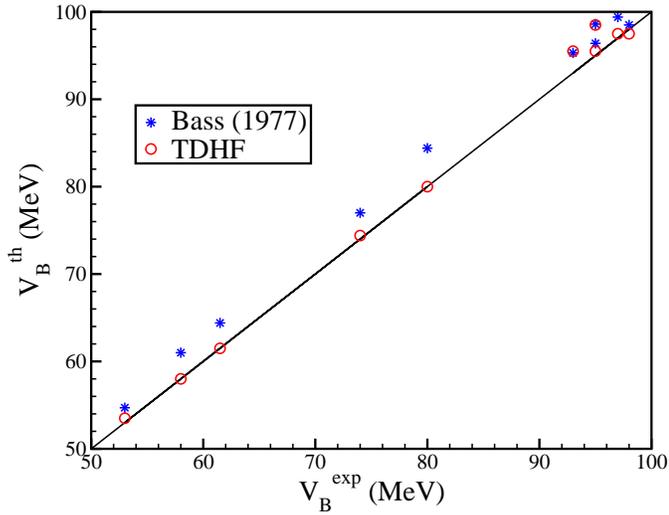} 
\caption{Bass barriers~\cite{bas77} (stars) and barriers 
extracted from TDHF calculations (circles) as a function of experimental barriers (centroids of fusion barrier distributions).  
}
\label{fig:th_exp}
\end{figure}

Systematic calculations of fusion barriers have been performed for medium mass systems involving spherical nuclei~\cite{sim08}. 
A summary of the results is shown in Fig.~\ref{fig:th_exp}.
A good reproduction of the barrier distribution centroids has been obtained (better than the Bass parametrisation) for all the studied systems. 
Other calculations with 3-dimensional TDHF codes confirmed the predictive power of the TDHF approach for the determination of fusion barriers~\cite{was08,guo12}. 

Above barrier fusion cross-sections have been computed for the $^{16}$O+$^{208}$Pb system in Ref.~\cite{sim08}.
The fact that fusion probabilities are either 0 or 1 implies that cross sections are obtained using the ''quantum sharp cutoff formula''~\cite{bla54}
\oeq
\si_{fus} (E) = \frac{\pi \hbar^2}{2\mu E} \sdf [l_{max}(E)+1]^2,
\ceq
where the fusion probability is $0$ for $l>l_{max}(E)$ and 1 for $l\le l_{max}(E)$.
To avoid discontinuities due to the integer values of  $l_{max}(E)$,
$[l_{max}(E)+1]\hbar$ is generally approximated by its semi-classical equivalent $\mL_c=\sqrt{2\mu E}\, b_c$.
The latter corresponds to the classical angular momentum threshold for fusion and $b_c$ denotes the maximum impact parameter below which fusion takes place \cite{bas80}.
We finally obtain the standard classical expression for fusion cross sections
 $\si_{fus}(E) \simeq \pi \mL_c^2/2\mu E = \pi b_c^2$.

\begin{figure}
\includegraphics[width=8.8cm]{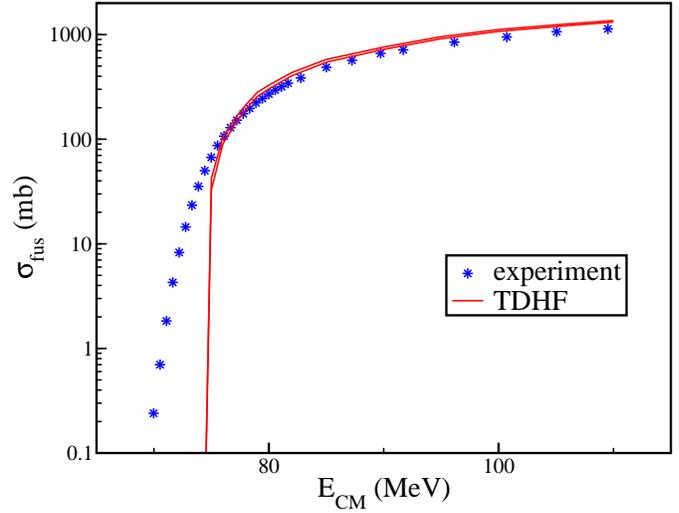} 
\caption{Experimental fusion cross sections from Ref.~\cite{mor99}
(stars) compared to cross sections deduced from TDHF calculations (lines) of $^{16}$O+$^{208}$Pb collisions.
The two lines correspond, respectively, to lower and upper limits of theoretical 
cross sections.}
\label{fig:fus}
\end{figure}

The results are shown in Fig.~\ref{fig:fus} for the $^{16}$O+$^{208}$Pb system. 
Fusion cross-sections are overestimated by about $16\%$ above the barrier. 
Although this discrepancy is small for a theory which has no parameter adjusted on reaction mechanisms, its origin is unclear. 

Finally, the calculations are not able to reproduce the sub-barrier energies. 
This is of course one of the main drawbacks of the TDHF approach. 
The inclusion of quantum tunnelling of the many-body wave function is clearly one of the biggest challenges in the microscopic treatment of low-energy nuclear reactions.

\subsubsection{Fusion barriers with a deformed nucleus}

\begin{figure}
\includegraphics[width=8.8cm]{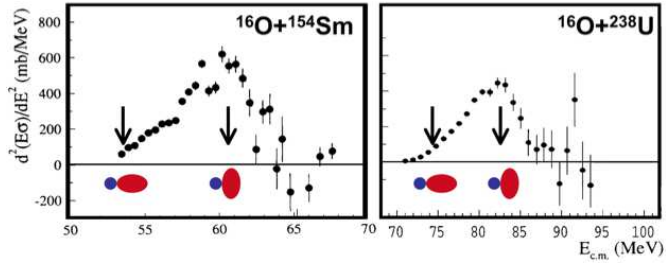} 
\caption{(left) Experimental barrier distributions for $^{16}$O+$^{154}$Sm~\cite{lei95}.
(right) Same for the $^{16}$O+$^{238}$U system~\cite{hin96}.
The arrows indicate the barriers obtained from TDHF calculations for 
central collisions with the tip (lower barriers) and with the side (higher barriers) of the deformed nucleus. 
From Ref.~\cite{sim08}.}
\label{fig:def}
\end{figure}

We now consider collisions of a spherical nucleus on a deformed one. 
In such a case, the barrier depends on the orientation of the deformed
nucleus at the touching point, leading to a wider barrier distribution 
than the single barrier case~\cite{das98,row91}.

Fig.~\ref{fig:def} shows two examples of experimental barrier distributions involving a prolatly deformed heavy target \cite{lei95,hin96}. 
Such barrier distributions are usually well reproduced in the framework of coupled channel calculations~\cite{das98}.
The standard interpretation is that the low (resp. high) energy part of the barrier distribution corresponds to collisions with the tip (side) of the deformed nucleus.
This interpretation has been confirmed with standard TDHF calculations~\cite{uma06c,sim08} and density-constrained TDHF (DC-TDHF) calculations~\cite{uma06b,uma06d,uma07}. The DC-TDHF technique allows for the computation of an energy-dependent nucleus-nucleus potential from TDHF trajectories (see Refs. \cite{cus85,uma85,uma06b} for details). The fusion thresholds for the two different orientations are represented with arrows in Fig.~\ref{fig:def}.
This confirms that collisions with the tip of the deformed nucleus have lower fusion barriers than collisions with the side. 
We conclude that, in addition to a good reproduction of the centroids,
modern TDHF calculations also reproduce the widths of the barrier distributions
generated by static deformations of heavy targets without any adjustment of parameters. 


The case of a light deformed projectile on a heavy spherical target has been investigated in Ref.~\cite{sim04} both within the TDHF approach and with a coupled channel framework~\cite{hag99}.
For such systems, the barrier distribution is affected by the reorientation of the deformed nucleus in the Coulomb field of the target. 
This induces an anisotropy of the orientation axis distribution and results into a fusion hindrance at low energies~\cite{sim04,uma06c,uma07}.
Possible experimental evidences of this effect have been reported~\cite{nay07}.
Note that the reorientation is proportional to $A_{spherical}/A_{total}$  and could be neglected in the systems studied in Fig.~\ref{fig:def}.

Finally, it is worth mentioning that, although TDHF calculations help to understand the structure of fusion barrier distributions, they would not be able to reproduce their detailed structure due to a lack of quantum effects associated to the collective degrees of freedom. 
In particular, the state of the deformed nucleus should be a coherent superposition of different orientations in the laboratory frame.  
This quantum coherence is lost at the mean-field level.
A possible improvement would be to restore this coherence using a time-dependent generator-coordinate method (TDGCM) \cite{rei83}, using the orientation of the nucleus as a collective coordinate.

\subsubsection{Charge equilibration in fusion \label{sec:charge-eq}}

The charge equilibration process in fusion reactions has been investigated with microscopic models such as  semi-classical~\cite{sur89,cho93,bar96,bar01} and TDHF \cite{bon81,sim01,sim07,uma07,obe12} approaches.
Charge equilibration occurs when two nuclei with different $N/Z$ collide. 
This difference induces a net dipole moment at contact which can oscillate. 
This isovector dipole oscillation is also called preequilibrium GDR, as it is a collective motion occuring in the preequilibrium stage of the compound system, i.e., before a complete equilibration of its degrees of freedom is reached.  

\begin{figure}
\begin{center}
\includegraphics[width=7cm]{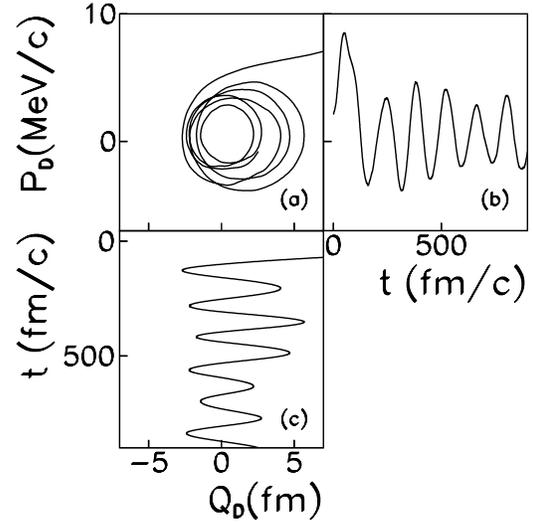} 
\caption{Evolution of the expectation value of the dipole moment $Q_D$ and its conjugated moment $P_D$ in the case of the $N/Z$ asymmetric reaction $^{40}$Ca+$^{100}$Mo at a center-of-mass energy of 0.83 MeV/nucleon.}
\label{fig:fli}
\end{center}
\end{figure}

An example of such preequilibrium dipole motion is shown in Fig.~\ref{fig:fli}, where the time evolution of the dipole moment $Q_D$ (proportional to the distance between the proton and neutron centers of mass) and its conjugated moment $P_D$ (proportional to their relative velocity) are shown in the case of the $N/Z$ asymmetric reaction $^{40}$Ca +$^{100}$Mo \cite{sim07}. 
$P_D$ and $Q_D$ oscillate in phase quadrature. They exhibit a spiral in the plot of $P_D$ as a function of $Q_D$ due to the damping of the dipole vibration. 

\begin{figure}
\begin{center}
\includegraphics[width=6cm]{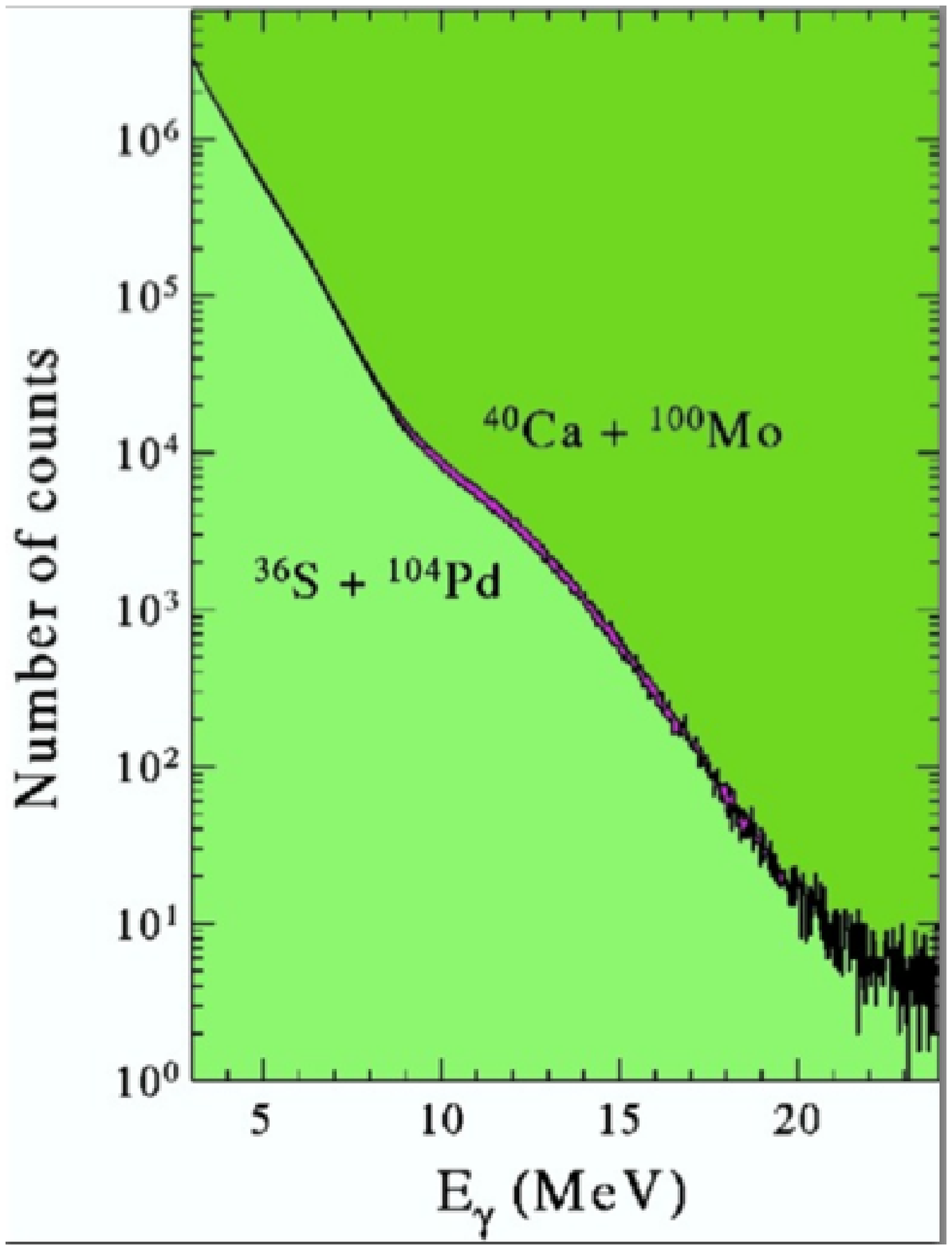} 
\hspace{0.5cm}
\includegraphics[width=6cm]{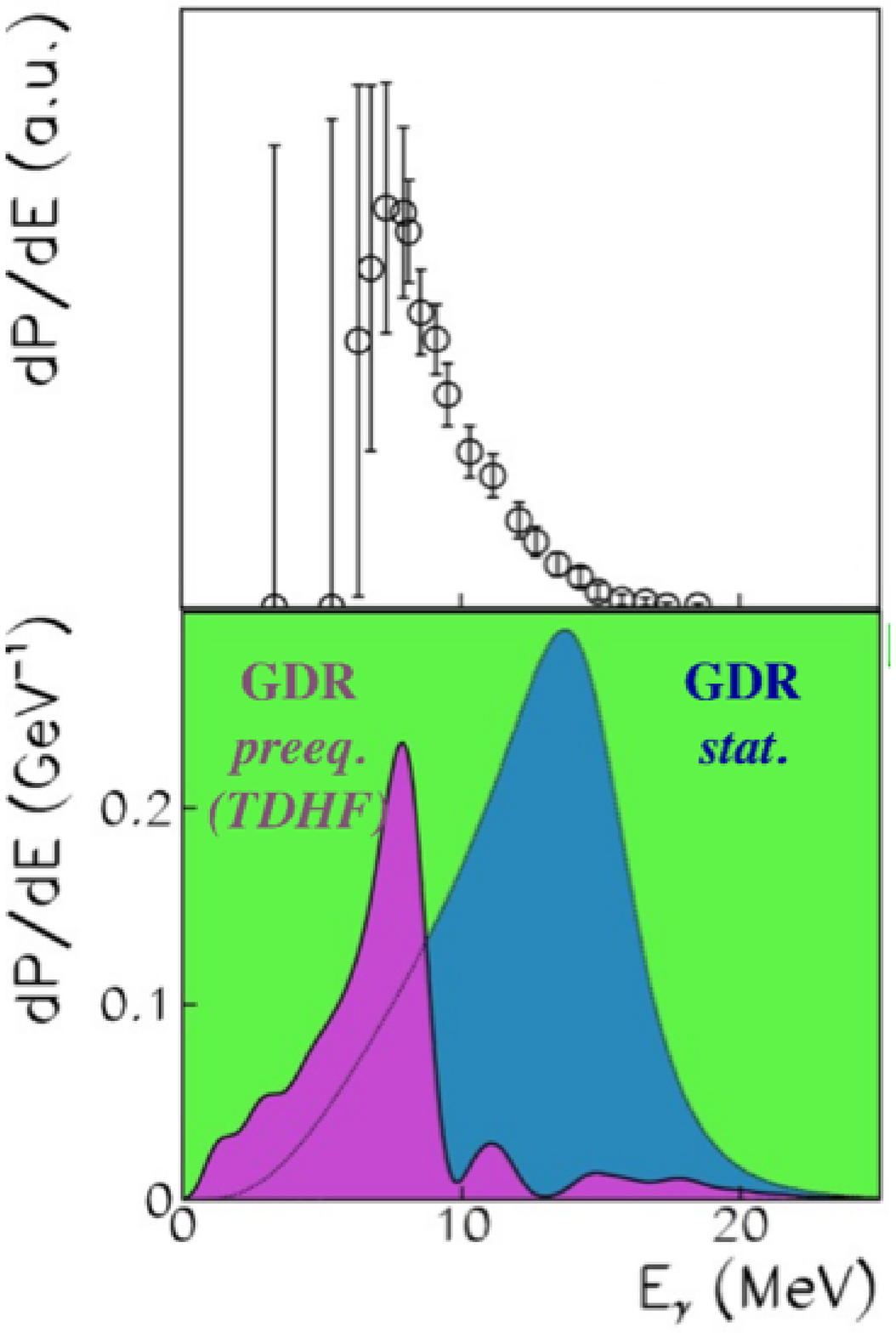} 
\caption{(top) $\gamma$-spectra measured in the $^{40}$Ca+$^{100}$Mo  and $^{36}$S+$^{104}$Pd reactions at a center-of-mass energy of 0.83 MeV/nucleon~\cite{fli96}. (middle) Preequilibrium GDR $\gamma$-decay spectrum obtained from the difference between the two $\gamma$-spectra in the top. (bottom) Theoretical $\gamma$ spectrum computed from the preequilibrium dipole moment evolution in Fig.~\ref{fig:fli} (solid line, purple area). The dotted line represents the first chance statistical $\gamma$-ray decay spectrum (blue area). Adapted from~\cite{sim07}.}
\label{fig:fli_gamma}
\end{center}
\end{figure}

It is possible to compute the spectrum of~$\gamma$ emitted by the preequilibrium GDR using laws of classical electrodynamics. 
The preequilibrium GDR $\gamma$-ray spectrum is computed from the Fourier transform of the acceleration of the charges \cite{jac62,bar01}
\begin{equation}
\frac{dP}{dE_\gamma}(E_\gamma)=\frac{2\alpha}{3\pi}\frac{|I(E_\gamma)|^2}{E_\gamma}
\label{fourier}
\end{equation}
where $\alpha$ is the fine structure constant and
$$I(E_\gamma)=\frac{1}{c}\int_0^\infty \!\!\! dt \,\, \frac{d^2Q_D}{dt^2}\exp\left(i\frac{E_\gamma t}{\hbar}\right).$$
Such a $\gamma$-spectrum is shown in the bottom of Fig.~\ref{fig:fli_gamma} (solid-line, purple area). 
A comparison with the first chance statistical GDR $\gamma$-ray decay spectrum is also shown in the bottom of  Fig.~\ref{fig:fli_gamma} (dotted line, blue area). (See Ref.~\cite{sim07} for details on the calculations of the latter). 
We observe that the preequilibrium GDR $\gamma$ are emitted at a lower energy than the statistical component. 
This is interpreted in terms of a large deformation of the compound nucleus in its preequilibrium phase~\cite{sim07}. Indeed, the preequilibrium dipole motion occurs along the prolate deformation axis of the compound system associated to a lower energy (see also Fig.~\ref{fig:U-GDR}).

Experimental $\gamma$-spectra are shown in the upper panel of Fig.~\ref{fig:fli_gamma} for the $^{40}$Ca+$^{100}$Mo $N/Z$ asymmetric reaction and for the  $^{36}$S+$^{104}$Pd  reaction which is quasi-symmetric in $N/Z$. 
Only the first reaction is expected to exhibit a preequilibrium dipole motion. 
Indeed, more $\gamma$ are observed in this reaction. 
The difference (purple area in Fig.~\ref{fig:fli_gamma}-top) is interpreted in terms of $\gamma$-decay from the preequilibrium GDR~\cite{fli96}.
Subtracting the two $\gamma$-spectra, one obtains the preequilibrium GDR $\gamma$-spectrum which is shown in the middle panel of Fig.~\ref{fig:fli_gamma}.
The energy of the peak is in good agreement with the spectrum computed from the TDHF response.
As mentioned above, this lowering of the preequilibrium GDR energy, by comparison to the hot GDR, is interpreted  as a signature of a strong deformation of the preequilibrium nucleus.
In particular, this means that the  equilibration of the shape is much slower than the charge equilibration.
We will discuss charge equilibration times more quantitatively in section~\ref{sec:QF}.

To conclude, we see that the preequilibrium GDR contains informations on the structure of the preequilibrium compound system and, then, on the path to fusion. 
Here, the example of deformation has been discussed.
In Refs.~\cite{sim01,sim07}, the preequlibrium GDR is also used to investigate other properties of the path to fusion, such as the role of rotation, the coupling with collective shape vibrations, the role of mass asymmetry...
It is also proposed that the decay of a preequilibrium GDR could serve as a cooling mechanism increasing the survival probability of the heaviest compound nuclei~\cite{bar01,sim07}. 
This effect needs further theoretical and experimental investigations. 

\subsection{Transfer reactions \label{sec:transfer}}

We discussed in section~\ref{sec:spher} the interplay between fusion and transfer reactions in the $^{16}$O+$^{208}$Pb system around the barrier. 
In particular, we observed just below the barrier (see top of Fig.~\ref{fig:dens}) an average transfer of two protons. 
We now discuss such transfer reactions in more details.

\begin{figure}
\includegraphics[width=8.8cm]{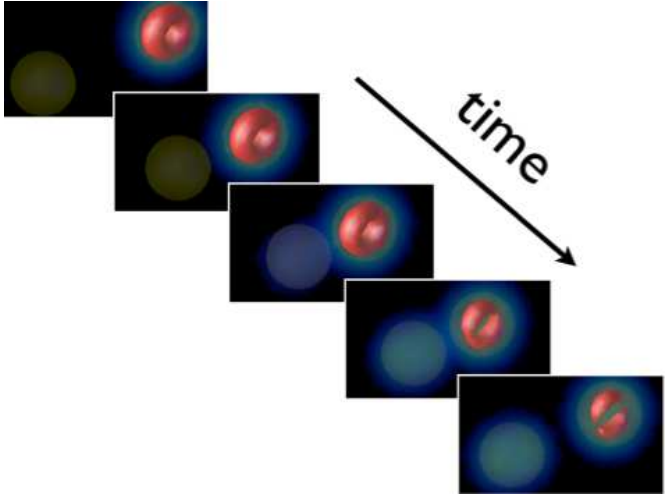} 
\caption{TDHF calculations of central sub-barrier 
collision of two $^{16}$O nuclei. The nuclei approach each 
other, and re-separate back-to-back due to the 
Coulomb repulsion. The evolution of a $p_{3/2}$ single-particle wave-function belonging initially to the nucleus in the right is 
shown. After the collision, part of this wave function 
has been transferred to its collision partner. Since this 
reaction is symmetric, a similar transfer occurs from 
the left to the right, and both fragments have the same 
particle number distributions. 
}
\label{fig:transfer_sp}
\end{figure}

Several recent attempts of describing nucleon transfer in heavy-ion collisions within the TDHF framework have been made in medium mass systems~\cite{uma08a,sim08,was09b,sim10a,sim10b,yil11}.
The TDHF equation describes the evolution of single-particle wave-functions.
The latter, initially localised within one collision partner, may be partially transferred to the other fragment during the collision, as illustrated in Fig.~\ref{fig:transfer_sp}

\begin{figure}
\includegraphics[width=8.8cm]{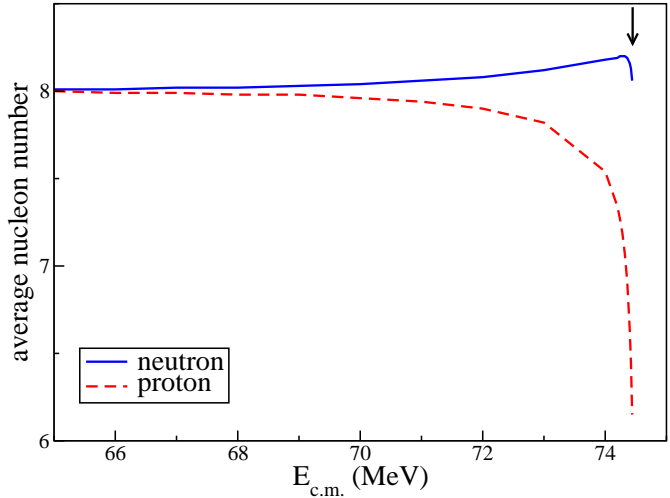} 
\caption{Transfer in $^{16}$O+$^{208}$Pb sub-barrier central collisions. 
The average number of protons and neutrons of the small fragment in the exit channel
are plotted as a function of $E_{c.m.}$. The arrow indicates the TDHF fusion barrier.}
\label{fig:average}
\end{figure}

In the case of an asymmetric collision, a change of the average particle number in the fragments in the exit channel is a clear signature that a transfer mechanism occurred in the reaction.
Figure~\ref{fig:average} gives the evolution of the expectation value of $\oZ$ and $\oN$ of the small fragment in the exit channel of $^{16}$O+$^{208}$Pb sub-barrier central collisions. 
At the barrier, $\sim2$~protons and no neutron, in average, are transferred (the corresponding evolution of the density is shown in the top of Fig.~\ref{fig:dens}), while at $\sim10\%$ below the barrier, $Z\simeq N\simeq8$ is obtained in average, indicating a dominance of (in)elastic scattering.
We see that the probability for proton stripping (transfer from the light to the heavy nucleus) is higher than for proton pickup (transfer from the heavy to the light nucleus), while neutron pickup is more probable 
than neutron stripping.
This qualitative observation is in agreement with experimental data~\cite{vid77,eve11}.

\begin{figure}
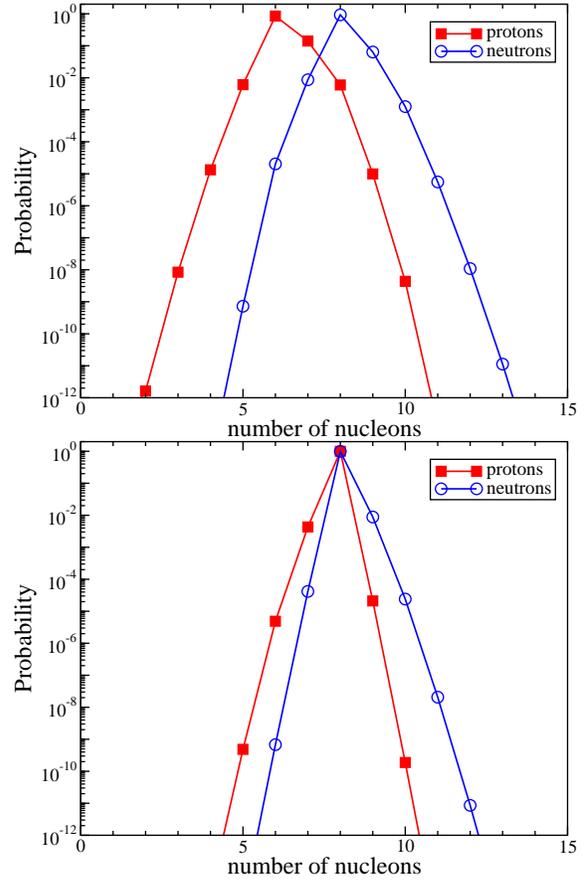

\begin{center}
\includegraphics[width=7.5cm]{proba_B.eps} 
\includegraphics[width=7.5cm]{proba_subB.eps} 
\caption{Neutron (circles) and proton (squares) number probability distributions 
of the lightest fragment in exit channel of a head-on 
$^{16}$O+$^{208}$Pb collision at $E_{c.m.}=74.44$~MeV (top) and $65$~MeV (bottom). Adapted from Ref.~\cite{sim10b}.}
\label{fig:proba}
\end{center}
\end{figure}

To get a deeper insight into this transfer mechanism, the transfer probabilities are extracted at the TDHF level \cite{koo77,sim10b} thanks to a projection onto a good particle number technique\footnote{This technique is standard in beyond-mean-field models for nuclear structure when the number of particles is only given in average~\cite{rin80}.} applied on the outgoing fragments\footnote{One could question this approach as the TDHF wave functions are not used for the calculation of expectation values of one-body operators. In particular, the width of the distributions should be underestimated~\cite{das79}. However, as we will see in section~\ref{sec:DIC}, the TDHF and BV widths are similar for non violent collisions such as sub-barrier transfer, justifying {\it a posteriori} this approach.}.
It is possible to extract the component of the wave function associated to a specific transfer channel 
using a particle number projector onto $N$ protons or neutrons in the $x>0$ region where one fragment is located at the final time, the other one being in the $x<0$ region.
Such a projector is written~\cite{bender03}
\oeq
\oP_R(N)=\frac{1}{2\pi}\int_0^{2\pi} \stb \d \tet \stf e^{i\tet(\oN_R-N)},
\label{eq:projector}
\ceq
where
\oeq
\oN_R = \sum_{s} \sdf \int \stb \d \vr \stf \oad(\vr s) \sdf \oa(\vr s) 
\sdf \Theta(x)
\label{eq:NG}
\ceq
counts the number of particles in the $x>0$ region ($\Theta(x)=1$ if $x>0$ and 0 elsewhere).
Isospin is omitted to simplify the notation. 

The projector defined in Eq.~(\ref{eq:projector}) can be used to compute the probability to find $N$ nucleons in $x>0$ in the final state $\kfi$,
\oeq
\left|\oP_R(N)\kfi\right|^2=\frac{1}{2\pi}\int_0^{2\pi} \stb \d \tet \stf e^{-i\tet{N}}\bfi\phi_R(\tet)\>,
\label{eq:proba}
\ceq
where $|\phi_R(\tet)\>=e^{{i\tet\oN_R}}\kfi$ 
 represents a rotation of $\kfi$ by a gauge angle $\tet$ 
in the gauge space associated to the particle number degree of freedom.
Note that $|\phi_R(\tet)\>$ is an independent particle state. 
The last term in Eq.~(\ref{eq:proba}) is then the determinant of the matrix of the occupied single particle state overlaps:
\oeq
\bfi\phi_R(\tet)\>=\det (F)
\ceq
with
\oeq
F_{ij}= \sum_{s} \int \stb\d \vr \sdf{\az_i^s}^*(\vr) {\az_j^{s}}(\vr) e^{i\tet\Theta(x)}.
\ceq
The integral in Eq.~(\ref{eq:proba}) is discretised using $\tet_n=2\pi{n}/M$ with the integer $n=1\cdots{M}$.
Choosing $M=300$ ensures numerical convergence for the $^{16}$O+$^{208}$Pb system. 
Fig~\ref{fig:proba} shows the resulting transfer probabilities at (top) and well below the barrier at $E_{c.m.}=65$~MeV (bottom).
As expected from the average values (see Fig.~\ref{fig:average}), the most probable channels are $Z=6$ and $N=8$ at the barrier, and $Z=N=8$ with a small probability of neutron pickup or proton stripping (of the order of $10^{-2}$) well below the barrier. 

A standard representation of experimental sub-barrier energy transfer data is to plot transfer probabilities as a function of the distance of closest approach $R_{min}$ between the collision partners~\cite{cor09}. 
$R_{min}$ is computed assuming a Rutherford trajectory~\cite{bro91}:
\oeq
R_{min}={Z_1Z_2e^2}[1+\mbox{cosec}(\theta_{c.m.}/2)]/{2E_{c.m.}}
\label{eq:R_min}
\ceq
where $\theta_{c.m.}$ is the center of mass scattering angle.

\begin{figure}
\includegraphics[width=8.8cm]{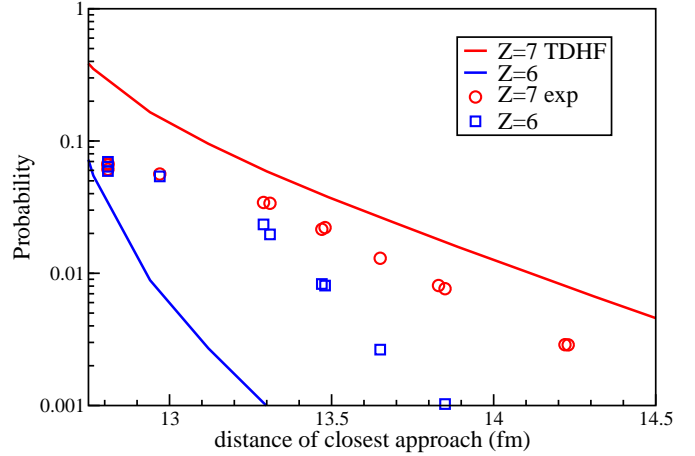} 
\caption{Proton number probability  
as function of the distance of closest approach  in the small outgoing fragment of the $^{16}$O+$^{208}$Pb reaction. TDHF results are shown with lines. Experimental data (open symbols) are taken from Ref.~\cite{eve11}.}
\label{fig:proba_TDHF+exp}
\end{figure}

A comparison of TDHF calculations with recent data from Ref.~\cite{eve11} is shown in Fig.~\ref{fig:proba_TDHF+exp} for sub-barrier one and two-proton transfer channels in the $^{16}$O+$^{208}$Pb reaction.
We see that TDHF overestimates the one-proton transfer probabilities and underestimates the 2-proton transfer channel.
This discrepancy is interpreted as an effect of pairing interactions~\cite{sim10b,eve11}.
Indeed, paired nucleons are expected to contribute to two-nucleon transfer channels.
As a result, the two-nucleon (resp. single-nucleon) transfer probability increases (decreases). 
Note, however, that for $R_{min}>13$~fm, the TDHF calculations reproduce reasonably well the sum of one and two-proton transfer channels\footnote{For $R_{min}<13$~fm,  sub-barrier fusion, not included in the TDHF calculations, reduces transfer probabilities~\cite{sim10b,eve11}.}~\cite{sim10b}.

These studies emphasize the role of pairing interactions in heavy-ion collisions.
The recent inclusion of pairing interactions in 3-dimensional microscopic codes~\cite{ass09,eba10,ste11} gives hope in our ability to describe such data with more details in a near future. 
Finally, microscopic theories should also be used to determine, not only the transfer probabilities, but also the excitation energy of the final fragments. 
In particular, this would be of great help to understand how transfer reactions induce an energy dissipation, possibly hindering fusion at deep sub-barrier energies~\cite{das07,eve11}.

\subsection{Deep-Inelastic Collisions \label{sec:DIC}}

Deep-Inelastic Collisions (DIC) occur essentially well above the barrier.
The main characteristics of DIC exit channels are:
\begin{itemize}
\item A strong damping of the initial kinetic energy,
\item Large fluctuations of the fragment proton and neutron numbers around their initial value, 
\item  An angular distribution of the fragments following a $1/\sin\theta_{c.m.}$ behaviour.
\end{itemize}
The last point is due to a large orbiting of the fragments at contact. All $\theta_{c.m.}$ are then equiprobable in DIC. 
However the emission is not isotropic as it occurs essentially in the collision plane.
As a result, the differential cross-sections for DIC events obey 
$$\frac{d\sigma}{d\theta_{c.m.}}=2\pi\sin\theta_{c.m.}\frac{d\sigma}{d\Omega}\simeq\mbox{ constant}.$$
This leads to the $1/\sin\theta_{c.m.}$ behaviour, as 
$$\frac{d\sigma}{d\Omega}\propto \frac{1}{\sin \theta_{c.m.}}.$$

Early TDHF calculations were able to reproduce fragment kinetic energies, mean masses, and scattering angles, but they failed to reproduce the observed large fluctuations of the fragment $Z$ and $N$ distributions~\cite{koo77,dav78}.
This failure was seen as a necessity to include a collision term in the dynamics. 
However, Balian and V\'en\'eroni showed that the TDHF theory was in fact not optimised for the prediction of such fluctuations~\cite{bal81}. Instead, one should use the prescription given in Eq.~(\ref{eq:CiiBV}) which is derived from their variational principle and optimised for one-body fluctuations~\cite{bal84}.

The BV prescription was used to compute particle number fluctuations in DIC in Refs. \cite{bon85,mar85,bro09,sim11,sim11b}, and within a semi-classical approximation in Ref. \cite{zie88}.
Comparing with standard TDHF calculations, these works showed, indeed, an increase of the fluctuations. 
However, the first realistic calculations allowing for a direct comparison with experimental data\footnote{The authors of Ref.~\cite{mar85} claim that their calculations are in good agreement with  experimental data. However, they only compute the fluctuations of $A=N+Z$ and compare with measured fluctuations of $Z$. In addition, their calculations do not include the spin-orbit interaction. They are performed at angular momenta leading to fusion when spin-orbit terms are included (see discussion in Ref.~\cite{bro09}).} was only recently performed in Ref.~\cite{sim11}. We now discussed the main results of this paper.

\begin{figure}
\includegraphics[width=8.5cm]{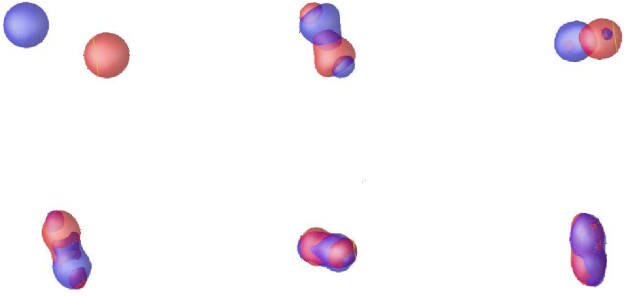} 
\caption{Density evolution for a $^{40}$Ca+$^{40}$Ca collision at $E_{c.m.}=128$~MeV and $L=60\hb$. Each snapshot is separated by 1.5~zs.}
\label{fig:40Ca+40Ca_L60}
\end{figure}

\begin{figure}
\includegraphics[width=8.5cm]{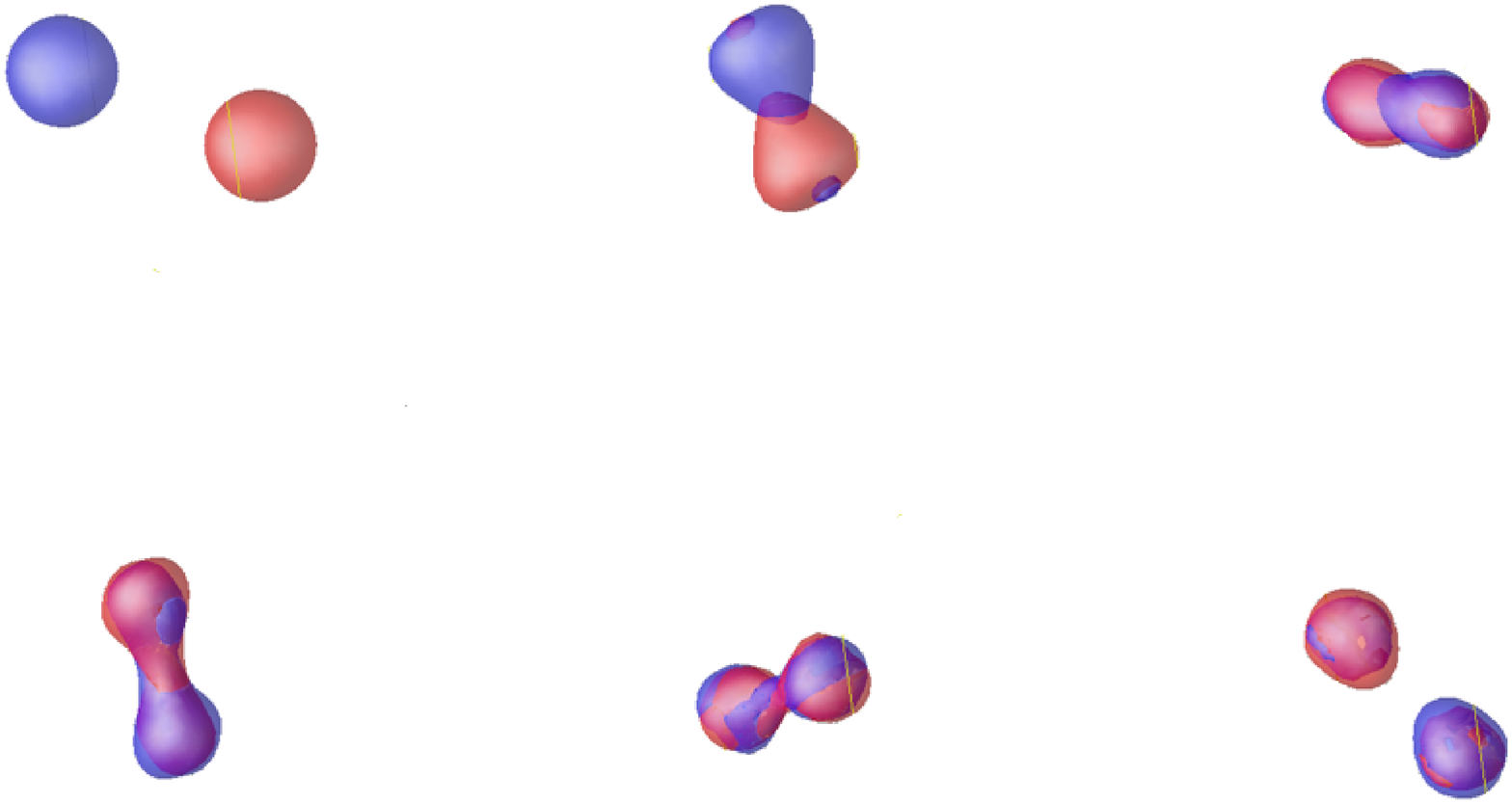} 
\caption{Same as Fig.~\ref{fig:40Ca+40Ca_L60} with $L=70\hb$.}
\label{fig:40Ca+40Ca_L70}
\end{figure}

\begin{figure}
\includegraphics[width=8.5cm]{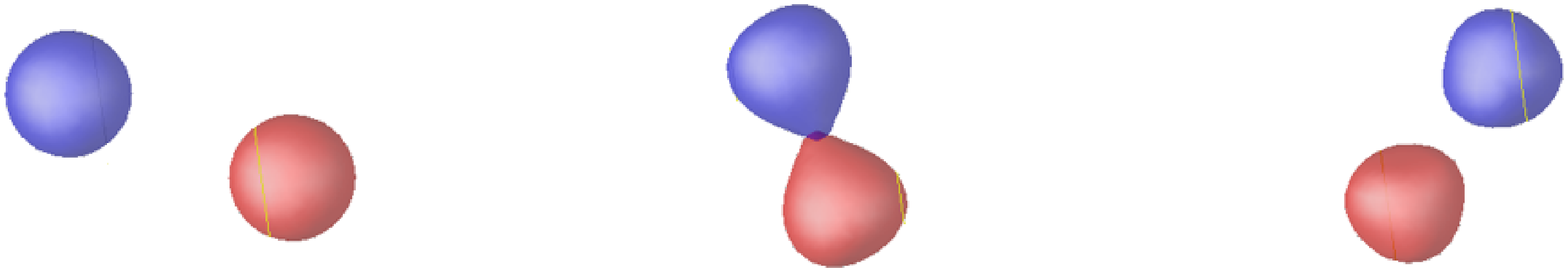} 
\caption{Same as Fig.~\ref{fig:40Ca+40Ca_L60} with $L=80\hb$.}
\label{fig:40Ca+40Ca_L80}
\end{figure}

\begin{figure}
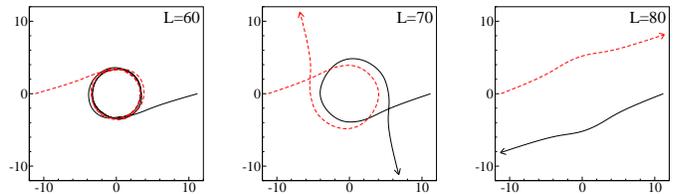

\includegraphics[width=2.6cm]{traj60.eps}
\hspace{0.3cm} 
\includegraphics[width=2.6cm]{traj70.eps} 
\hspace{0.3cm} 
\includegraphics[width=2.6cm]{traj80.eps} 
\caption{Trajectories of the centers of mass of the fragments in $^{40}$Ca+$^{40}$Ca collisions at $E_{c.m.}=128$~MeV.}
\label{fig:40Ca+40Ca_traj}
\end{figure}

$^{40}$Ca+$^{40}$Ca collisions at $E_{c.m.}=128$~MeV ($\sim2.4$ times the barrier height) have been studied with the \textsc{tdhf3d} code.  
Figs.~\ref{fig:40Ca+40Ca_L60}, \ref{fig:40Ca+40Ca_L70}, and \ref{fig:40Ca+40Ca_L80} show density plots obtained with angular momenta L=60, 70, and 80 (in units of $\hb$), respectively.
Three reaction mechanisms are observed: capture (at $L=60$), DIC (at $L=70$), and a partially damped collision (at $L=80$).
In particular, the $L=70$ case leads to an orbiting trajectory (see middle panel of Fig.~\ref{fig:40Ca+40Ca_traj}) which is characteristic of a DIC. 
This trajectory also corresponds to a strongly damped collision, as can be seen from the upper panel of Fig.~\ref{fig:sigma} where the total kinetic energy loss (TKEL) is plotted as a function of the initial angular momentum. 
Indeed, around $L\simeq70$, the TKEL is $\sim60-70$~MeV.
These values are slightly below the Viola systematics~\cite{vio85} which predicts $TKEL_{Viola}\simeq76$~MeV, indicating that these collisions are almost fully damped. 

\begin{figure}
\includegraphics[width=8.8cm]{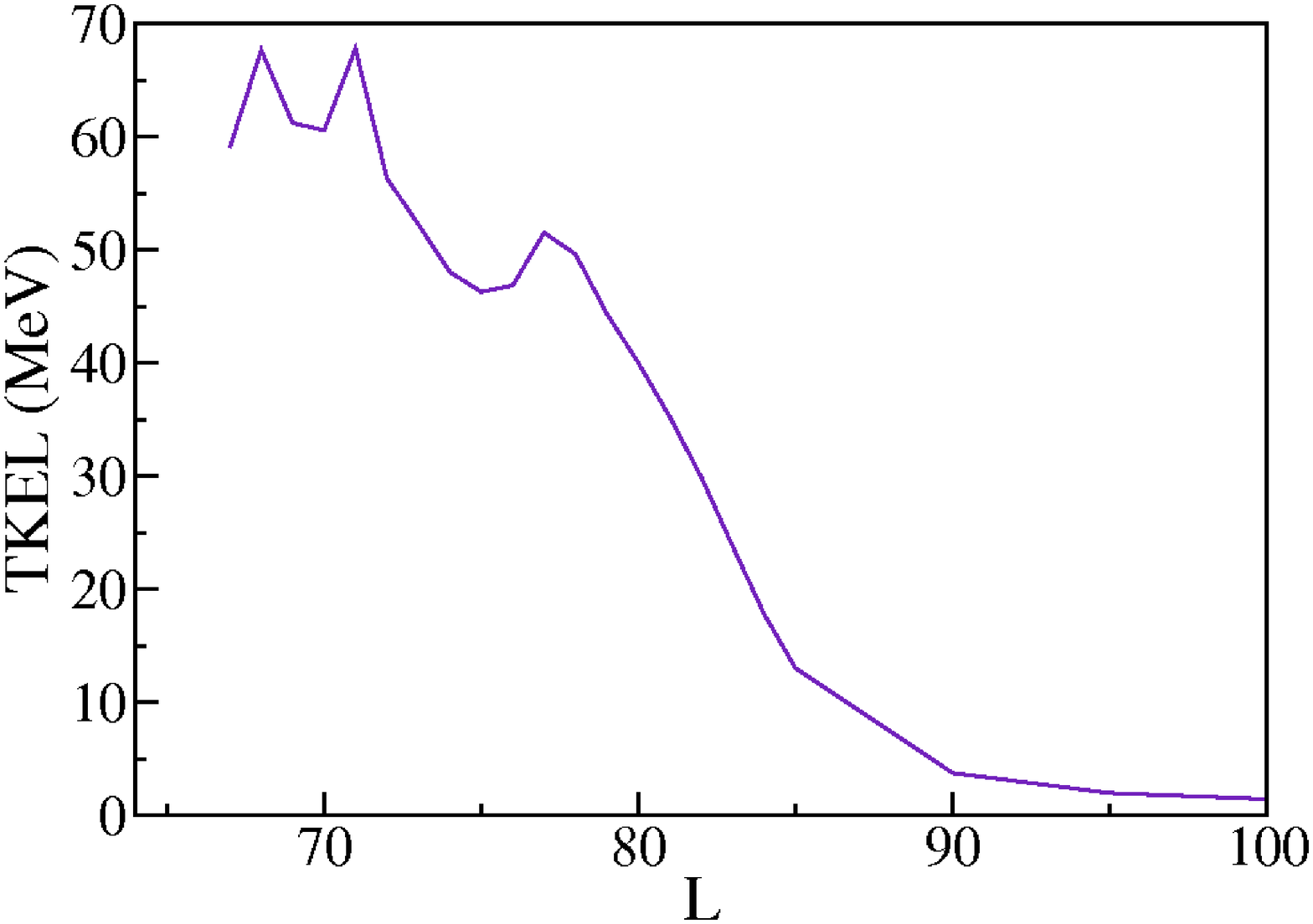} 
\includegraphics[width=8.8cm]{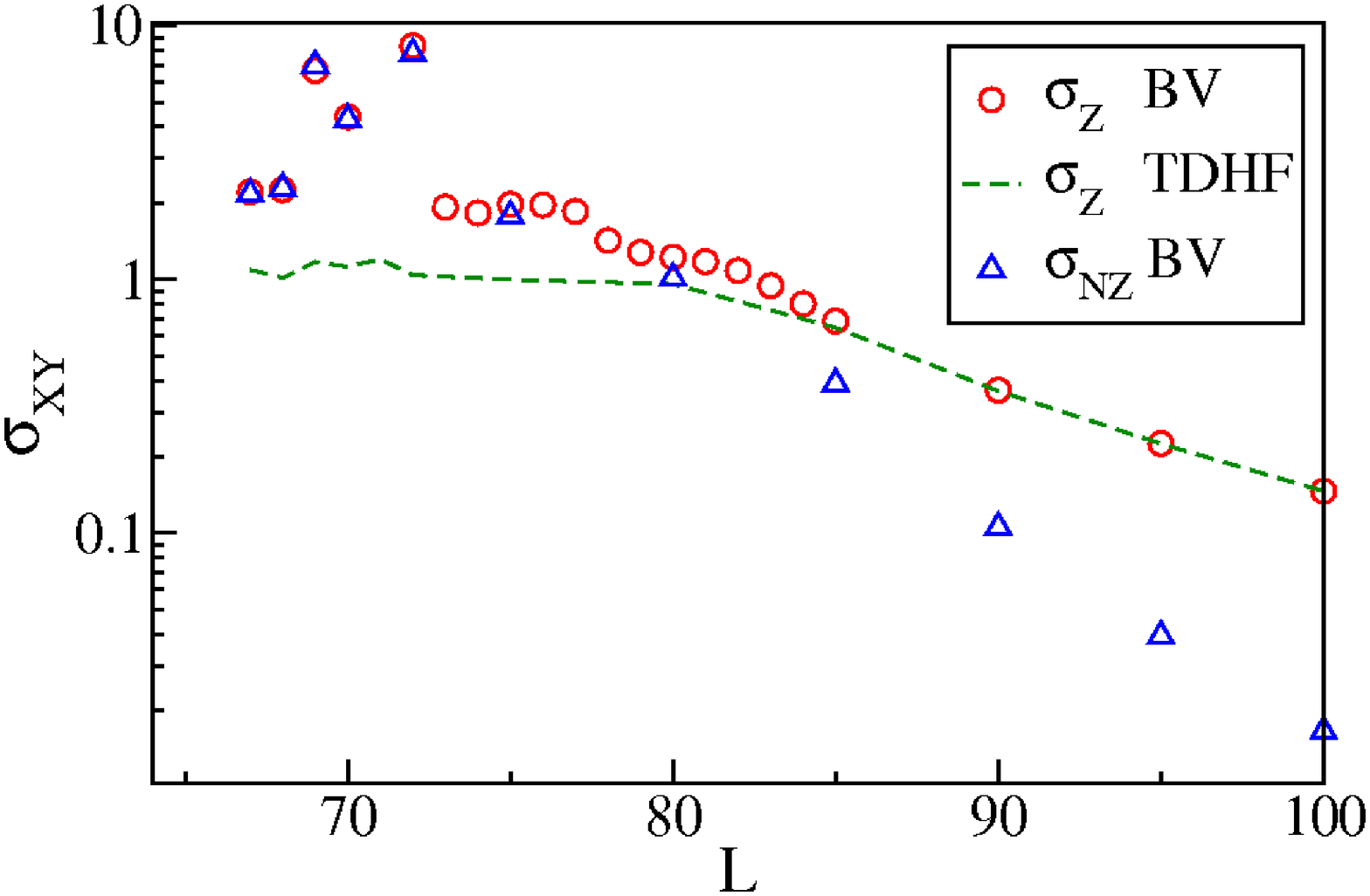} 
\caption{Properties of the exit channel of the $^{40}$Ca+$^{40}$Ca collisions at $E_{c.m.}=128$~MeV as a function of the initial angular momentum $L$. (top) Total kinetic energy loss from TDHF.
(bottom) TDHF (dashed line) and BV (circles) fluctuations of $Z$, 
and BV correlations between $N$ and $Z$ (triangles) of the outgoing fragments.
}
\label{fig:sigma}
\end{figure}

The lower panel of Fig.~\ref{fig:sigma} shows the evolution of the fluctuations $\sigma_Z$ of the number of protons\footnote{Neutron fluctuations are very close to the proton ones for this system. This is due to the fact that the collision partners are $N=Z$ nuclei.} in the outgoing fragments as a function of $L$.  
As expected\footnote{See early applications of the BV prescription for fluctuations where greater fluctuations than their TDHF counterparts were obtained~\cite{mar85,bon85,bro09}.}, the BV prescription [Eq.~(\ref{eq:CiiBV})] leads to larger fluctuations than standard TDHF fluctuations obtained from Eq.~(\ref{eq:CiiTDHF}).
This is particularly true for DIC.
However, TDHF and BV prescriptions converge at large $L$, e.g., for $L>90$ where the TKEL is less than 5~MeV, indicating a dominance of quasi-elastic scattering.
We conclude that TDHF calculations of particle number distributions give reasonable estimates for such non-violent collisions.
In particular, this justifies the calculation of transfer probabilities for sub-barrier collisions with TDHF (see section~\ref{sec:transfer}). 

Fig.~\ref{fig:sigma} also shows an application of the BV prescription for the correlations between $N$ and $Z$ distributions. 
These correlations are determined from Eq.~(\ref{eq:CijBV}).
We observe increasing correlations with decreasing $L$. 
In particular, in DIC, the BV prescription predicts\footnote{$\sigma_N$ is not shown in Fig.~\ref{fig:sigma} for the clarity of the figure.} $\sigma_{N}\simeq\sigma_{Z}\simeq\sigma_{NZ}$, while, for quasi-elastic scattering, correlations are negligible. 

Let us recall the signification of such correlations. 
Uncorrelated distributions mean that the probability to transfer $z$ protons is independent of the probability to transfer $n$ neutrons, i.e., $P(z,n)\equiv P(z)P(n)$.
On the other side, strongly correlated distributions mean that if we measure $n$ (resp. $z$), then we know what is $z$ ($n$).
This is the case, for instance, if all fragments have $N=Z$.
In this case, we would have $P(z,n)\equiv P(z)\delta_{n,z}\equiv P(n)\delta_{n,z}$.
The reality is usually in between and, assuming Gaussian distributions, we have
\oeqn
\stb\stb\stb P(z,n)&=&\(2\pi\sigma_N\sigma_Z\sqrt{1-\rho^2}\)^{-1} \nonumber\\
&&\exp\[ -\frac{1}{1-\rho^2} \( \frac{n^2}{\sigma_N^2} + \frac{z^2}{\sigma_Z^2} -\frac{2\rho nz}{\sigma_N\sigma_Z} \) \],
\label{eq:Gauss}
\ceqn
where $|\rho|=\frac{\sigma_{NZ}^2}{\sigma_N\sigma_Z}$.
The case $\rho=0$ means no correlations between $N$ and $Z$ distributions, while the limit $|\rho|\rightarrow1$ corresponds maximum correlations.  

In the calculations of $^{40}$Ca+$^{40}$Ca at $E_{c.m.}=128$~MeV shown in Fig.~\ref{fig:sigma}, we have $\sigma_{NZ}\simeq\sigma_{N}\simeq{\sigma_Z}$ for DIC ($L<80$).
This means that $N$ and $Z$ distributions of the fragments are strongly correlated in DIC.
However, quasi-elastic reactions ($L>90$) have $\sigma_N\simeq\sigma_Z\gg\sigma_{NZ}$, meaning almost independent $N$ and $Z$ distributions in this case. 

\begin{figure}
\begin{center}
\includegraphics[width=8cm]{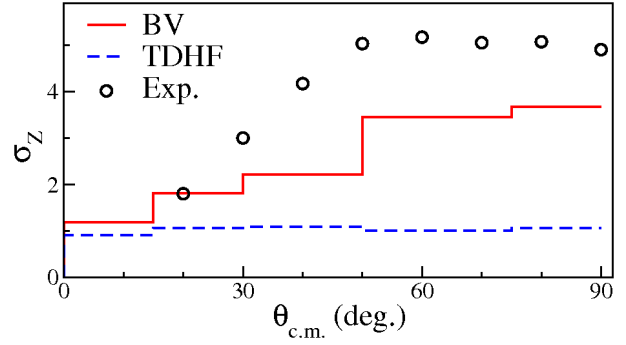} 
\caption{Comparison between BV (solid line) and TDHF (dashed line) predictions of $\sigma_{ZZ}$ for damped events (see text) as a function of $\theta_{c.m.}$ with data (circles) from~\cite{roy77}.
}
\label{fig:sig_tet}
\end{center}
\end{figure}

We now compare these results with the experimental data of Roynette {\it et al.}~\cite{roy77}.
The angle $\theta_{c.m.}$ between the fragments in the outgoing channel and the collision axis have been determined for each $L$. 
Fig.~\ref{fig:sig_tet} shows theoretical and experimental evolutions of the charge fluctuations for damped events (defined, as in Ref.~\cite{roy77}, by a $TKEL>30$~MeV) as a function of $\theta_{c.m.}$. 
Due to  orbiting, only DIC are expected to emit fragments at large angles, and the experimental plateau at $\theta_{c.m.}\ge50$~deg is then attributed to DIC. 
We see that TDHF fluctuations underestimate experimental results at all angles, except at very forward angles where quasi-elastic reactions dominate.
The results of the BV prescription are in better agreement, although they still underestimate the experimental data. 
This is probably due to fusion-fission events (not included in the calculations) leading to large fluctuations and, to a less extent, to the cooling down of the fragments by nucleon emission~\cite{sim11}. 

In addition, calculations with exotic nuclei have been performed to study the role of isospin asymmetry in the entrance channel on the $Z$ and $N$ fluctuations and correlations~\cite{sim11}. 
The $^{80,92}$Kr+$^{90}$Zr systems have been investigated at a beam energy $E/A=8.5$~MeV. 
Charge equilibration is observed in the  $^{92}$Kr+$^{90}$Zr due to an initial $N/Z$ asymmetry, inducing larger correlations between $N$ and $Z$ distributions, while  fluctuations are only slightly affected.
This increase of the correlations between $N$ and $Z$ distributions due to charge equilibration could be tested in DIC experiments with exotic beams~\cite{Lem12}.
Note, finally, that other fluctuations should also be computed with the BV approach, such as the width of TKE distributions of the fragments. 

\subsection{Fusion and quasi-fission in heavy systems\label{sec:QF}}

In section~\ref{sec:fusion}, we restricted the study to $Z_1Z_2<1600$ systems for which no fusion hindrance is usually observed\footnote{This threshold is empirical. Based on his extra-push model, Swiatecki proposed an effective fissility, depending on both charges and masses of the nuclei, above which extra-push energy is needed to fuse~\cite{swi82}. Note that this should not be confounded with a threshold for quasi-fission which, in fact, may occur in lighter systems.
Indeed, quasi-fission has been observed in, e.g., $^{16}$O,$^{32}$S+$^{238}$U \cite{hin96,itk11,nas07}, and even in lighter systems such as $^{32}$S+$^{208}$Pb~\cite{nas07}.}. 
We now investigate the reaction mechanism in heavy systems with possible fusion hindrance.
We first illustrate the fusion hindrance with TDHF calculations of fusion thresholds~\cite{ave09,sim09b}.  
Then, we present recent results on a study of the quasi-fission process with the TDHF approach. 

\subsubsection{TDHF calculations of fusion hindrance}

As an example, we consider the $^{90}$Zr+$^{124}$Sn system which has a charge product $Z_1Z_2=2000$, and, then, is expected to exhibit a fusion hindrance. 
The proximity model~\cite{blo77} predicts a barrier for this system $V^{prox.}\simeq215$~MeV.
Fig.~\ref{fig:ZrSn} shows the TDHF evolution of the relative distance between the fragments as a function of time for central collisions at different energies~\cite{ave09}. 
We see that the system encounters a fast re-separation at the energy of the barrier predicted by the proximity model. 
Long contact times possibly leading to fusion are observed at $E_{c.m.}\ge240$~MeV.

\begin{figure}
\begin{center}
\includegraphics[width=7.5cm]{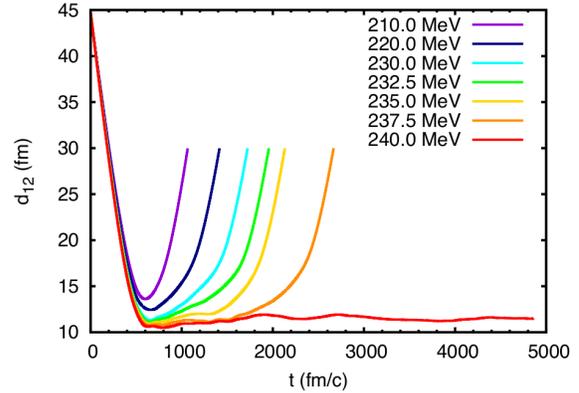} 
\caption{Distance between the centers of mass of the fragments as a function of time in head-on 
$^{90}$Zr+$^{124}$Sn collisions for different center of mass energies. From Ref.~\cite{ave09}.}
\label{fig:ZrSn}
\end{center}
\end{figure}

Density profiles for this system at $E_{c.m.}=235$~MeV are shown in Fig.~\ref{fig:ZrSn_dens}.
A rapid neck formation is observed.
However the system keeps the shape of two fragments in contact during $\sim1400$~fm/$c$ before re-separation in two fission-like fragments. 
This reaction mechanism differs from fusion followed by statistical fission as a compound system is not formed in the present case. 
In particular, the system is expected to keep the memory of its entrance channel. 
Such a process is called quasi-fission. 

\begin{figure}
\begin{center}
\includegraphics[width=8.8cm]{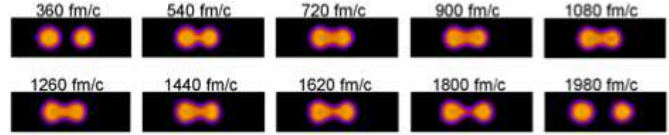} 
\caption{Density profile in the $^{90}$Zr+$^{124}$Sn  head-on collision at $E_{c.m.}=235$~MeV. From Ref.~\cite{ave09}.}
\label{fig:ZrSn_dens}
\end{center}
\end{figure}

Similar calculations have been performed for other systems with $Z_1Z_2>1600$~\cite{sim09b}. 
The TDHF fusion thresholds are shown in Fig.~\ref{fig:xpush}.
A comparison with the interaction barriers predicted by the proximity model~\cite{blo77} shows that dynamical effects included in TDHF induce a strong increase of the fusion threshold, in particular for the more heavy and symmetric systems. 
The order of magnitude of the additional energy needed to fuse is similar to the one predicted with the extra-push phenomenological model~\cite{swi82}. 

\begin{figure}
\begin{center}
\includegraphics[width=8.8cm]{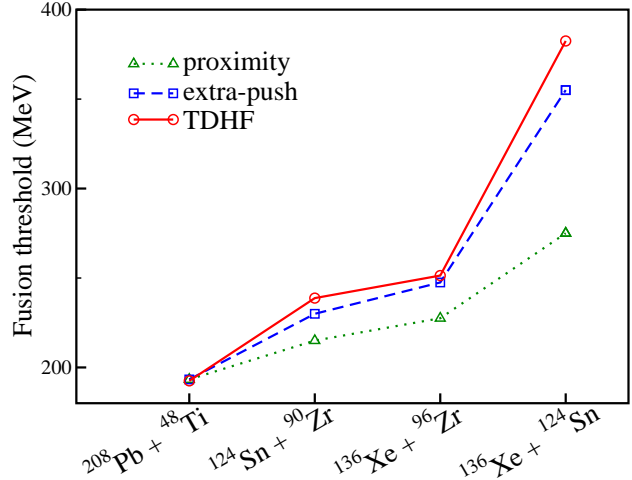} 
\caption{TDHF fusion thresholds for several heavy systems are compared with the proximity barrier~\cite{blo77} and with results from the extra-push model~\cite{swi82}.
}
\label{fig:xpush}
\end{center}
\end{figure}

These results motivate a deeper study of the origin of the fusion hindrance in such heavy systems.
In particular, the quasi-fission mechanism which is mostly responsible for this hindrance is studied in section~\ref{sec:QFTDHF} with the TDHF approach.

\subsubsection{TDHF calculations of the $^{40}$Ca+$^{238}$U reaction\label{sec:QFTDHF}}

The previous studies showed the importance of the quasi-fission process as a mechanism in competition with fusion, hindering the formation of heavy systems. 
A quantum and microscopic theoretical framework able to describe properly the quasi-fission properties would be of great importance to get a deep insight into the interplay between structure properties and this mechanism.

Here, our goals are to show that quasi-fission may appear in the outgoing channel within the TDHF approximation, and that TDHF calculations can then be used to predict quasi-fission properties.
An example of initial condition of the $^{40}$Ca+$^{238}$U collision is shown in Fig.~\ref{fig:sdvision} where we can clearly see internal structures in the $^{238}$U nucleus. 

\begin{figure}
\begin{center}
\includegraphics[width=8cm]{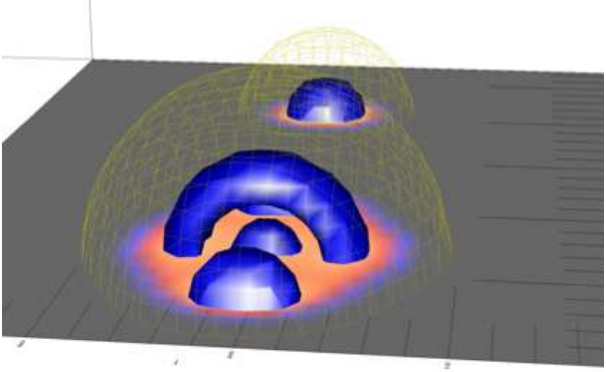}
\caption{Initial condition of a $^{40}$Ca+$^{238}$U collision visualised with the SDVision code~\cite{sdvision}. 
 Two isodensities are shown (yellow grid and blue area). A projection of the density is also shown on the $z=0$ plane.
\label{fig:sdvision}}
\end{center}
\end{figure}

\begin{figure}
\begin{center}
\includegraphics[width=7cm]{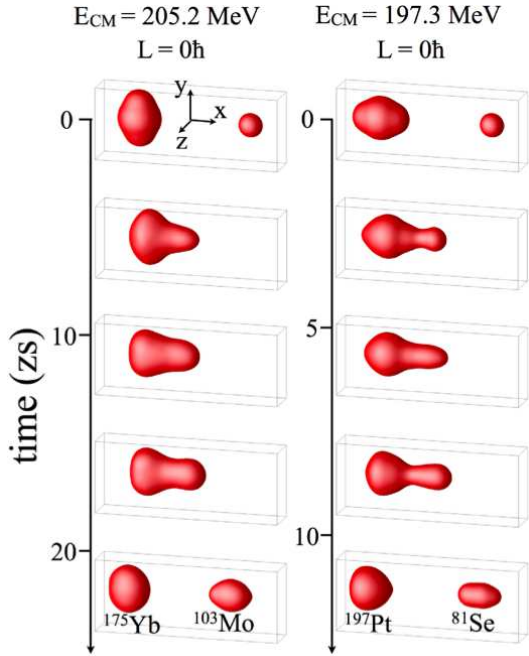}
\caption{Snapshots of the TDHF isodensity at half the saturation density in the $^{40}$Ca+$^{238}$U system for different initial orientations and $E_{c.m.}$.}
\label{fig:densQF}
\end{center}
\end{figure}

Examples of density evolutions obtained with the \textsc{tdhf3d} code are shown in Fig.~\ref{fig:densQF} for the $^{40}$Ca+$^{238}$U system around the barrier~\cite{wak12}.
These configurations lead to quasi-fission as we observe an important multi-nucleon transfer from the heavy fragment toward the light one. 
We also see that the mass equilibration (i.e., the formation of two fragments with symmetric masses) is not complete and may depend on the initial conditions.

\begin{figure}
\begin{center}
\vspace{1cm}
\includegraphics[width=7cm]{mass.eps}
\hspace{0.5cm}
\includegraphics[width=7cm]{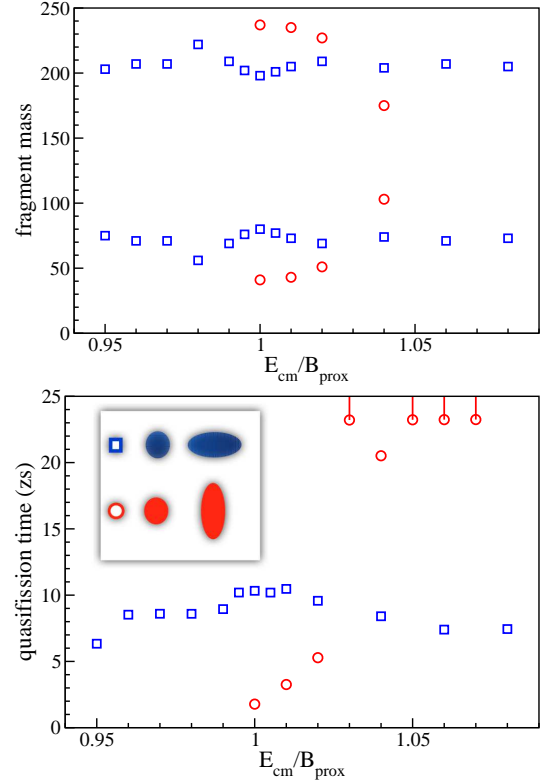}
\caption{TDHF calculations of the mass of the fragments (top) and of the quasi-fission time (bottom) in $^{40}$Ca+$^{238}$U central collisions as a function of the center of mass energy (divided by the proximity barrier~\cite{blo77}). For quasi-fission times larger than 23~zs, only a lower limit is given.  Two different orientations of the $^{238}$U are considered (see inset). From Ref.~\cite{wak12}.}
\label{fig:QFtime}
\end{center}
\end{figure}

Fig.~\ref{fig:QFtime} presents results for $^{40}$Ca+$^{238}$Ca central collisions.
Final fragment masses (top) and quasi-fission times (bottom) are shown for two different orientations of the $^{238}$U. 
We see that all the calculations with the $^{238}$U deformation axis aligned with the collision axis lead to a quasi-fission with partial mass equilibration and quasi-fission times smaller than 10~zs. 
Shell effects may affect the final outcome of the reaction by favouring the production of fragments in the $^{208}$Pb region.
In particular, this orientation never leads to fusion, while the other orientation produces long contact time above the barrier which may be associated to fusion. 
We also see that longer quasi-fission times lead to larger mass equilibrations.
Calculations of non-central $^{40}$Ca+$^{238}$U collisions are ongoing in order to compare with experimental data.

\subsection{Actinide collisions \label{sec:actinides}}

The collision of actinides form, during few zs, the heaviest nuclear systems available on Earth. 
In one hand, such systems are interesting to study the stability of the QED vacuum under strong electric fields~\cite{rei81,ack08,gol09}. 
In the other hand, they might be used to form neutron-rich heavy and super-heavy elements via multi-nucleon transfer reactions, as discussed in section~\ref{sec:introSHE}.

\begin{figure}
\begin{center}
\includegraphics[width=7.5cm]{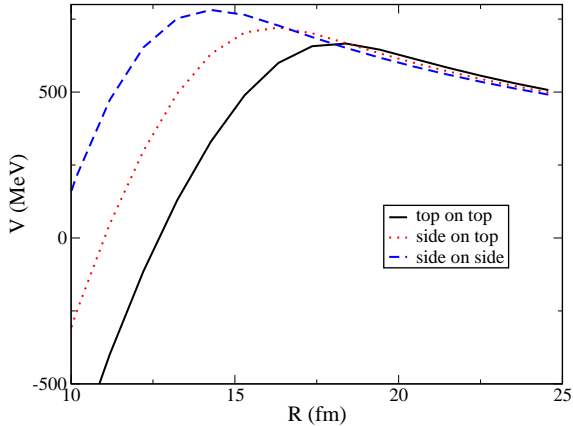}
\caption{Nucleus-nucleus potentials of the $^{238}$U+$^{238}$U system obtained with HF ground-states and the frozen approximation for different orientations of the actinides as a function of the distance between their centers of mass. }
\label{fig:pot_frozen}
\end{center}
\end{figure}

\begin{figure}
\begin{center}
\includegraphics[width=7.5cm]{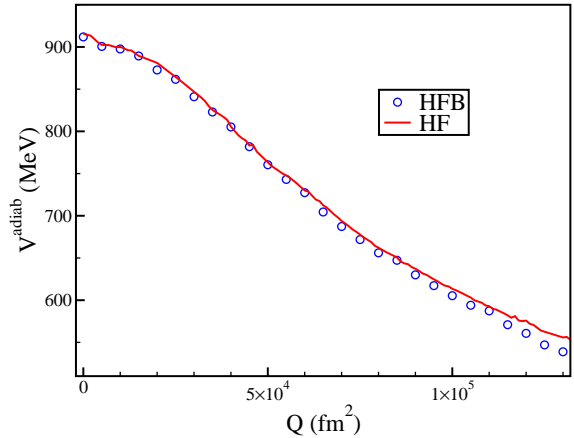}
\caption{The adiabatic potential is computed with the \textsc{ev8} code~\cite{bon05} as a function of the total quadrupole moment.  Calculations are performed with (HFB) and without (HF) pairing residual interaction.}
\label{fig:pot}
\end{center}
\end{figure}

As we saw in section~\ref{sec:QF}, there is a link between the collision time and the amount of transferred nucleons~\cite{tok85}. 
It is then important to optimize collision times in order to favour the formation of heavy systems. 
It was initially believed that the potentials between two actinides have a barrier (and then a pocket)~\cite{sei85}, leading to possible long contact times at energies close to the barrier. 
Such barriers are predicted by frozen models, as shown in Fig.~\ref{fig:pot_frozen} where the frozen potentials have been computed with two $^{238}$U HF ground states with different orientations. 
However, Berger and collaborators  showed with constrained HFB calculations of the composite system of $^{238}$U+$^{238}$U that there is, in fact, no barrier in their (adiabatic) nucleus-nucleus potential~\cite{ber90}.
This result  is confirmed with modern Skyrme-HF calculations using the \textsc{ev8} code~\cite{bon05}, as shown in Fig.~\ref{fig:pot}.

\begin{figure}
\includegraphics[width=8.8cm]{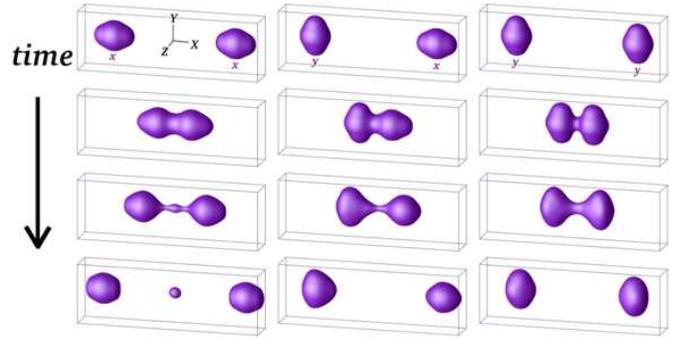}
\caption{Snapshots of the isodensity at half the saturation density in $^{238}$U+$^{238}$U central collisions at $E_{c.m.}=900$~MeV from TDHF calculations. Snapshots are given at times $t=0$, 1.5, 2.7, and $4.2$~zs from top to bottom. From Ref.~\cite{gol09}.}
\label{fig:density}
\end{figure}

The dynamics of actinide collisions have been studied with the TDHF approach~\cite{cus80,gol09,ked10,sim11b} and the quantum molecular dynamics (QMD) formalism~\cite{tia08,zha09}. 
These microscopic studies complement other works with a quantum master equation~\cite{sar09}, the dinuclear system (DNS) model \cite{ada05,fen09}, multidimensional Langevin equations~\cite{zag06}, and the constrained molecular dynamics model~\cite{mar02}. 

Fig.~\ref{fig:density} shows zeptosecond snapshots of the density obtained in TDHF calculations of $^{238}$U+$^{238}$U central collisions at $E_{c.m.}=900$~MeV. 
We see that the initial orientation of the nuclei plays a crucial role on the reaction mechanism~\cite{gol09}. 
For instance, we observe the formation of a third fragment in the left column.
A net transfer is also obtained in the middle column.
Indeed, integration of  proton and neutron 
densities in each reactant indicates an average transfer of $\sim6$~protons and $\sim11$~neutrons from the right to the left nucleus. In this case, transfer occurs from the 
tip of the aligned nucleus to the side of the other. 
This configuration is then expected to favor the formation of nuclei heavier than~$^{238}$U. 

Pursuing this idea, a new ''inverse quasi-fission'' mechanism has been identified~\cite{ked10}.
The inverse process of quasi-fission corresponds to a transfer of nucleons from the light collision partner to the heavy one. 
Such a mechanism may occur in actinide collisions due to shell effects in the $^{208}$Pb region~\cite{vol78,zag06}. 
The mechanism proposed in Ref.~\cite{ked10} is different (but may be complementary).
It occurs for specific orientations of the actinides, where the tip of the lighter one is in contact with the side of the heavier one. 

An example of inverse quasi-fission reaction is shown in the right panels of Fig.~\ref{fig:dist}, where a $^{232}$Th+$^{250}$Cf collision produces a $^{265}$Lr fragment at the end of the calculation\footnote{This fragment is not exactly a primary fragment as at the final time of the calculation, about 3 neutrons have been emitted in the entire system (see Fig.~4 of~\cite{ked10}). However, it is still excited (its excitation energy may be estimated from the TKE) and may cool down by the emission of 1-2 additional neutrons or, of course, by fission.}.
The $^{265}$Lr heavy fragment indicates the average $N$ and $Z$ of a distribution. 
The fluctuations and correlations of these distributions have been computed with the BV prescriptions [Eqs.~(\ref{eq:CiiBV}) and~(\ref{eq:CijBV})] in Ref.~\cite{sim11b}.
Fig.~\ref{fig:dist}(left) shows the resulting probabilities assuming Gaussian distributions of the form given in Eq.~(\ref{eq:Gauss}).
We see that many $\beta-$stable and neutron-rich transfermium {\it primary} fragments could be produced thanks to this inverse quasi-fission mechanism. 
In particular, these nuclei are more neutron-rich than those formed in fusion-evaporation reactions. 
This inverse quasi-fission process needs further studies to determine the role of the shape, orientation, 
shell closures, mass and charge asymmetries, beam energy, and angular momentum on the transfermium 
production yields.  
Associated cross-sections need to be determined to estimate the experimental possibility of neutron-rich transfermium and SHE productions. 

Only few early experimental data exist on heavy element production in actinide collisions, with chemical identification techniques leading to limited sensitivity in terms of cross-section measurements. 
For example, the smallest cross-sections obtained for the production of transfermiums (in this case, mendelevium isotopes with $Z=101$) were of the order of $\sim20$~nb in the $^{238}$U+$^{248}$Cm reaction~\cite{sch82}, while modern fusion experiments using spectrometers for the fragment separation 
have measured cross-sections down to 30~fb~\cite{mor07}, i.e. $\sim6$ orders of magnitude smaller. 
The use of modern experimental techniques and equipments to identify
multi-nucleon transfer products with actinide collisions is then expected to be very fruitful in terms of transfermium studies~\cite{dvo11}. 

A non exhaustive list of relevant heavy nuclei to be searched for is given below: 
\begin{itemize}
\item SHE in the island of stability (e.g., $^{298}$114). 
\item $^{261-263}$Fm which have a predicted $\beta$-decay. If observed, they would be the heaviest nuclei with this decay mode. 
\item $^{264}$Fm could undergo symmetric spontaneous fission to form two doubly magic $^{132}$Sn. 
\item From $^{271}$Rf$_{104}$ to $^{284}$Mt$_{109}$, $\beta-$stable nuclei and their neighbours are predicted to decay by spontaneous fission. Verifying this prediction would constrain the models predicting a spontaneous fission region in the decay path of r-process progenitors with mass $\sim300$~\cite{lan11}. The existence of the latter is crucial to determine if long-lived SHE can be produced in the r-process. 
\item $^{264}$No$_{102}$ has 162 neutrons which corresponds to a deformed shell gap in the $Z=108$ region. The question of the robustness of the N=162 shell gap away from Z=108 could be answered by measuring the mass and life-time of $^{264}$No$_{102}$ and its neighbours. 
\item $^{266}$Rf$_{104}$ and its neighbours should be studied to ''fill in'' the blank spot between cold and hot fusion decay chains. 
\item $^{290}$114, $^{287}$113, $^{286}$112 should be produced to confirm the increase of life time with neutron number attributed to the closeness of the island of stability\footnote{  $^{277}112$ and $^{285}112$ have $T_{1/2}=0.7$~ms and 34~s measured half-lives, respectively. A gain 
of 8 neutrons increases the life-time by 5 orders of magnitude!}. 
\item Elements $Z=109-111$ still need chemical characterisation~\cite{gag11}. Present chemical techniques tag the isotopes thanks to their $\alpha-$decay and $T_{1/2}>0.5$~s is required. $^{278}$Mt$_{109}$ and $^{282}$Rg$_{111}$ are good candidates. Ds  ($Z=110$) element is more problematic as no isotope decaying by $\alpha-$emission with $T_{1/2}>0.5$~s is known. Decay modes of $^{280}$Ds$_{110}$ should then be measured. If the latter does not match experimental requirements for chemistry studies, isotopes on the other side of the spontaneous fission region (see Ref.~\cite{lan11}), i.e. $^{287}$Ds and its neighbours, might be considered as they are expected to decay by $\alpha$ or $\beta^-$ emission. 
\item $^{290}115$, which has $T_{1/2}=0.7$~s is the next milestone in chemical studies of SHE. 
\end{itemize}

\begin{figure*}
\includegraphics[width=18cm]{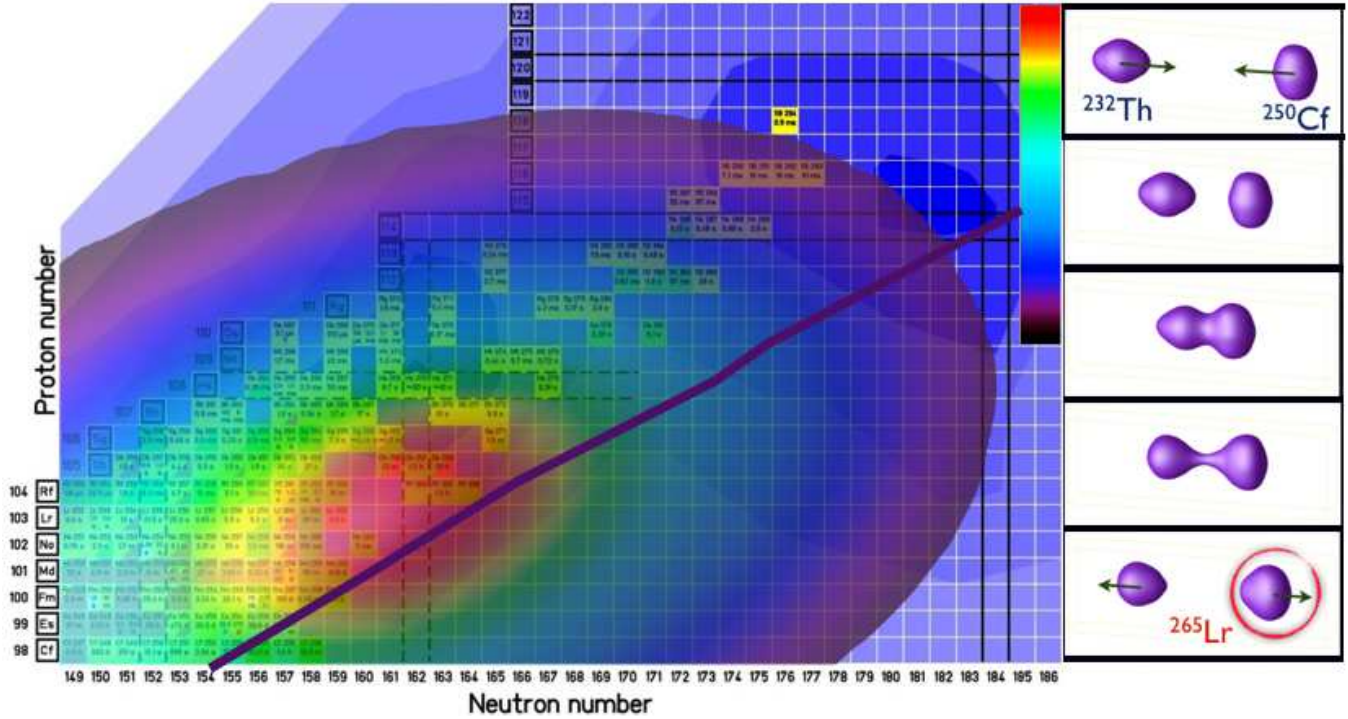}
\caption{(right) Snapshots of the isodensity at half the saturation density in $^{232}$Th+$^{250}$Cf central collisions at $E_{c.m.}=916$~MeV. (left) Gaussian distributions of $N$ and $Z$ of the heavy outgoing fragment with widths and correlations computed with the BV prescription (linear color scale). The purple line shows the expected $\beta$-stability line.}
\label{fig:dist}
\end{figure*}

\subsection{Conclusions and perspectives}

This section was devoted to the study of heavy-ion reaction mechanisms in order to investigate quantum dynamical effects in complex systems and to predict the formation of new nuclei.

The TDHF theory was shown to reproduce energy thresholds for fusion for systems spanning the entire  nuclear chart.
The effect of deformation and transfer on fusion barriers is well treated.
The competition with quasi-fission leading to a fusion hindrance in heavy systems is also included. 
The possibility to study the quasi-fission mechanism with a fully microscopic quantum approach is promising. 
It will help to understand the strong fusion hindrance  in quasi-symmetric heavy systems. 

It was also shown that the charge equilibration process in $N/Z$ asymmetric collision affects the reaction mechanism: excitation of a preequilibrium GDR in the compound system (which can be used to study the path to fusion), enhancement of proton stripping and neutron pickup, and correlations between fragment $N$ and $Z$ distributions in DIC (which should be sensitive to the symmetry energy).

Actinide collisions have been investigated both within the TDHF approach and with the BV prescription. 
A new inverse quasi-fission mechanism associated to specific orientations was found in numerical simulations.
This mechanism might produce $\beta-$stable and neutron-rich heavy and super-heavy nuclei.
A systematic investigation of this effect (cross-sections, angular and energy distributions) is mandatory to help the design of future experimental equipments dedicated to the study of fragments produced in actinide collisions.

Quantum effects at the single-particle level are well treated in the TDHF approach, allowing for realistic predictions of independent particle transfer probabilities using a particle-number projection technique. 
A comparison with experimental data shows the importance of pairing correlations on transfer.
The latter could be studied with 3-dimensional TDHFB codes in the future. 
In particular, one could answer the question on the origin of these pairing correlations: Are they present in the ground-states or are they generated dynamically during the collision? 

A strong limitation of the TDHF approach is that it does not allow for quantum tunnelling of the many-body wave function. 
As a result, sub-barrier fusion cannot be studied. 
Recently, suppression of tunnelling was observed at deep sub-barrier energies, which could not be reproduced by standard coupled channel calculations~\cite{jia04,das07,esb10}. 
This is one of the biggest puzzles in low-energy heavy ion collision physics, attracting the interest of physicists from various fields including astrophysics, where changed quantum tunnelling rates can have drastic effects~\cite{gas07}. 
In the future, microscopic models should be developed to study such low-energy reactions including the tunnelling of the many-body  wave-function. 
A possible approach is to use an extension of the TDHF theory based on a path integral technique and on the stationary phase approximation~\cite{neg82} (see discussion in section~\ref{sec:persp}).

\section{Conclusions}

Nuclei are ideal to investigate fundamental  aspects of the quantum many-body problem. 
They exhibit collective motions built from coherent superpositions of the states of their constituents.
Examples range from collective vibrations to the formation of a compound system in collisions.
These features are common to other composite systems (atomic clusters, molecules, Bose-Einstein condensates...).
Their study in nuclear systems is obviously part of a wider physics field. 

The Balian-V\'en\'eroni variational principle offers a powerful approach to the many-body quantum dynamics. 
Both the observable and the state of the system are considered as variational quantities. 
In the limit of independent particles, one obtains the time-dependent Hartree-Fock formalism for the expectation of one-body observables, and the time-dependent RPA for their fluctuations and correlations. 

Different studies of nuclear dynamics, from collective vibrations to heavy-ion collisions have been presented in this review. 
A particular attention was devoted to the interplay between collective motions and internal degrees of freedom within a unified theoretical description. 

We showed that strongly interacting systems such as nuclei could exhibit collective vibrations in the continuum, and that their direct decay could be used to infer their microscopic structure. 
We also questioned the harmonic nature of these vibrations.
In particular, we identified a source of anharmonicity from the coupling between different vibrational modes.
Nuclei are known to exhibit superfluidity due to pairing residual interaction and it is natural to wonder if the pairing field could also vibrate. 
We then extended the theoretical approach to study such pairing vibrations. 

Large amplitude collective motions were investigated in the framework of heavy-ion collisions. 
We described the mechanism of the formation of a compound system. 
In particular we discussed how fusion is affected by the internal structure of the collision partners. 
We then investigated the other mechanisms in competition with fusion, and responsible for the formation of  fragments which differ from the entrance channel: transfer reactions, deep-inelastic collisions, and quasi-fission. 
We finally studied actinide collisions forming, during very short times of few zeptoseconds, the heaviest nuclear systems available on Earth. 
In particular, we identified a new reaction mechanism occurring in these collisions which could be used to study the upper part of the nuclear chart.

\section*{Acknowledgements}

This work is dedicated to P. Bonche, who is the main author of the \textsc{tdhf3d} code and was one of the principal contributors to this field. 
An essential part of the work presented in this review article as been carried out by B. Avez, C. Golabek, D. J. Kedziora and A. Wakhle who are thanked for their important contributions.
Motivating long term collaborations on nuclear dynamics with Ph. Chomaz and D. Lacroix are acknowledged. 
Collaborations with experimentalists are crucial to this field and M. Dasgupta, D. J. Hinde, M. Evers, G. de France, A. Drouart, Ch. Theisen, B. Sulignano and many others from CEA Saclay, GANIL and ANU are warmly thanked. 
Advances in nuclear dynamics strongly benefit from the work of nuclear structure theoreticians, and collaborations with M. Bender, K. Bennaceur, and T. Duguet are acknowledged. 
F. Gulminelli, J. Maruhn, N. Rowley, F. S\'ebille, and E. Suraud are thanked for stimulating discussions as well as a careful reading of the manuscript. 
Useful discussions with M. V\'en\'eroni, S. Umar, and P.D. Stevenson are also acknowledged. 
K. Bennaceur and V. Yu. Denisov are thanked for their corrections.

The calculations have been performed on the Centre de Calcul Recherche et Technologie of the Commissariat \`a l'\'Energie Atomique, France, and on the NCI National Facility in Canberra,
Australia, which is supported by the Australian Commonwealth Government.
Partial support from ARC Discovery grants DP0879679,  
DP110102858, and DP110102879 is acknowledged.

{
\appendix 
\section*{APPENDICES}
\addcontentsline{toc}{section}{APPENDICES}
\section{TDHF approach from the standard variational principle\label{annexe:standardTDHF}}

Consider a state of $N$ independent particles described by the Slater determinant $\kfi$.
The action defined in Eq.~(\ref{eq:SDirac}) reads
\oeq
S \equiv S_{t_0,t_1}[\phi] = \int_{t_0}^{t_1} \stb d t \stf \<\phi (t)| \( i\hb \frac{d }{{d} t} - \oH \) |\phi(t)\>
\ceq
where $\oH$ is the Hamiltonian of the system. 
In such a state, every $M$-body density matrix ($1\le M\le N$) is simply expressed as a function of the one-body density matrix $\ro$~(see appendix~C of Ref.~\cite{sim10a}).
Then, all the information on the system is contained in $\ro$.
The expectation value of the Hamiltonian on the state $\kfi$ may then be written as a functional of~$\ro$: $E[\ro] = \bfi \oH \kfi$.
In addition, we have
\oeq
\bfi \frac{d}{d t} \kfi = \sum_{i=1}^N \<\az_i|  \frac{d}{d t} |{\az}_i\>
\ceq
and, then,
\oeqn
S &=& \int_{t_0}^{t_1} \stb d t \stf \( i\hb\sum_{i=1}^N \<\az_i| \frac{ d}{ d t} |{\az}_i\>
-E[\ro(t)]\) \nonumber \\
&=& \int_{t_0}^{t_1} \stb  d t \stf \(i\hb \sum_{i=1}^N \sdf \int \stb  d x \stf \az_i^*(x,t) \sdf \frac{ d}{ d t} {\az}_i(x,t)
-E[\ro(t)]\) \nonumber \\
\ceqn
where $x\equiv (\vr s q)$ describes all the single-particle degrees of freedom (position $\vr$, spin $s$ and isospin $q$). 
We introduced the notation $\int \sdb d x = \sum_{q s} \int \sdb  d \vr$.

The variational principle reads $\del S = 0$. 
The variation must be done on each independent variable.
Here, these variables are the real part $\az^{\rm Re}_\al$ and the imaginary part $\az_\al^{\rm Im}$ of each occupied single particle state $\az_\al$.
We must then consider 
\oeq
\frac{\delta S}{\delta  \az^{\rm Re}_\al(x,t)} = 0 \mbox{\hspace{1cm} and \hspace{1cm}}
\frac{\delta S}{\delta  \az^{\rm Im}_\al(x,t)} = 0 
\label{eq:delRe}
\ceq
for each  $\al \in \{1...N\}$, for all $t$ such as $t_0\le t \le t_1$, and for all $x$. 

However, the calculation is more straightforward if we use $\az_\al$ and $\az^*_\al$ as {\it independent} variables instead of $\az^{\rm Re}_\al$ and $\az_\al^{\rm Im}$.
We can then consider the variations over $\az$ and over $\az^*$ independently. 
Note, however, that we loose the property ''$\az^*$ 
being complex conjugated of $\az$'' that we should restore later.

Equations (\ref{eq:delRe}) are then replaced by
\oeq
\frac{\delta S}{\delta  \az_\al(x,t)} = 0 \mbox{\hspace{1cm} and \hspace{1cm}}
\frac{\delta S}{\delta  \az^*_\al(x,t)} = 0 .
\label{eq:delphiphistar}
\ceq
The variation over $\az^*$  gives
\oeq
\frac{\delta S}{\delta  \az^*_\al(x,t)} = i\hb \sdf \frac{ d}{ d t} \az_\al(x,t) - 
 \int_{t_0}^{t_1} \stb  d t' \stf \frac{\delta E\[\ro(t')\]}{\delta  \az^*_\al(x,t)}.
\ceq
The functional derivative of $E$ can be re-written thanks to a change of variable
\oeq 
 \frac{\delta E\[\ro(t')\]}{\delta  \az^*_\al(x,t)}
 =  \int \stb  d y \,  d y' \stf \frac{\delta E\[\ro(t')\]}{\delta  \ro(y,y';t')}
\sdf  \frac{\delta  \ro(y,y';t')}{\delta  \az^*_\al(x,t)}.
\ceq
Using
\oeq 
  \frac{\delta  \ro(y,y';t')}{\delta  \az^*_\al(x,t)} =  \az_\al(y,t') \sdf \delta(y'-x) \sdf \delta(t-t')
\ceq
and noting the single-particle Hartree-Fock Hamiltonian $h$ with matrix elements
\oeq
h(x,y;t) = \frac{\delta E\[\ro(t)\]}{\del \ro(y,x;t)},
\label{eq:hEHF}
\ceq
we get the TDHF equation for the set of occupied states
\oeq
\fbox{$ \displaystyle i\hb \sdf \frac{ d }{ d t} \az_\al(x,t) = \int \stb  d y \stf h(x,y;t) \sdf \az_\al(y,t)$}\sdf. \label{eq:TDHF_az}
\ceq

The variation over $\az$ gives, after integrating by part the term with the time derivative, 
\oeqn
\frac{\delta S}{\delta  \az_\al(x,t)} &=& \frac{\delta }{\delta  \az_\al(x,t)}  \int_{t_0}^{t_1} \stb  d t' \nonumber\\
&&\stb\stb\stb\stb\stb\stb\stb \left[ i \hb \frac{ d}{ d t'}
 \(\sum_\be \int \stb  d y \stf \az^*_\be(y,t')\sdf \az_\be(y,t')\) \right. \nonumber \\
&&\stb\stb\stb\stb\stb\stb\stb\left.
-i\hb \sdf  \sum_\be \int \stb  d y \stf \(\frac{ d}{ d t'} \az^*_\be(y,t')\) \sdf \az_\be(y,t')-  E\[\ro(t')\]\right].\nonumber\\
\label{eq:var_az}
\ceqn
The first term in the r.h.s. cannot  be replaced {\it  a priori} by  $\Tr [\ro] = N$
because we considered $\az$ and $\az^*$ to be independent variables. 
Let us note the variation of $\az$ at a specific ''point'' of the Hilbert space at time~$t$ and affecting only the state $\az_\al$. This variation can be defined as 
\oeq
\delta_{\al x t} \, \az_\be(y,t') = \Delta \az \sdf f(t) \sdf \delta(t-t') \sdf \delta(x-y) \sdf \delta_{\al \be}.
\ceq
Using the usual definition of the functional derivative, we may write
\oeqn
&& \stb\frac{\delta }{\delta  \az_\al(x,t)}
 \int_{t_0}^{t_1} \stb  d t' \stf  \frac{ d}{ d t'}
 \(\sum_\be \int \stb  d y \stf \az^*_\be(y,t')\sdf \az_\be(y,t')\) \nonumber \\
&&\stb =\lim_{\Delta \az \rightarrow 0} 
 \int_{t_0}^{t_1} \stb  d t' \stf  \frac{ d}{ d t'}
\sdf \frac{1}{\Delta \az} \(\sum_\be \int \stb  d y \stf \az^*_\be(y,t')\sdf \delta_{\al x t} \az_\be(y,t')\) \nonumber \\
&&\stb= \[ f(t) \sdf \delta(t-t') \sdf \az^*_\al(x,t') \]_{t'=t_0}^{t_1}.
\ceqn
We have to choose $f(t_0)=f(t_1)=0$ in oder to cancel this term at $t=t_0$ or $t=t_1$. 
It is equivalent to forbid variations of $\az$ at $t_0$ and $t_1$. 
As a result, Eq.~(\ref{eq:var_az}) leads to the complex conjugated of the TDHF equation~(\ref{eq:TDHF_az}), 
restoring the fact that $\az$ and $\az^*$ are complex conjugated. 
This last property is also necessary for energy and particle number conservations.

\section{Fluctuations and correlations of one-body observables with the Balian-V\'en\'eroni variational principle \label{annexe:BV}}

In this appendix, we detail the derivation of the Balian-V\'en\'eroni prescription for fluctuation and correlation of one-body observables, given in Eqs.~(\ref{eq:CiiBV}) and~(\ref{eq:CijBV}), respectively. 
We consider the particular case where the independent particle state of the system is a pure state described by a unique Slater determinant at all time. 
The more general derivation with mean-field states of the form given in Eq.~(\ref{eq:defD}) can be found in Refs.~\cite{bal92,bro09}.

As mentioned in section~\ref{sec:fluccor}, fluctuations and correlations of the one-body observables 
\oeq
\oQ_i = \sum_{\al\be} Q_{i_{\al\be}} \oad_\al \oa_\be
\ceq
are obtained by evaluating 
\oeq
\oA_1=\exp \(-\sum_i \varepsilon_i \oQ_i\) 
\label{eq:boundA1}
\ceq
at the final time $t_1$, and using Eq.~(\ref{eq:lnA}) for small $\varepsilon_i$. 
The expectation value of $\oA_1$ at $t_1$ is obtained from
\oeq
\<\oA_1\>(t_1)=\Tr \[ \oA(t_1) \oD(t_1) \],
\ceq
where the observable is assumed to be a time-dependent operator $\oA(t)$ with the boundary condition $\oA(t_1) = \oA_1$. The state of the system is assumed to be known at the initial time $t_0$, with the boundary condition $\oD(t_0)=\oD_0$. 
As a result, the variations of the observable and the state obey 
\oeq
\delta \oA(t_1) = \delta \oD (t_0) = 0.
\ceq

\subsection*{Variational space and parametrisation of variational quantities}

The BV variational principle is solved by requiring the stationarity of the action-like quantity $J$ defined 
 in Eq.~(\ref{eq:JA}), or, equivalently, in Eq.~(\ref{eq:JD}). 
 In the present application, the variational space for the observable is restricted to exponential of one-body operators. The observable is then parametrised as 
\oeq
\oA(t)=\exp[-\oL(t)]
\ceq
with 
\oeq 
\oL(t)=\sum_{\al\be}L_{\al\be}(t) \oad_\al \oa_\be
\ceq
The mean-field density-matrix is parametrised as
\oeq
\oD(t)=\exp[-m(t)-\oM(t)]
\ceq
with 
\oeq 
\oM(t)=\sum_{\al\be}M_{\al\be}(t) \oad_\al \oa_\be.
\ceq
As discussed in section~\ref{sec:fluccor}, the case of a pure Slater determinant corresponds to the limit where the eigenvalues of the matrix $M(t)$ tend to $\pm \infty$ and $m(t)\rightarrow +\infty$ in such a way that $z(t)=\Tr\oD(t)=1$~\cite{bal85}. 

According to Eqs~(\ref{eq:rhoM}) and (\ref{eq:mz1}), we have the following relationships 
\oeqn
\rho(t)&=&\frac{1}{1+e^{M(t)}}\mbox{ and}\label{eq:roeM}\\
m(t)&=&\tr \ln (1+e^{-M(t)}).
\ceqn
where $\rho$ is the one-body density matrix with elements
\oeq
\rho_{\al\be}(t) = \Tr[\oD(t)\oad_\be\oa_\al].
\label{eq:rhoalbe}
\ceq 
The inverse relationships read
\oeqn
m(t)&=&-\tr [\ln (1-\rho(t))] \mbox{ and}\label{eq:m}\\
e^{-M(t)} &=& \frac{\rho(t)}{1-\rho(t)}.\label{eq:e-M}
\ceqn
As expected, we see that the Slater determinant can be entirely parametrised by $\rho(t)$.

By analogy, we introduce a similar parametrisation of the observable using Eqs.~(\ref{eq:rhoM}) and~(\ref{eq:TrD}). We get 
\oeqn
\sigma(t)&=&\frac{1}{1+e^{L(t)}}\mbox{ and}\\
y(t)&=&\exp \[\tr \ln (1+e^{-L(t)})\], 
\ceqn
with the inverse relationships
\oeqn
\ln y(t)&=&-\tr[\ln(1-\sigma(t))] \mbox{ and}\label{eq:lny}\\
e^{-L(t)} &=& \frac{\sigma(t)}{1-\sigma(t)}.\label{eq:e-L}
\ceqn

Eqs.~(\ref{eq:JA}) and (\ref{eq:JD}) contain the products $\oA\oD$ and $\oD\oA$. 
It is convenient to define a similar parametrisation of these quantities, using the fact that the product of the exponential of one-body operators is also an exponential of a one-body operator
\oeqn
\oA(t)\oD(t)&=&e^{-\oL(t)}e^{-\oM(t)}=e^{-\oL'(t)}Ê\mbox{ and}\\
\oD(t)\oA(t)&=&e^{-\oM(t)}e^{-\oL(t)}=e^{-\oM'(t)}.\label{eq:DA}
\ceqn
$\oL'$ and $\oM'$ are one-body operators which are parametrised as 
\oeqn
\oL'(t)&=&m(t)+\sum_{\al\be}L'_{\al\be}(t) \oad_\al\oa_\be \mbox{ and}\\
\oM'(t)&=&m(t)+\sum_{\al\be}M'_{\al\be}(t) \oad_\al\oa_\be,
\ceqn
with
\oeqn
e^{-L'(t)}&=&e^{-L(t)}e^{-M(t)} \mbox{ and}\\
e^{-M'(t)}&=&e^{-M(t)}e^{-L(t)}.\label{eq:defM'}
\ceqn
By analogy with Eq.~(\ref{eq:rhoalbe}), we introduce two new matrices, $\rho'$ and $\sigma'$, with elements
\oeqn
\rho'_{\al\be}(t)&=& \frac{\Tr[\oD(t)\oA(t)\oad_\be\oa_\al]}{\omega(t)} \mbox{ and}\\
\sigma'_{\al\be}(t)&=& \frac{\Tr[\oA(t)\oD(t)\oad_\be\oa_\al]}{\omega(t)},
\ceqn
and obeying the following relationships:
\oeqn
\rho'(t)&=&\frac{1}{1+e^{M'(t)}}=\frac{1}{1+e^{L(t)}e^{M(t)}} \mbox{ and}\label{eq:rho'}\\
\sigma'(t)&=&\frac{1}{1+e^{L'(t)}}=\frac{1}{1+e^{M(t)}e^{L(t)}}.
\ceqn
Using Eqs.~(\ref{eq:DA}) and (\ref{eq:TrD}), the normalisation factor 
\oeq
\omega(t)=\Tr[\oD(t)\oA(t)]
\label{eq:defomega}
\ceq
becomes
\oeq
\omega(t)=\exp\[-m(t)+\tr\(\ln(1+e^{-M(t)}e^{-L(t)})\)\].
\label{eq:omegamML}
\ceq
Using Eqs.~(\ref{eq:m}), (\ref{eq:e-M}), (\ref{eq:lny}) and (\ref{eq:e-L}), we get
\oeq
\omega(t)=y(t)\exp\[\tr\(\ln(1-\rho(t)-\sigma(t)+2\sigma(t)\rho(t))\)\].
\label{eq:omega}
\ceq

The action contains a term with a time derivative, like $\Tr[\oA\frac{d\oD}{dt}]$ in Eq.~(\ref{eq:JA}), which, starting from Eq.~(\ref{eq:omega}), becomes
\oeq
\Tr\[\oA\frac{d\oD}{dt}\] = \omega \sdf\tr\[ \( \frac{2\sigma -1}{1-\ro-\sigma+2\sigma\rho} \) \frac{d\rho}{dt}\]
\ceq
The other terms of the action contain the Hamiltonian $\oH$.
For mean-field states, all the information on the system is contained in the one-body density matrix $\rho$ and we can write the expectation value of the Hamiltonian as an energy density functional, i.e.,
\oeq
\Tr\[\oD(t)\oH\]= E[\rho(t)].
\ceq
Similarly, we have
\oeqn
\Tr\[ \oD(t)\oA(t)\oH \] &=& \omega(t)E[\rho'(t)]\mbox{ and}\\
\Tr\[ \oA(t)\oD(t)\oH \] &=& \omega(t)E[\sigma'(t)].
\ceqn
As a result, the action defined in Eq.~(\ref{eq:JA}) becomes
\oeqn
 \stb\stb\stb \stb J&=& \omega(t_1) - \int_{t_0}^{t_1} dt \, \omega(t) \[\frac{}{} iE[\rho'(t)] -iE[\sigma'(t)]\right. \nonumber\\
&+& \left.\tr \( \frac{2\sigma(t) -1}{1-\ro(t)-\sigma(t)+2\sigma(t)\rho(t)}  \frac{d\rho(t)}{dt}\)\].
\ceqn

\subsection*{Equations of motion for $\rho$, $\rho'$ and $\sigma'$}

We now seek for an equation of motion for $\rho(t)$. 
This is obtained by requiring the stationarity of $J$ when $y$ and $\sigma$ vary. 
$J$ depends on $y$ due to  $\omega$  which is linear in $y$, i.e., 
\oeq
\frac{d\omega}{dy}=\frac{\omega}{y}.
\ceq
Requiring the stationarity of the action when $y(t)$ varies implies
\oeq
0=iE[\rho'(t)] -iE[\sigma'(t)]+ \tr \( \eta(t) \frac{d\rho(t)}{dt}\)
\label{eq:vary}
\ceq
where we have introduced a new matrix 
\oeq
\eta(t) = \frac{2\sigma(t) -1}{1-\ro(t)-\sigma(t)+2\sigma(t)\rho(t)} .
\ceq
A variation of $\sigma$ implies a variation of $\eta$ according to 
\oeq
\delta \eta=\frac{d\eta}{d\sigma}\delta \sigma .
\ceq 
Then, instead of considering the variation of $\sigma$, we equivalently study the variation of $J$ with $\eta$.  
The $\eta$ matrix can be re-written in such a way that it contains an explicit dependence on $\rho'$ and $\sigma'$:
\oeqn
\eta &=& \rho^{-1}(\rho'-\rho)(1-\ro)^{-1}\\
&=&(1-\ro)^{-1}(\sigma'-\rho)\rho^{-1}.
\ceqn
This tells us how $\rho'$ and $\sigma'$ vary with a variation of $\eta$:
\oeqn
\delta \rho' &=& \rho \, \delta \eta  \, (1-\rho) \mbox{ and}\\
\delta \sigma' &=& (1-\rho) \, \delta \eta \, \rho.
\ceqn
These relations are used to determine the variation of the energies $E[\rho']$ and $E[\sigma']$. Note that the latter can be written as a function of the self-consistent HF Hamiltonian $h$ using Eq.~(\ref{eq:hEHF}):
\oeqn
E[\rho']&=& \tr\(h[\rho'] \rho' \)\mbox{ and}\\
E[\sigma']&=& \tr\(h[\sigma'] \sigma' \).
\ceqn
A variation of $\eta$ induces, then, a variation of Eq.~(\ref{eq:vary}) which can be written as
\oeq
\tr\[\delta \eta \( \frac{d\ro}{dt} +i(1-\ro)\,h[\rho']\,\ro-i\ro \,h[\sigma'] \,(1-\ro)\)\]=0.
\ceq
Requesting this equation to hold for any variation $\delta \eta$ gives the equation of motion for $\rho$:
\oeq
i\frac{d\ro}{dt} =(1-\ro)\,h[\rho']\,\ro-\ro \,h[\sigma'] \,(1-\ro).
\label{eq:deltaeta}
\ceq

To get the equations of motion for $\ro'$ and $\sigma'$, it is convenient to express the time derivative term with the $L$ and $M$ matrices.
From Eq.~(\ref{eq:e-M}), we have
\oeq
\frac{d}{dt} e^{-M} = \frac{1}{1-\ro} \, \frac{d\ro}{dt} \, \frac{1}{1-\ro}.
\ceq
Using, Eq.~(\ref{eq:deltaeta}), we get
\oeq
i\frac{d}{dt}e^{-M} = h[\ro']\, e^{-M} - e^{-M}\, h[\sigma'].
\ceq
Similarly, we have
\oeq
i\frac{d}{dt}e^{-L} = h[\sigma']\, e^{-L} - e^{-L}\, h[\rho'].
\ceq
We can then write
\oeqn
\frac{d }{dt} e^{-M'} &=& \frac{de^{-M}}{dt} e^{-L} + e^{-M} \frac{de^{-L}}{dt} \nonumber \\
&=& i \[e^{-M'},h[\ro']\].
\label{eq:de-Mdt}
\ceqn
From Eq.~(\ref{eq:rho'}), we have
\oeq
1-\ro'=\frac{1}{1+e^{-M'}}.\label{eq:1-ro'}
\ceq
This leads to
\oeq
\frac{d}{dt} \ro' = (1-\ro') \, \frac{de^{-M'}}{dt} \, (1-\ro').
\ceq
Using Eq.~(\ref{eq:de-Mdt}) we get the equation of motion for $\ro'$:
\oeq
i\frac{d\rho'}{dt}=\[h[\ro'],\ro'\].
\label{eq:drho'}
\ceq
Similarly, we have
\oeq
i\frac{d\sigma'}{dt}=\[h[\sigma'],\sigma'\].
\label{eq:dsigma'}
\ceq

\subsection*{Development in powers of $\varepsilon$}

From the definition of $\omega(t)$ in Eq.~(\ref{eq:defomega}), we see that the expectation of $\oA_1$ at the final time $t_1$ is equal to $\omega(t_1)$. 
We now show that, when the action is stationary, $\omega(t)$ is in fact constant in time.
The one-body density matrix associated to a Slater determinant has eigenvalues 1 for the occupied states and 0 for the others. As a result, we see from Eq.~(\ref{eq:trln1+A}) that $m(t)$ is equal to an infinite constant and, then, $dm/dt=0$. 
Using this property and Eqs.~(\ref{eq:defM'}), (\ref{eq:rho'}), (\ref{eq:omegamML}), and (\ref{eq:de-Mdt}), we get
\oeqn
\frac{d}{dt}\omega &\propto& \frac{d}{dt} \tr\(\ln (1+e^{-M'})\)\nonumber \\
&\propto&\tr \frac{\frac{d}{dt}e^{-M'}}{1+e^{-M'}}\nonumber \\
&\propto& \tr\[ h[\ro'],\ro' \]=0.
\ceqn
$\omega(t)$ is then a constant we need to evaluate to get the fluctuations and correlations from Eq.~(\ref{eq:lnA}). 

$\ln \omega$ can be expressed as a function of $\rho$ and $L$ from Eq.~(\ref{eq:omegamML}) and using Eqs.~(\ref{eq:m}) and (\ref{eq:e-M}):
\oeq
\ln  \omega = \tr \[ \ln \( 1-\ro+\ro\, e^{-L} \)\].
\label{eq:lnomega}
\ceq
Deriving $\ln \omega$ according to the parameter $\varepsilon_i$, we get
\oeqn
\frac{d}{d\varepsilon_i}\ln \omega &=& -\<\oQ_i\>+\sum_j \varepsilon_jC_{ij}+ \cdots \label{eq:epsilondel} \\
&=&\tr\[\ro'e^L\frac{de^{-L}}{d\varepsilon_i}\]\label{eq:dode}
\ceqn
where the first identity is obtained from Eq.~(\ref{eq:lnA}) and the second from Eqs.~(\ref{eq:lnomega}), (\ref{eq:roeM}) and (\ref{eq:rho'}).
Let us expand $\rho$ and $L$ in powers of the $\varepsilon$:
\oeqn
\rho(t)&=& \rho^{(0)}(t) + \rho^{(1)}(t) +\cdots \mbox{ and}\\
L(t)&=&  L^{(1)}(t) +\cdots 
\ceqn
where the superscript denotes the power in $\varepsilon$. 
The fact that $L^{(0)}=0$ is due to the boundary condition in Eq.~(\ref{eq:boundA1}) implying
\oeq
L(t_1)=\sum_i \varepsilon_i Q_i.
\label{eq:bcL}
\ceq 
We now expand $\rho'$ up to first order in $\varepsilon$ using Eqs.~(\ref{eq:rho'}) and  (\ref{eq:e-M}):
\oeq
\rho'=\rho^{(0)} + \rho^{(1)} - \rho^{(0)}L^{(1)} (1-\rho^{(0)}) +O(\varepsilon^2).
\ceq
Separating the zeroth and first order contributions, i.e.,  $\rho'\simeq\rho'^{(0)}+\rho'^{(1)}$, we get
\oeqn
\rho'^{(0)}&=&\rho^{(0)} \nonumber\\
\rho'^{(1)}&=&\rho^{(1)} - \rho^{(0)}L^{(1)} (1-\rho^{(0)}) \label{eq:rho'exp}
\ceqn

\subsection*{Expectation value of one-body observables}

At the zeroth order, Eq.~(\ref{eq:dode}) becomes
\oeqn
\<\oQ_i\>&=& \tr\[\rho^{(0)}(t)\frac{dL^{(1)}(t)}{d\varepsilon_i}\] \nonumber \\
&=& \tr\[\rho^{(0)}(t_1)Q_i\],
\label{eq:Qtdhf}
\ceqn
where we have used the boundary condition in Eq.~(\ref{eq:bcL}).
From Eqs.~(\ref{eq:drho'}) and~(\ref{eq:rho'exp}) we see that $\rho^{(0)}$ obeys the TDHF equation. 
Eq.~(\ref{eq:Qtdhf}) is then exactly the TDHF result for the expectation value of one-body observables. 

\subsection*{Fluctuations and correlations}

We now expand Eq.~(\ref{eq:dode}) up to first order in $\varepsilon$, using Eqs.~(\ref{eq:rho'exp}) and (\ref{eq:bcL}):
\oeqn
&&\frac{d\ln \omega}{d\varepsilon_i} =\nonumber\\
&& \tr\[\(\ro^{(0)}(t_1)+\ro^{(1)}(t_1)-\ro^{(0)}(t_1)\sum_j\varepsilon_jQ_j(1-\ro^{(0)}(t_1))\)
\right.\nonumber\\
&&\left.  \( 1+\sum_n\varepsilon_nQ_n\)\(-Q_i+ \sum_m \varepsilon_mQ_iQ_m\) \],
\ceqn
where we assumed $[Q_i,Q_j]=0$ for simplicity.
Identifying with Eq.~(\ref{eq:epsilondel}), we get
\oeqn
\sum_j\varepsilon_jC_{ij} &=& \sum_j \varepsilon_j \tr\[Q_i\rho^{(0)}(t_1) Q_j \(1-\rho^{(0)}(t_1)\) \] \nonumber \\
&&- \tr\(\rho^{(1)}(t_1)Q_i\).\label{eq:epsCij}
\ceqn
The first term in the right hand side gives the usual fluctuations and correlations computed at the TDHF level [see Eq.~(\ref{eq:CiiTDHF})]. 
The present approach, which is more general as it uses a larger variational space for the evolution of the observable than in the TDHF formalism, is expected to improve the fluctuations and correlations thanks to the second term in the right hand side. 

We see in Eq.~(\ref{eq:epsCij}) that the time evolution of $\rho^{(1)}$ is needed in addition to $\rho^{(0)}$. 
Unfortunately, $\rho^{(1)}$ does not obey a simple equation and its determination is not trivial. 
It is however possible to find an equation for $C_{ij}$ which is easier to implement. 
In fact, although Eq.~(\ref{eq:epsCij}) has been obtained by evaluating the matrices at time $t_1$, it is possible to show that a similar equation holds at any time. 
Let us first introduce the quantity
\oeqn
F_{ij}(t) &=& \frac{1}{2} \tr \( \rho^{(0)} [L_i^{(1)},L_j^{(1)}]\)+ \tr \[ L_i^{(1)}\ro^{(0)} L_j^{(1)}(1-\rho^{(0)})\]\nonumber \\
&&-  \tr\(\rho^{(1)}L_i^{(1)}\),\label{eq:Fij}
\ceqn
where the matrices in the right hand side are expressed at time~$t$ and the $L_i^{(1)}$ obey the boundary condition
\oeq
L_i^{(1)}(t_1)=\varepsilon_iQ_i.
\ceq
We have also the boundary condition 
\oeq
\sum_j F_{ij}(t_1)=\sum_j\varepsilon_i\varepsilon_jC_{ij}.
\ceq 
Note that a similar term than the first term in the right hand side of Eq.~(\ref{eq:Fij}) appears in Eq.~(\ref{eq:epsCij}) when we consider the general case $[Q_i,Q_j]\ne0$. 
We now show that, in fact, $F_{ij}$ does not depend on time. 
We first re-write $F_{ij}$ as a function of the matrices $L$, $\sigma'$ and $\rho'$:
\oeq
F_{ij}=\frac{1}{2}\tr \[ L_i^{(1)} (\sigma_j'^{(1)}+\rho_j'^{(1)}) \],
\ceq
where we have used Eq.~(\ref{eq:rho'exp}) and its equivalent for $\sigma'$.
The matrices $\rho'^{(1)}$ and $\sigma'^{(1)}$ are obtained by taking the first order in $\varepsilon$ in the equations (\ref{eq:drho'}) and (\ref{eq:dsigma'}), respectively.
We obtain the time-dependent RPA equations
\oeqn
i\frac{d\rho'^{(1)}}{dt}&=&\[h[\rho^{(0)}] , \rho'^{(1)}\] + \[ \tr_2(\bar{v}  \rho'^{(1)}), \rho^{(0)}\]\label{eq:idrho'}\\
i\frac{d\sigma'^{(1)}}{dt}&=&\[h[\rho^{(0)}] , \sigma'^{(1)}\] + \[ \tr_2(\bar{v}  \sigma'^{(1)}), \rho^{(0)}\],\label{eq:idsigma'}
\ceqn
 where $\bar{v}$ is the antisymetrised two-body interaction defined in Eq.~(\ref{eq:vbar}). 
We also get a similar equation for $L$:
\oeq
i\frac{dL^{(1)}}{dt}=\[h[\rho^{(0)}] , L^{(1)}\] + \tr_2\[ \bar{v} [L^{(1)}, \rho^{(0)}]\].\label{eq:idL}
\ceq
The time derivative of $F_{ij}$ obeys
\oeqn
\frac{d}{dt}F_{ij} &=& -\frac{1}{2}\tr\[ \frac{dL_{i}^{(1)}}{dt}{\rho'}_j^{(1)} + L_{i}^{(1)} \frac{d{\rho'}_j^{(1)}}{dt} \right.\nonumber \\
&&+ \left.\frac{dL_{i}^{(1)}}{dt}{\sigma'}_j^{(1)} + L_{i}^{(1)} \frac{d{\sigma'}_j^{(1)}}{dt} \].
\ceqn
 Replacing with Eqs.~(\ref{eq:idrho'}), (\ref{eq:idsigma'}) and (\ref{eq:idL}), we get 
\oeq 
\frac{d}{dt} F_{ij}(t) = 0.
\ceq
$F_{ij}$ is then constant with time. 

It is convenient to write it with the matrices at the time $t_0$ instead of $t_1$. Indeed, in this case, we have $\rho^{(1)}(t_0)=0$ due to the boundary condition at initial time. This simplifies the expression for the fluctuations and correlations:
\oeqn
\stb\stb C_{ij} &=& -\lim_{\varepsilon_i,\varepsilon_j\rightarrow0} \frac{1}{2\varepsilon_i\varepsilon_j} \nonumber \\
&&\tr \[ [L_i^{(1)}(t_0),\rho^{(0)}(t_0)] [L_j^{(1)}(t_0),\rho^{(0)}(t_0)]  \] \label{eq:CijLro}
\ceqn
Let us introduce a new matrix $\eta_i(t,\varepsilon_i)$ defined as
\oeqn
\eta_i(t,\varepsilon_i) &=& \rho^{(0)}(t) + i \[ L_i^{(1)}(t) , \rho^{(0)}(t) \].\label{eq:etafull}\\
&\approx&  e^{iL_i^{(1)}(t)}\rho^{(0)}(t) e^{-iL_i^{(1)}(t)},\label{eq:etaapp}
\ceqn
where the last equation holds for small $\varepsilon$. 
We can show that $\eta_i(t,\varepsilon_i)$ follows also a TDHF equation.
Note that the fact that $\rho^{(0)}$ is a Slater determinant implies that 
 $\eta_i$ is also a Slater determinant according to 
Eq.~(\ref{eq:etaapp}).

From Eqs.~(\ref{eq:CijLro}) and~(\ref{eq:etafull}) we get our final result
\oeqn
C_{ij} = \lim_{\varepsilon_i,\varepsilon_j\rightarrow0} \frac{1}{2\varepsilon_i\varepsilon_j}
&\tr&\[\(\rho^{(0)}(t_0)-\eta_i(t_0,\varepsilon_i)\)\right.\nonumber\\
&&\stb \stb \stb\left.\(\rho^{(0)}(t_0)-\eta_j(t_0,\varepsilon_j)\)\],
\ceqn
with the boundary condition  at the final time $t_1$
\oeq
\eta_i(t_1,\varepsilon_i)=e^{i\varepsilon_iQ_i}\rho^{(0)}(t_1)e^{-i\varepsilon_iQ_i}.
\ceq

\section{From Fock to single-particle space with exponential of one-body operators \label{annexe:formula}}

Several relations are derived which are used in the derivation of the Balian-V\'en\'eroni prescription for fluctuation and correlation of one-body operators in appendix~\ref{annexe:BV}.
They relate operators or their associated matrices in Fock space to matrices expressed in the single-particle space.

\subsection*{Link between the density matrix of an independent-particle state and the one-body density matrix}

Our goal is to show the relation
\oeq
\boxed{\rho=\frac{1}{1+e^M}}
\label{eq:rhoM}
\ceq
where $\rho$ is the one-body density matrix with elements 
\oeq
\rho_{\al\be}=\<\oad_\be\oa_\al\>=\Tr\(\oD\oad_\be\oa_\al\),
\label{eq:rho_albe}
\ceq
and
\oeq
\oD=e^{-m-\oM}
\label{eq:D}
\ceq
is the density matrix associated to an independent-particle state described by a Slater determinant.
The later obeys
\oeq
\Tr\oD=1
\ceq
and $\oM$ is a one-body operator of the form 
\oeq
\oM=\sum_{\al\be}M_{\al\be}\oad_\al\oa_\be.
\label{eq:oM}
\ceq
We first show the following property: 
\oeq
\oad_\al\oD=\sum_\be\(e^M\)_{\be\al}\oD\oad_\be
\label{eq:oadD}
\ceq

{
Define $\oF(x)=e^{x\oA}\oB e^{-x\oA}.$
Its Taylor development reads
$$\oF(x)=\sum_{n=0}^\infty\frac{x^n}{n!}\of_n \stf \mbox{with}\stf \of_n=\left.\frac{\partial^n\oF}{\partial x^n}\right|_{x=0}$$
\oeqn
\oF(x)&=&\sum_{n=0}^\infty\frac{(x\oA)^n}{n!}\sdf\oB\sdf\sum_{m=0}^\infty\frac{(x\oA)^m}{m!}\nonumber\\
&=&\oB+x\[\oA,\oB\]+\frac{x^2}{2!}\[\oA,[\oA,\oB]\]+\cdots\nonumber
\ceqn
$\Rightarrow \of_0=\oB, \stf \of_{n+1}=[\oA,\of_n]$. 
Using this relation with $x=1$, we get
$$
\oD^{-1}\oad_\al\oD=\sum_{n=0}^\infty \frac{1}{n!} \of_n \stf \mbox{with}\stf 
\left\{\begin{array}{cl}
\of_0&=\oad_\al \\
\of_{n+1}&=\[\oM,\of_n\],
\end{array}\right.
$$
where the $\of_n$ have an implicit label $\al$.
Using Eq.~(\ref{eq:oM}) and the commutation rules for creators and annihilators of fermions, we find
$$\of_n=\sum_\be\(M^n\)_{\be\al} \oad_\be.$$
We finally get
$$\oD^{-1}\oad_\al\oD=\sum_\be\(e^M\)_{\be\al}\oad_\be,$$
leading to Eq.~(\ref{eq:oadD}).
}

Using $\oad_\al\oa_\be=\delta_{\al\be}-\oa_\be\oad_\al$ and Eqs.~(\ref{eq:rho_albe}) and~(\ref{eq:oadD}) leads directly to
\oeq
\rho_{\be\al} = \delta_{\be\al} - \Tr\(\oad_\al\oD\oa_\be\) =  \delta_{\be\al} - \sum_\ga \(e^M\)_{\ga\al}\rho_{\be\ga}. 
\ceq
We get $\rho=1-\rho e^M$ and finally Eq.~(\ref{eq:rhoM}).

\subsection*{Trace of the exponential of a one-body operator}
Our goal is to show the relation
\oeq
\boxed{\Tr e^{-\oA}=\det(1+e^{-A})}
\label{eq:e-A}
\ceq
where $\oA$ is a one-body operator.

We first show that
\oeq
\Tr\oB=\int dz^*dz \sdf e^{-z^*z}\<z|\oB|-z\>
\label{eq:TrB}
\ceq
where $\oB$ contains only even powers of $\oad$ and $\oa$ (this is the case, for instance, when $\oB$ is the exponential of a one-body operator), and $z$ is an element of the Grassmann algebra (see appendix~\ref{annexe:Grassmann}). 

{
With $\{|\xi\>\}$ a basis of the Fock space, the trace of $\oB$ reads
$ \Tr\oB=\sum_\xi \<\xi|\oB|\xi\>.$
Inserting the closure relation from Eq.~(\ref{eq:zclosure}), we get
\oeq
\Tr\oB= \int dz^*dz e^{-\mathbf{z}^*\mathbf{z}}\sum_\xi \<\xi|\oB|z\>\<z|\xi\>.
\label{eq:TrB2}
\ceq
From Eq.~(\ref{eq:zAz'}), we have
$$\<\xi|\oB|z\>= B(\overleftarrow{\partial}_\mathbf{z},\mathbf{z})\<\xi|z\>.$$
$\{|\xi\>\}$ can be chosen to be an ensemble of Slater determinants with different particle numbers $N_\xi$:
$$|\xi\>=\(\prod_{i=1}^{N_\xi} \oad_{\xi_i}\) |-\>.$$ 
Using Eq.~(\ref{eq:az}) we get
$$\<\xi|z\>=\(\prod_{i=1}^{N_\xi} z_{\xi_i}\) \<-|z\>=\prod_{i=1}^{N_\xi} z_{\xi_i}.$$
Indeed, from Eq.~(\ref{eq:coh_state}), we have 
$$\<-|z\>=\<-|e^{\sum_\al z_\al\oad_\al}|-\>=1.$$
Then, using Eq.~(\ref{eq:z_alz_be}), we get
$$\<\xi|z\>\,\<z|\xi\> = (-1)^{N_\xi}\<z|\xi\> \<\xi|z\>=\<z|\xi\> \<\xi|-z\>.$$
Using the closure relation $\sum|\xi\>\<\xi|=\hat{1}$, Eq.~(\ref{eq:TrB2}) finally becomes
\oeqn
\Tr\oB&=& \int dz^*dz \sdf e^{-\mathbf{z}^*\mathbf{z}}\, B(\overleftarrow{\partial}_\mathbf{z},\mathbf{z})\,\<z|-z\>\nonumber\\
&=& \int dz^*dz \sdf e^{-\mathbf{z}^*\mathbf{z}}\, \<z|\oB|-z\>,\nonumber
\ceqn
}
i.e., Eq.~(\ref{eq:TrB}).

The second step is to show the relation
\oeq
e^{-\oA}|z\> = |e^{-A}z\>.
\label{eq:eA}
\ceq
{
Similarly to Eq.~(\ref{eq:oadD}), we have
$$\oa_\al e^{-\oA}= e^{-\oA} \sum_\be \(e^{-A}\)_{\al\be}\, \oa_\be $$
and then, using Eq.~(\ref{eq:az}), we get
\oeqn
\oa_\al e^{-\oA}|z\>&=& e^{-\oA} \sum_\be \(e^{-A}\)_{\al\be}\, z_\be |z\> \nonumber \\
&=&  \(e^{-A}\mathbf{z}\)_\al e^{-\oA}|z\> \equiv \oa_\al|e^{-A}z\>.\nonumber
\ceqn
}

Finally, we write Eq.~(\ref{eq:TrB}) with $\oB=e^{-\oA}$ and we use Eqs.~(\ref{eq:eA}) and (\ref{eq:zz'}) to get
\oeqn 
\Tr e^{-\oA} &=& \int dz^*dz \sdf e^{-\mathbf{z}^*\mathbf{z}}\<z|-e^{-A}z\>\nonumber \\
&=&\int dz^*dz \sdf e^{-\mathbf{z}^* (1+e^{-A})\mathbf{z}}.\nonumber 
\ceqn
Using Eq.~(\ref{eq:detM}) we obtain our final result in Eq.~(\ref{eq:e-A}).

\subsection*{Trace of the density matrix}

Our goal is to show the relation
\oeq
\boxed{\Tr \oD = e^{-m+\tr [\ln (1+e^{-M})]}}
\label{eq:TrD}
\ceq
where $\Tr$ denotes the trace in the Fock space while $\tr$ denotes the trace of a single-particle matrix.
$\oD$ is the density matrix of an independent particle system, as defined by Eqs.~(\ref{eq:D}) and~(\ref{eq:oM}). However we do not assume a normalised state, i.e., $\Tr\oD$ is not necessarily equal to 1. 

From Eq.~(\ref{eq:e-A}), we have
\oeq\Tr \oD= e^{-m} \det (1+e^{-M}) \label{eq:TrD2}.\ceq
We now show the property 
\oeq
\tr \ln (1+A) = \ln \det (1+A) 
\label{eq:trln}
\ceq
where $A$ is a  matrix which can be diagonalised.
{
In its diagonal form, $A_{ij} = a_i \delta_{ij}$, we have
$\tr A^n=\sum_i a_i^n.$
Then we get 
\oeqn
\tr \ln (1+A) &=& \tr (A -\frac{1}{2} A^2 + \frac{1}{3} A^3 - \cdots) \nonumber \\
&=& \sum_i (a_i -\frac{1}{2}  a_i^2 + \frac{1}{3} a_i^3 - \cdots )\nonumber \\
&=&  \sum_i \ln (1+a_i) \nonumber\\
&=& \ln \prod_i (1+a_i) \label{eq:trln1+A}\\
&=&\ln \det (1+A). 
\ceqn}
Taking the exponential of Eq.~(\ref{eq:trln}) with $A=e^{-M}$ and replacing in Eq.~(\ref{eq:TrD2})  gives Eq.~(\ref{eq:TrD}). 
 
 In the particular case where the system is described by a Slater determinant, the state is normalised ($\Tr \oD=1$) and one gets 
 \oeq
 m=\tr \ln (1+e^{-M}).
\label{eq:mz1}
 \ceq
 
 \section{Grassmann algebra \label{annexe:Grassmann}}

A brief introduction to Grassmann algebra can be found in p.~21-27 of Ref.~\cite{bla86}.
Here, we recall mostly the main relations which are used in appendix~\ref{annexe:formula}. 
Similarly to the imaginary number $i$ which has been introduced to satisfy $i^2=-1$, 
the Grassmann algebra has been developed such that its elements vanish when squared.
Elements $\{z_\al\}$ of the Grassmann algebra and their complex conjugated elements obey
\oeq
z_\al^2\equiv{z_\al^*}^2=0 
\label{eq:z_al}
\ceq
\oeq
z_\al z_\be + z_\be z_\al\equiv z_\al z_\be^* + z_\be^* z_\al=0.
\label{eq:z_alz_be}
\ceq
They are eigenvalues of annihilation and creation operators such that
\oeq
\oa_\al |z\>=z_\al|z\> \stf\mbox{ and }\stf  \<z|\oad_\al = \<z|z_\al^*
\label{eq:az}
\ceq
where $|z\>$ and $\<z|$ are coherent states of the form 
\oeq
|z\> = e^{\sum_\al z_\al\oad_\al}|-\> \stf \mbox{ and } \stf \<z|=\<-|e^{\sum_\al z_\al^*\oa_\al},
\label{eq:coh_state}
\ceq
and $|-\>$ is the particle vacuum. 

Elements of the Grassmann algebra are polynomial of second degree at most, as, for example, 
\oeq
P(z_\al,z_\al^*)=P_0+P_1z_\al+P_2z_\al^*+P_{12}z_\al z_\al^*.
\ceq
Derivation is defined  as
\oeqn
\overrightarrow{\partial}_{z_\al}P  &=& P_1+P_{12}z_\al^* \nonumber\\
\overrightarrow{\partial}_{z_\al^*}P  &=& P_2-P_{12}z_\al \nonumber\\
(\mbox{indeed }\overrightarrow{\partial}_{z_\al^*}z_\al z_\al^*&=&-(\overrightarrow{\partial}_{z_\al^*}z_\al^*) z_\al=-z_\al)\nonumber\\
\overleftarrow{\partial}_{z_\al}P  &=& P_1-P_{12}z_\al^* \nonumber\\
 \overleftarrow{\partial}_{z_\al^*}P  &=& P_2+P_{12}z_\al. 
\ceqn
We can show the following properties
\oeq 
\oad_\al|z\>=\overleftarrow{\partial}_{z_\al}|z\> \stf \mbox{ and } \stf \<z|\oa_\al=\overrightarrow{\partial}_{z_\al^*}\<z|.
\ceq
Integration is defined as
\oeq
\int dz_\al = \int dz_\al^* = 0
\label{eq:intdz}
\ceq
and
\oeq
\stf \int dz_\al z_\al= \int dz_\al^* z_\al^* = 1.
\ceq

The overlap of two coherent states reads
\oeq
\<z|z'\>=e^{\sum_\al z_\al^*z_\al'}=e^{\mathbf{z}^*\mathbf{z}'}.
\label{eq:zz'}
\ceq
{
Indeed, using  Eqs.~(\ref{eq:z_al}) and (\ref{eq:coh_state}),  and the Wick theorem, we have
\oeqn
\<z|z'\>&=&\<-|\sum_n\frac{1}{n!}\(\sum_\al z_\al^*\oa_\al\)^n\sum_m\frac{1}{m!}\(\sum_\be {z'_\be}\oad_\be\)^m|-\>\nonumber\\
&=& \sum_n\frac{1}{n!^2}\<-|\(\sum_\al z_\al^*\oa_\al\)^n\( \sum_\be z_\be'\oad_\be\)^n|-\>\nonumber\\
&=&1+\sum_{\al\be} z_\al^*\<-|\oa_\al\oad_\be|-\>z_\be'\nonumber\\
&&+\frac{1}{2!^2}\sum_{\al\be\ga\delta}z_\al^*z_\be^*\<-|\oa_\al\oa_\be\oad_\ga\oad_\delta|-\>z'_\ga z'_\delta+\cdots\nonumber\\
&=& \!1\!+\!\sum_{\al} z_\al^*z_\al'+\frac{1}{2!^2}\sum_{\al\be}(z_\al^*z_\be^*z'_\be z'_\al\!-\!z_\al^*z_\be^*z'_\al z'_\be)\!+\!\cdots\nonumber\\
&=& 1+\sum_{\al} z_\al^*z_\al'+\frac{1}{2!}\sum_{\al}z_\al^*z'_\al\sum_\be z_\be^*z'_\be +\cdots\nonumber\\
&=&e^{\sum_\al z_\al^*z'_\al}\nonumber
\ceqn
}
Eq.~(\ref{eq:zz'}) can be used to define the metric entering the following closure relation
\oeq
\int dz^*dz \sdf e^{-\mathbf{z}^*\mathbf{z}} |z\>\, \<z| = \hat{1},
\label{eq:zclosure}
\ceq
where we introduced the notation $ \int dz^*dz  \equiv \int \prod_\al dz_\al^*dz_\al$.

An operator $\oA(\oad,\oa)\equiv \oA(\cdots,\oad_\al,\cdots,\oa_\be,\cdots)$ may be represented by the differential operator $A$ defined as
\oeq
\<z|\oA(\oad,\oa)|z'\>= A(\overleftarrow{\partial}_{\mathbf{z}'},\mathbf{z}')\,\<z|z'\>
\label{eq:zAz'}
\ceq
or
\oeq
\<z|\oA(\oad,\oa)|z'\>= A(\mathbf{z}^*,\overrightarrow{\partial}_{\mathbf{z}^*})\,\<z|z'\>
\ceq
where we used the notation $\overleftarrow{\partial}_\mathbf{z}\equiv\{\cdots,\overleftarrow{\partial}_{z_\al},\cdots\}$ and $\overrightarrow{\partial}_\mathbf{z}\equiv\{\cdots,\overrightarrow{\partial}_{z_\al},\cdots\}$.

We now show the relation
\oeq
\int  dz^*dz\sdf e^{-\mathbf{z}^*M\mathbf{z}} = \det M
\label{eq:detM}
\ceq
{
Starting from the Taylor development
$$
\int  dz^*dz\sdf e^{-\mathbf{z}^*M\mathbf{z}} = \int  dz^*dz \sum_{n=0}^\infty \frac{1}{n!}(-\mathbf{z}^*M\mathbf{z})^n ,
$$
and using Eq.~(\ref{eq:intdz}), we see that the only non-vanishing term of the Taylor development is for $n=N$ where $N$ is the dimension of the Fock space:
$$
\int  dz^*dz\sdf e^{-\mathbf{z}^*M\mathbf{z}} = \frac{(-1)^N}{N!} \int  dz^*dz \sdf (\mathbf{z}^*M\mathbf{z})^N 
$$
with
$$ (\mathbf{z}^*M\mathbf{z})^N=\sum_{\al_1\be_1}z_{\al_1}^*M_{\al_1\be_1}z_{\be_1}\cdots\sum_{\al_N\be_N}z_{\al_N}^*M_{\al_N\be_N}z_{\be_N}.
$$
From Eq.~(\ref{eq:z_al}), we see that all terms with $\al_i=\al_{j\neq i}$ or $\be_i=\be_{j\neq i}$ vanish. It leads to two sums over all possible permutations of the $\al$ and of the $\be$ indices:
\oeqn
 (\mathbf{z}^*M\mathbf{z})^N=\sum_{\al,\be\in Perm_N}&&z_{\al(1)}^*M_{\al(1)\be(1)}z_{\be(1)}\cdots \nonumber\\
&&z_{\al(N)}^*M_{\al(N)\be(N)}z_{\be(N)}.\nonumber
\ceqn
Noting the following relations:
\oeqn
\int dz^*dz&=&(-1)^{N(N-1)/2} \nonumber\\
&&\int dz_1^*\cdots dz_N^* dz_1 \cdots dz_N,\nonumber \\
z^*_{\al(1)} z_{\be(1)} \cdots z^*_{\al(N)} z_{\be(N)}
&=& (-1)^{N(N+1)/2}\nonumber\\
&&z_{\be(1)}\cdots z_{\be(N)}z_{\al(1)}^*\cdots z_{\al(N)}^*,\nonumber\\
z_{\be(1)}\cdots z_{\be(N)} &=& (-1)^{N(N-1)/2} \mbox{sign}(\be) z_N\cdots z_1,\nonumber
\ceqn
and the similar equation for the $z_\al^*$, we get
$$
\int \sdb dz^*dz\sdf {z_{\al(1)}}^*z_{\be(1)}\cdots {z_{\al(N)}}^*z_{\be(N)}=(-1)^N \mbox{sign}(\al) \mbox{sign}(\be) 
$$
 where sign$(\al)$ is the sign of the permutation $\al$, i.e., +1 for an even number of permutations and -1 for an odd one. This leads to Eq.~(\ref{eq:detM})
\oeqn
 \int  dz^*dz e^{-\mathbf{z}^*M\mathbf{z}} &=& \frac{1}{N!}\sum_{\al,\be\in Perm_N} \mbox{sign}(\al) \mbox{sign}(\be)  \nonumber\\
&& \stf\stf\stf\stf\stf\stf\stf\stf\stf\stf\stf\stf M_{\al(1)\be(1)}\cdots M_{\al(N)\be(N)} \nonumber \\
&=&  \sum_{\al\in Perm_N} \mbox{sign}(\al)  M_{\al(1)1}\cdots M_{\al(N)N} \nonumber \\
&=& \det M.\nonumber
  \ceqn
}

}

\bibliographystyle{epj}
\bibliography{biblio}

\end{document}